\documentclass[AMA,STIX1COL]{WileyNJD-v2}

\articletype{Article Type}%

\received{}
\revised{}
\accepted{}

\usepackage{bm}		
\usepackage{subcaption}

\hypersetup{
    colorlinks=true,    
    linkcolor=blue,
    citecolor=blue,
    urlcolor=blue,
    bookmarks=true,
    pdfusetitle=true   
}

\definecolor{corr}{rgb}{0,0,0}

\begin{document}

\title{Robust interpolation for dispersed gas-droplet flows using statistical learning with the Fully Lagrangian Approach} 

\author[1]{C. P. Stafford*}

\author[1]{O. Rybdylova}

\authormark{C. P. Stafford \textsc{et al}}

\address[]{\orgdiv{Advanced Engineering Centre, School of Architecture, Technology and Engineering}, \orgname{University of Brighton}, \orgaddress{\state{Brighton BN2 4GJ}, \country{UK}}}

\corres{*\email{c.stafford@brighton.ac.uk}}


\abstract[Summary]{A novel methodology is presented for reconstructing the Eulerian number density field of dispersed gas-droplet flows modelled using the Fully Lagrangian Approach (FLA). In this work, the nonparametric framework of kernel regression is used to accumulate the FLA number density contributions of individual droplets in accordance with the spatial structure of the dispersed phase. The high variation which is observed in the droplet number density field for unsteady flows is accounted for by using the Eulerian-Lagrangian transformation tensor, which is central to the FLA, to specify the size and shape of the kernel associated with each droplet. This procedure enables a high level of structural detail to be retained, and it is demonstrated that far fewer droplets have to be tracked in order to reconstruct a faithful Eulerian representation of the dispersed phase. Furthermore, the kernel regression procedure is easily extended to higher dimensions, and inclusion of the droplet radius within the phase space description using the generalised Fully Lagrangian Approach (gFLA) additionally enables statistics of the droplet size distribution to be determined for polydisperse flows. The developed methodology is applied to a range of one-dimensional and two-dimensional steady-state and transient flows, for both monodisperse and polydisperse droplets, and it is shown that kernel regression performs well across this variety of cases. A comparison is made against conventional direct trajectory methods to determine the saving in computational expense which can be gained, and it is found that $10^3$ times fewer droplet realisations are needed to reconstruct a qualitatively similar representation of the number density field.}



\maketitle


\section{Introduction}

Gas-droplets flows are ubiquitous within science and engineering, and are central to many industrial applications and environmental phenomena, including facilitating the decarbonisation of propulsion systems for transportation \cite{Costa2018}, determining the spread of airborne droplets containing a pathogen \cite{Bourouiba2014}, and modelling droplet condensation within atmospheric clouds \cite{Sidin2009}. Of primary interest in the characterisation and description of such systems is the variation of droplet number density throughout the flows, with this being a key determinant in physical factors such as reaction rates and probability of infection. The mechanisms that are responsible for the variation in droplet number density have received much attention, with studies considering both the influence of the carrier flow structures and droplet physics upon the clustering and dispersion of droplets.

Whilst experimental studies continue to be an important informant of droplet behaviour through the provision of quantities such as the distribution of droplet velocities and sizes, it is useful to be able to study specific flow configurations without having to use an experimental setup. Mathematical and numerical modelling is able to fulfil this role with the aid of computational simulations, and provides a wealth of detailed information that can be difficult to obtain experimentally, such as knowledge of the instantaneous droplet number density field. Such modelling is able to provide detailed insight into the physical phenomena which occur in droplet-laden flows, and constitutes a topic of growing research interest.

Within the realm of modelling gas-droplet flows two distinct approaches exist for handling the droplet phase. The first treats the droplets as a continuum in a similar manner to the gas carrier flow, and solves a set of coupled partial differential equations which govern the evolution of the droplet phase field variables, specifically the number density, mean velocity, and kinetic stresses. By its nature, the resolution of this approach is limited by the fidelity of the computational grid upon which the equations are solved, and consequently the resultant Eulerian description of the droplet phase is restricted in the level of detail that can be represented \cite{Laurent2012}. In contrast, droplets can instead be treated individually, with the equation of motion for each droplet being solved along their separate Lagrangian trajectories. This method is able to capture a considerable level of detail, with the behaviour of droplets in different regions of the flow being accurately accounted for through their individual treatment, however the drawback to the Lagrangian approach lies in construction of the Eulerian droplet number density field from the trajectories. In order to obtain statistical convergence of the resulting Eulerian description, the number of droplets required is often prohibitive, being of the order of $\mathcal{O}(10^3)$ in every grid cell for a well resolved flow \cite{Healy2005}.

A point of contention between the Eulerian and Lagrangian descriptions of the droplet phase is in the capturing of important physical phenomena such as the crossing trajectories effect \cite{Yudine1959}, in which the non-negligible inertia of droplets can cause their trajectories to intersect in physical space. This has the consequence of the droplet phase field variables becoming multi-valued; a consideration that is important to account for in the modelling approach. Such instances of multi-valuedness can be observed in the structure of the dispersed phase continuum where they manifest as the envelope of intersecting trajectories, which is referred to as a \textit{fold}. Eulerian descriptions of the droplet field are by construction single-valued at a given point in space, and therefore unable to include the full detail of the crossing trajectories effect in their standard form. To address this, studies have sought to account for this behaviour directly within the Eulerian representation of the droplet phase field variables by means of developments such as a family of moment methods \cite{Laurent2012} and multi-fluid models \cite{DeChaisemartin2009}. In contrast, the Lagrangian description of droplets along their trajectories is naturally able to handle the intersection of trajectories, and therefore also the ensuing multi-valued nature of the droplet field, making it an ideal candidate for the development of further models.

To utilise the detail intrinsic to the Lagrangian description and address the issue of needing to simulate prohibitively high numbers of droplets, a body of work has been developed upon the assumption that the droplet phase can also be treated as a continuum with no self-stresses \cite{Marble1970}. This has been pioneered by Osiptsov \cite{Osiptsov2000}, and involves tracking the number density along individual trajectories by utilising the conservation of mass in Lagrangian form, a procedure known as the Fully Lagrangian Approach (FLA) (some works refer to this as the Fully Lagrangian Method, or Osiptsov's Method). A number of subsequent studies have built upon this foundation, including its application to a gas flow through a turbine \cite{Healy2005}, examination of the moments of droplet number density \cite{IJZERMANS2010}, comparison to experimental spray data \cite{Zaripov2017}, applicability to turbulent flows \cite{Papoutsakis2018}, inclusion of momentum coupling effects \cite{Rybdylova2019}, and most recently, extension to the consideration of polydisperse flows \cite{Li2021}. Despite these advances, the FLA remains a developing concept which presents a variety of challenges within its use, chief of which is the accurate reconstruction of the droplet number density field from trajectory data.

In this paper, the use of a statistical learning approach is proposed to accumulate the contributions from individual droplets, and reproduce the Eulerian number density field in a robust manner which is applicable across a range of different flows. To accomplish this, kernel regression is used, with the averaged nature of this procedure enabling it to account for the multi-valued nature of the contributions along folds, and thereby construct a single-valued representation of the number density field. This extends the linear interpolation approach used in a previous study \cite{Li2021}, and makes accurate representation of the dispersed phase-field possible for a wider class of flows. The developed methodology is applied to a range of one-dimensional and two-dimensional steady-state and transient flows, for both monodisperse and polydisperse droplets, and it is shown that kernel regression performs well across this variety of cases.

The remainder of the paper is structured as follows. Section \ref{sec:fla} outlines the formulation of the Fully Lagrangian Approach, and details the assumptions involved. Section \ref{sec:interpolation} introduces the concept of kernel regression, how it can be applied to the calculation of  number density, and how the information provided by the Fully Lagrangian approach can be used to ensure contributions from trajectories are accumulated in a physically consistent manner. It is further demonstrated how the procedure can be generalised into a higher-order phase space that includes droplet radius to enable modelling of polydisperse flows. Additionally, aspects of the computational implementation are detailed. Section \ref{sec:results} describes application of the kernel regression approach to a wide class of flow configurations and presents the numerical results. Section \ref{sec:conclusions} appraises the observed simulation results, and discusses the potential of the procedure for application to more complex flows in terms of computational efficiency.

\section{The Fully Lagrangian Approach} \label{sec:fla}

\subsection{General formulation for polydisperse droplets} \label{sec:gfla}

Representation of the dispersed phase in a Lagrangian sense is considered through the use of an appropriate governing equation for individual droplets or particles. Henceforth, in this paper droplets are taken to be liquid and can be subject to evaporation, whilst particles are assumed to be solid and non-evaporating; the more general case of droplets is usually focused upon. In this work, droplets are considered to be spherical, within a dilute suspension, and are also assumed to act as point masses. Under these conditions, droplets with position $\bm{x}_d(t)$, velocity $\bm{v}_d(t)$ and radius $r_d(t)$ are modelled individually along trajectories according to the arbitrary governing laws and initial conditions at time $t_0$
\begin{subequations}
    \label{eq:part-rad-eom}
    \begin{align}
        \ddot{\bm{x}}_d(t) & = \bm{f} (\bm{x}_d(t),\bm{v}_d(t),r_d(t),t) \, ,
        & \bm{x}_d(t_0) = \bm{x}_0 \, , \, \dot{\bm{x}}_d(t_0) = \bm{v}_0 \, ,
        \label{eq:part-eom} \\
        \dot{r}_d(t) & = \varphi (\bm{x}_d(t),\bm{v}_d(t),r_d(t),t) \, ,
        & r_d(t_0) = {r}_0 \, ,
        \label{eq:rad-eom}
    \end{align}
\end{subequations}
where the subscript $d$ denotes variables that are sampled along the Lagrangian droplet trajectories, $\bm{f}(\bm{x},\bm{v},{r},t)$ is the force per unit mass acting upon a droplet, and $\varphi(\bm{x},\bm{v},{r},t)$ is the rate of radial change for a droplet with Eulerian position $\bm{x}$, velocity $\bm{v}$, and radius ${r}$ at time $t$, and droplets have initial position $\bm{x}_0$, velocity $\bm{v}_0$, and radius ${r}_0$ at time $t_0$. Without loss of generality, both $\bm{f}$ and $\varphi$ are assumed to be dependent upon all of $\bm{x}$, $\bm{v}$, ${r}$, and $t$, although $\varphi$ may also be dependent upon other thermodynamic parameters which are related to the droplet evaporation.

The FLA is based on the assumption that the dispersed phase can be represented as a continuum \cite{Osiptsov2000}, with trajectories characterised by their initial position $\bm{x}_0$ and time $t$. Recent work has generalised this concept by extending the phase space to include the dependence on initial droplet radius ${r}_0$, referred to as the \textit{generalised Fully Lagrangian Approach} (gFLA) \cite{Li2021}. Then the dispersed phase probability density $p (\bm{x}_d,r_d,t)$ can be interpreted as the number density in this wider phase space, and is calculated directly along the trajectories of individual droplets by considering the Lagrangian form of conservation of mass \cite{Osiptsov2000}
\begin{equation} \label{eq:Lagrangian-COM}
p (\bm{x}_d,r_d,t) = \frac{p (\bm{x}_0,{r}_0,t_0)}{ \lvert \det (\bm{J} (\bm{x}_0,{r}_0,t)) \rvert} \, ,
\end{equation}
where the Jacobian tensor $\bm{J} (\bm{x}_0,{r}_0,t)$ is specified in block matrix form as \cite{Li2021}
\begin{equation} \label{eq:Jacobian-gFLA}
\bm{J} (\bm{x}_0,{r}_0,t) =
\left[
\begin{array}{cc}
\bm{J}^{\bm{x}\bm{x}} & \bm{J}^{\bm{x}{r}} \\
\bm{J}^{r\bm{x}} & {J}^{r{r}}
\end{array}
\right]
=
\left[
\fontsize{14}{12}
\begin{array}{cc}
\vspace{0.1em}
\frac{ \partial \bm{x}_d}{\partial \bm{x}_0} &
\frac{\partial \bm{x}_d}{\partial {r}_0}
\vspace{0.1em}
\\
\frac{\partial r_d}{\partial \bm{x}_0} &
\frac{\partial r_d}{\partial {r}_0} \\
\end{array}
\right]
\, .
\end{equation}
The procedure for deriving Eq.~\eqref{eq:Lagrangian-COM} directly from the behaviour of individual droplets described by Eqs.~\eqref{eq:part-rad-eom} is outlined in Appendix \ref{sec:gfla-derivation}, along with the implications of the physical assumptions involved. The physical interpretation of $\bm{J}$ is the local elemental deformation along a trajectory with respect to its initial state at $(\bm{x}_0,{r}_0,t)$. This is equivalent to the Eulerian-Lagrangian transformation that describes the influence of variables transported by droplets on the local Eulerian field, meaning that $\bm{J}$ is able to provide information about the phase space deformation of the droplet field. The blocks of the Jacobian given by Eq.~\eqref{eq:Jacobian-gFLA} represent this deformation in the associated part of phase space; $\bm{J}^{\bm{x}\bm{x}}$ is the spatial deformation given by the standard FLA, ${J}^{rr}$ is the deformation in radial space, and the remaining blocks provide the corresponding cross components. The evolution equations for each block can be obtained by taking partial derivatives of the governing equations \eqref{eq:part-rad-eom} with respect to $(\bm{x}_0,{r}_0)$, yielding the system
\begin{subequations}
    \label{eq:Jacobian-evolution}
    \begin{align}
        \ddot{\bm{J}}^{\bm{x}\bm{x}} & = \frac{\partial \bm{f}_d}{\partial \bm{x}} \cdot \bm{J}^{\bm{x}\bm{x}}
        + \frac{\partial \bm{f}_d}{\partial \bm{v}} \cdot \dot{\bm{J}}^{\bm{x}\bm{x}}
        + \frac{\partial \bm{f}_d}{\partial {r}} \bm{J}^{r\bm{x}}
        \, ,
        \label{eq:Jacobian-evolution-physical-space}
        \\
        \ddot{\bm{J}}^{\bm{x}{r}} & = \frac{\partial \bm{f}_d}{\partial \bm{x}} \cdot \bm{J}^{\bm{x}{r}}
        + \frac{\partial \bm{f}_d}{\partial \bm{v}} \cdot \dot{\bm{J}}^{\bm{x}{r}}
        + \frac{\partial \bm{f}_d}{\partial {r}} {J}^{r{r}}
        \, ,
        \\
        \dot{\bm{J}}^{r\bm{x}} & = \frac{\partial \varphi_d}{\partial \bm{x}} \cdot \bm{J}^{\bm{x}\bm{x}}
        + \frac{\partial \varphi_d}{\partial \bm{v}} \cdot \dot{\bm{J}}^{\bm{x}\bm{x}}
        + \frac{\partial \varphi_d}{\partial {r}} \bm{J}^{r\bm{x}}
        \, ,
        \\
        \dot{{J}}^{r{r}} & = \frac{\partial \varphi_d}{\partial \bm{x}} \cdot \bm{J}^{\bm{x}{r}}
        + \frac{\partial \varphi_d}{\partial \bm{v}} \cdot \dot{\bm{J}}^{\bm{x}{r}}
        + \frac{\partial \varphi_d}{\partial {r}} {J}^{r{r}}
        \, .
    \end{align}
\end{subequations}
The general form of the initial conditions at time $t_0$ pertaining to Eqs.~\eqref{eq:Jacobian-evolution} are
\begin{subequations}
    \label{eq:Jacobian-initial-conditions}
    \begin{align}
        \bm{J}^{\bm{x}\bm{x}}(\bm{x}_0,{r}_0,t_0) & = \bm{I} \, ,
        & \dot{\bm{J}}^{\bm{x}\bm{x}}(\bm{x}_0,{r}_0,t_0) & = \partial \bm{v}_0 / \partial \bm{x}_0 \, , \\
        \bm{J}^{\bm{x}{r}}(\bm{x}_0,{r}_0,t_0) & = \bm{0} \, , & \dot{\bm{J}}^{\bm{x}{r}}(\bm{x}_0,{r}_0,t_0) & = \bm{0} \, , \\
        \bm{J}^{r\bm{x}}(\bm{x}_0,{r}_0,t_0) & = \bm{0} \, , \\
        {J}^{r{r}}(\bm{x}_0,{r}_0,t_0) & = 1 \, ,
    \end{align}
\end{subequations}
where $\bm{I}$ is the identity matrix. The system \eqref{eq:Jacobian-evolution} then describes the evolution of the blocks of $\bm{J}$ whilst accounting for the coupling that exists between the different blocks. Note that this coupling only occurs between blocks in the same column of the Jacobian in Eq.~\eqref{eq:Jacobian-gFLA}, with the different columns being associated with independent systems of equations. This formulation can also be extended to provide a more complete description of the droplet thermophysical behaviour by including temperature and mass as phase space variables instead of only the droplet radius. The initial conditions \eqref{eq:Jacobian-initial-conditions} are found by applying the Jacobian as defined in Eq.~\eqref{eq:Jacobian-gFLA} to the initial conditions associated with the governing equations \eqref{eq:part-rad-eom}. It is assumed here that the initial droplet position $\bm{x}_0$ and velocity $\bm{v}_0$ are independent of radius ${r}_0$, and further that the initial droplet radius ${r}_0$ is also independent of position $\bm{x}_0$. Following \cite{Healy2005}, however, it is noted that the initial droplet velocity $\bm{v}_0$ varies with position $\bm{x}_0$ depending upon both the nature of the carrier flow and how droplets are introduced into the flow, and identification of the correct initial condition for $\partial \bm{v}_0 / \partial \bm{x}_0$ is crucial to obtaining the correct behaviour for the evolution of $\bm{J}$.

For the standard FLA, in which the droplet radius $r_d$ is not considered, the probability density $p (\bm{x},r,t)$ reduces to the number density $n (\bm{x},t)$, which is a function of only physical space $\bm{x}$. For the monodisperse cases in Section \ref{sec:results-mono} the analysis is accordingly focused upon $n (\bm{x},t)$, whilst the polydisperse flows in Section \ref{sec:results-poly} consider the more general case of $p (\bm{x},r,t)$.

\subsection{Formulation for simplified physical models}

Whilst the gFLA formulation in Section \ref{sec:gfla} is applicable to general equations of motion and evaporation for spherical droplets modelled as point masses in a dilute suspension, by further assuming that the carrier flow density is much less than that of the dispersed phase, the applicability is then restricted to gas-droplet flows. Making the additional assumption of a low droplet Reynolds number enables the droplet momentum to be modelled using a linear drag law. For a simplified physical model of evaporation, all heat at the droplet surface is taken to be spent on evaporation, then the droplet evaporation rate $\varphi = \varphi ({r},t)$ is only dependent on the radius ${r}$, and independent of position $\bm{x}$ and velocity $\bm{v}$. Under these conditions, the associated nondimensional forms of $\bm{f}$ and $\varphi$ from Eqs.~\eqref{eq:part-rad-eom} are given by
\begin{subequations}
    \label{eq:simplified-models}
    \begin{align}
        \bm{f}(\bm{x},\bm{v},{r},t) & = \frac{1}{St_0^* \, {r}^2} \left( \bm{u}(\bm{x},t) - \bm{v} \right) \, , \\
        \varphi ({r},t) & = - \frac{\delta}{2 {r}} \, ,
    \end{align}
\end{subequations}
where $St_0^*$ is a reference Stokes number corresponding to droplets of characteristic radius $r_{d0}^*$, $\bm{u}(\bm{x},t)$ is the carrier flow velocity and $\delta$ is the rate of change of the droplet surface area. In this case the equations of evolution for system \eqref{eq:Jacobian-evolution} simplify to
\begin{subequations}
    \label{eq:Jacobian-evolution-simplified}
    \begin{align}
        \ddot{\bm{J}}^{\bm{x}\bm{x}} & = \frac{1}{St_0^* \, r_d^2} \frac{\partial \bm{u}_d}{\partial \bm{x}} \cdot \bm{J}^{\bm{x}\bm{x}}
        - \frac{1}{St_0^* \, r_d^2} \dot{\bm{J}}^{\bm{x}\bm{x}}
        - \frac{2}{St_0^* \, r_d^3} \left( \bm{u}_d - \dot{\bm{x}}_d \right) \cdot \bm{J}^{r\bm{x}}
        \, ,
        \\
        \ddot{\bm{J}}^{\bm{x}{r}} & = \frac{1}{St_0^* \, r_d^2} \frac{\partial \bm{u}_d}{\partial \bm{x}} \cdot \bm{J}^{\bm{x}{r}}
        - \frac{1}{St_0^* \, r_d^2} \dot{\bm{J}}^{\bm{x}{r}}
        - \frac{2}{St_0^* \, r_d^3} \left( \bm{u}_d - \dot{\bm{x}}_d \right) {J}^{r{r}}
        \, ,
        \\
        \dot{\bm{J}}^{r\bm{x}} & = \frac{\delta}{2 r_d^2} \bm{J}^{r\bm{x}} \, ,
        \\
        \dot{{J}}^{r{r}} & = \frac{\delta}{2 r_d^2} {J}^{r{r}} \, ,
    \end{align}
\end{subequations}
where $\bm{u}_d (t) = \bm{u} (\bm{x}_d(t),t)$ and  $\partial \bm{u}_d (t) / \partial \bm{x} =  \partial \bm{u} (\bm{x}_d(t),t) / \partial \bm{x}$ respectively denote that the fluid velocity and fluid velocity gradient are evaluated along the droplet trajectories. The initial conditions for the system \eqref{eq:Jacobian-evolution-simplified} remain the same as those given by Eqs.~\eqref{eq:Jacobian-initial-conditions}.

\subsection{Computational advantage}

The most salient advantage of using the FLA framework to compute the number density field is the increase in computational efficiency over conventional box-counting methods which can be realised, principally due to effective use of the information provided by the Eulerian-Lagrangian transformation. This aspect of the method has been focused upon in several works, and stems from the observation that an accuracy of $< 0.1\%$ in calculation of the Eulerian number density field is only achieved using box-counting methods when $\mathcal{O}(10^4)$ droplets are present in each Eulerian cell within a simulation \cite{Healy2005,Papoutsakis2018}. In contrast, due to the ability of the FLA to provide the number density along trajectories, and the facet of the Eulerian-Lagrangian transformation that trajectory data can be meaningfully extrapolated onto the local spatial region, the FLA only requires one trajectory per Eulerian cell to achieve the same level of accuracy \cite{Papoutsakis2018}. This advantage means that despite the additional computation of the Jacobian as it evolves along each trajectory according to Eqs.~\eqref{eq:Jacobian-evolution}, the FLA is still approximately 20 times more efficient than box-counting methods for two-dimensional simulations \cite{Healy2005,Zaripov2017}. Another analysis found that a box-counting approach for number density requires $\sim10^6$ trajectories to be computed, whereas the FLA only needs $\sim10^3$ trajectories to obtain the same level of detail, however this is offset by the additional expense of calculating the Jacobian along trajectories which entails a fortyfold increase in computational cost for three-dimensional simulations, and also by linear interpolation of the trajectory number density data onto the Eulerian grid which takes almost as long as the numerical integration along trajectories \cite{Mishchenko2019}. Therefore, despite providing a $10^3$ saving on the number of trajectories computed when compared against the box-counting approach, the overall efficiency improvement of the FLA for calculation of the Eulerian number density field is reduced to around tenfold. It is also noted that in implementations of the FLA, neglecting droplet collisions in the dilute phase limit results in the independence of trajectories, which further enables effective parallelisation and computational efficiency \cite{Govindarajan2011}.

\subsection{Numerical considerations}

The gFLA formulation in Section \ref{sec:gfla} is valid for the regime of dilute droplet flows, and makes the important assumption that the droplet field is continuous between trajectories and exhibits smooth spatial variation. Such an assumption is, however, at odds with the multi-valued nature that characterises the trajectory crossing experienced by inertial droplets. The strength of the standard FLA formulation is that it is able to retain this detail that is inherent to individual trajectories within its description of the spatial droplet field, however this is manifested in the elemental volume of the Eulerian-Lagrangian transformation, given by $\det (\bm{J}^{\bm{x}\bm{x}} (\bm{x}_0,t))$, becoming zero at the point of trajectory crossings \cite{Healy2005}. By virtue of Eq.~\eqref{eq:Lagrangian-COM}, the droplet number density is then singular at these points. This phenomenon is in fact necessary to maintain consistency with the assumption of continuity for the dispersed phase, and is simply how the FLA framework incorporates the information of crossing trajectories into the concept of a smoothly defined field. It has further been demonstrated in previous work that such singularities in number density are integrable \cite{Osiptsov1984}, meaning that the local Eulerian number density field within a defined region remains finite \cite{Papoutsakis2018}.

Notwithstanding the fact that the FLA presents a theoretically sound approach for number density calculation, in practice the issue lies in how to accumulate the multi-valued number density contributions along trajectories into an Eulerian field representation in a physically correct manner. An interpolation procedure is only valid within a region of the droplet field which is single-valued, meaning that careful treatment of the multi-valued regions is paramount to ensuring that the calculation of an Eulerian number density field is meaningful. Previous works have addressed this by using the fact that the sign of the Jacobian determinant ${J}$ changes every time a droplet crosses the trajectory of another droplet, effectively providing a means of keeping track of which layer of the droplet field a given trajectory is in. Since each layer of the droplet field is single-valued with non-intersecting trajectories, interpolation can be used to calculate the Eulerian number density within each individual layer, and then due to the number density field being additive, the contributions from each layer at a given point are summed together to obtain the total number density \cite{Papoutsakis2018,Mishchenko2019,Lebedeva2013}. This requires that droplets are indexed in such a way that a distinction is made every time a fold is crossed, with the most straightforward means of doing this being to keep count of the number of times this occurs for each droplet, then all droplets with the same count index form one layer of the droplet field \cite{Lebedeva2013}. Such a procedure enables the reproduction of a number density field which although single-valued, is able to contain the detail of different layers within the droplet field; a feature which is difficult to replicate in Eulerian-based simulations \cite{Laurent2012}. The drawback of indexing droplets in this manner is that sufficiently many need to be present within each layer of the droplet field for the interpolation methods to be well defined and able to work correctly. In practice this is only an issue when trajectories initially cross and there are few droplets within the newly created layer of the droplet field, however this still presents a situation which requires appropriate treatment to ensure that the contribution of these droplets is accounted for in the number density calculation. Notwithstanding this, the physical accuracy which can be achieved through the use of a Lagrangian approach highlights the potential ability of the FLA to accurately simulate industrially relevant systems.

\section{Interpolation of the number density field} \label{sec:interpolation}

The primary point of contention in developing useful computational tools which make use of the FLA is that the number density provided by the method is along trajectories, whereas in many numerical methods and applied contexts it is most often the Eulerian field information which is required. In this sense, the scattered number density data which the FLA produces requires the use of an appropriate interpolation procedure for accumulation onto an Eulerian grid. It has been noted that simple linear interpolation is sufficient for a steady-state flow configuration, however for transient flows, that contain vortices, a more comprehensive accumulation algorithm is required \cite{Govindarajan2011}. This is a subject which has received surprisingly little attention to date, yet has the potential to open up the FLA methodology to a wider and more general class of flows.

\subsection{Existing procedures for number density calculation} \label{sec:existing-interpolation}
The classical box-counting approach to calculating the number density field in dispersed multiphase flows simply involves the accumulation of mass contributions from all the droplets within an Eulerian cell, then weighting them by the cell volume. This is also referred to as the nearest neighbour method \cite{hastie2009elements}, and can be interpreted as reassigning the contribution from a droplet to act at the centre of the cell in which it is located. Such an approach is associated with a discontinuous droplet weighting function, since it takes no account of how far a droplet is located from the cell centre, meaning that this information is lost in the accumulation process, and resulting in the generation of artificial noise. It is this noise which can cause a high variation in the computed number density field, and necessitates the requirement for such a high number of droplets using the box-counting approach in order to produce a stable value for the number density.

An improvement to the box-counting approach is achieved by considering a droplet weighting function with $C^{0}$ continuity which is able to account for the location of a droplet relative to the Eulerian grid on which the accumulation of contributions is made. One procedure that utilises such a methodology is the Cloud-In-Cell (CIC) approach \cite{Laux1996}, in which the weighting function takes the form of a triangular-shaped kernel. This is equivalent to assigning contributions from a droplet to the corner nodes of the Eulerian cell in which it resides by using linear interpolation, which importantly ensures that the constraints of continuity and momentum conservation are respected by the accumulation procedure, whilst also maintaining numerical stability. The CIC approach is a widely adopted method of accumulating contributions in dispersed multiphase flow simulations due to its simplicity of implementation, and, owing to the fact that the weighting function retains $C^{0}$ continuity of the droplet field, requires fewer droplets than the nearest neighbour method to achieve a stable result for the number density field on an identical Eulerian grid. The domain of influence for the droplet weighting function for a Cloud-In-Cell approach is, however, naturally limited to the Eulerian cell in which the droplet is located. This means that many droplets per cell are still required to achieve statistical convergence of the number density field, and the computational expense associated with this can be prohibitive within simulations of industrial scale processes.

Within the broad class of meshfree methods which exist for interpolation of scattered data, kernel-based methods have seen a variety of applications, in particular for fluid dynamics in the form of smoothed particle hydrodynamics (SPH) \cite{Price2012}. Whilst the rationale for SPH is well established, the normalisation condition required of the weighting kernel is such that a sufficient number of particles must be inside the kernel for the accumulation procedure to remain valid, which has implications for the simulation of compressible flows in which the spatial distribution of particles is non-uniform. Although the SPH framework has been extended to handle weakly-compressible multiphase flows through the use of adaptive spatial resolution, this has only been shown to be accurate in the case where the density ratio between the two phases is less than 10 \cite{Wang2016a}, whilst typical gas-droplet flows have a much higher density ratio. Consequently the ability to effectively treat the clusters and voids that are characteristic of dispersed multiphase flow using the SPH framework remains limited.

Previous work has highlighted the ability of Vorono\"{i} tessellations to calculate the droplet number density field from droplet location data, by recognising that the area of the Vorono\"{i} cells is the inverse of the local droplet number density \cite{Monchaux2010}. Notably, it was demonstrated that it is possible to identify regions of clusters and voids within the droplet field through categorisation of the area of Vorono\"{i} cells. This methodology enables information about the Lagrangian dynamics of individual droplets to be included within the representation of the Eulerian number density field, meaning that a certain level of detail about the structures in the droplet field can be retained. However, since the influence of a given droplet is taken to be uniform over its associated Vorono\"{i} cell, the resulting Eulerian number density field is discontinuous along the Vorono\"{i} cell boundaries. In this sense, the construction of the number density field using Vorono\"{i} tessellations requires a sufficiently large number of droplets to obtain a smooth result, an aspect shared with the CIC approach. Despite this, Vorono\"{i} tessellations have been incorporated into a solver as a full hydrodynamics scheme \cite{Serrano2005}, and subsequently developed into Vorono\"{i} particle hydrodynamics (VPH) \cite{Hess2010}. The advantages offered by such a solver are however offset by the additional complexity of the formulation, and need to generate the tessellation for the entire droplet field at every point in time.

An option which mirrors the thinking behind the use of Vorono\"{i} tessellations is that of the Delaunay triangulation, with these two procedures being closely related as the dual graphs of each other. This is motivated by the use of Delaunay triangulations in scattered data interpolation, with the triangulation effectively providing a mesh over which commonly employed interpolation procedures can be used between the known values at data points. The desirable property unique to the Delaunay triangulation is that it is constructed to satisfy the empty circumcircle criterion, which ensures maximization of the internal angles of triangles, with the reasoning that interpolation routines will exhibit greater accuracy and stability over more uniformly shaped triangles. These aspects have seen the Delaunay triangulation developed as a means of constructing density fields from scattered data points \cite{Schaap2000}, and extended to a hydrodynamical method with the Delaunay tessellation field estimator (DTFE) technique \cite{Pelupessy2003}. However, the shortcoming of such an approach is that the Delaunay triangulation necessarily changes discontinuously at certain instants due to the empty circumcircle criterion \cite{Hess2010}, meaning that the DTFE is not well suited as the basis of numerical solver.

\subsection{Requirements of a suitable interpolation procedure}

The immediate advantage offered by the FLA methodology is that the exact value of the droplet number density field is already known along trajectories before the use of any interpolation procedure, thereby eliminating the noise that is inherent in procedures such as nearest neighbour or CIC interpolation. The challenge then remains to find an appropriate means of interpolating these values that will retain sufficient accuracy in describing the structures of the droplet field. Existing work has demonstrated that linear interpolation between neighbouring trajectories is able to capture the smooth, and often monotonic, variation of the number density field in steady-state flows well, however the considerably more complex structure of the droplet field in transient flows, for which folds occur and multiple layers are inherent, means that such an approach is not applicable \cite{Li2021}. To effectively extend the concept of linear interpolation to transient flows requires the establishment of a more general framework.

Whilst the ability to extend the Delaunay triangulation to points in $n$-dimensional space does makes it a possible method for interpolating number densities obtained using the gFLA onto a mesh that spans both physical and radial space, using a triangulation-based approach to interpolate FLA data has a number of distinct disadvantages. Principally, the computational expense of having to construct a triangulation over all droplets within each layer of the droplet field, when coupled with the subsequent interpolation procedure of the number density data, would likely negate most of the saving offered by computing the number density along trajectories for the fewer droplets required using the FLA. To compound this, triangulation of droplets would not be able to distinguish and properly resolve voids in the droplet field when using linear interpolation within triangles, which would necessitate the use of a higher order interpolation method that is able to capture the spatial gradient of the number density field, and thereby further increase the computational expense. Finally, whilst knowledge of the Jacobian tensor provides information about the Eulerian-Lagrangian transformation of trajectory variables onto the local Eulerian field for each droplet, and thereby detail of the structures within the droplet field, a triangulation-based procedure is unable to utilise this information in its reconstruction of the number density field.

The potential drawbacks of triangulation-based methods for accumulating the FLA number density data help to elucidate the desirable properties that a suitable interpolation routine should possess. To maintain consistency with the Lagrangian nature of the FLA methodology, a meshfree method which accumulates contributions without prior knowledge of the spatial relationship between droplets would be most appropriate in terms of numerical efficiency, and realising the optimal computational saving which can be achieved using the FLA. In this sense, the contribution of a given droplet should not be dependent on that of other droplets, and the procedure for the accumulation of droplet contributions should also be valid regardless of the number of droplets that contribute to the Eulerian field at a given point. Such a requirement is further justified in the context of the FLA by considering that since droplet contributions are first accumulated within layers of the droplet field, then interpolating between droplets becomes an issue when a layer contains sufficiently low numbers of droplets, and is not possible in the case when the number of droplets is less than or equal to the dimensionality of the simulation. In particular, it is possible that a certain layer may contain only a single droplet, however it is still desirable that the contribution from this droplet is included in the reconstruction of the Eulerian number density field.

Additionally, an interpolation scheme should be able to account for the fact that the number density field for inertial droplets within transient flows has defined clusters and voids, and in particular that droplets in voids can be sparse. It is therefore desirable that the influence of those droplets that are present in voids is able to extend over the surrounding domain to a greater degree, in order to reflect that there are fewer other droplets in the immediate vicinity from which contributions to the Eulerian droplet field can be accumulated. The nature of this demand is most appropriately met through the use of a meshfree method in which a degree of control over the domain of influence that individual droplet contributions have can be exercised.

\subsection{Kernel regression}

Within the context of accumulating the number density data obtained from the FLA, the problem of capturing clusters and voids in the droplet field at a competitive computational cost is addressed here through the means of kernel regression. The motivation behind this choice is that, as a form of statistical learning, the averaged nature of such an approach enables it to effectively handle spatial locations at which only a single data point makes a contribution. This is seen by considering the form of the Nadaraya-Watson estimator \cite{hastie2009elements}
\begin{equation} \label{eq:kernel-smoother}
{{n}}(\bm{x},t) = \frac{\sum_{i=1}^{N} K_{\bm{H}}(\bm{x},\bm{x}_d^i) n(\bm{x}_d^i,t)}{\sum_{j=1}^{N} K_{\bm{H}}(\bm{x},\bm{x}_d^j)} \, ,
\end{equation}
in which $n(\bm{x}_d^i,t)$ is the instantaneous number density along the trajectory $\bm{x}_d$ associated with droplet $i$ at time $t$ as obtained from the FLA, ${{n}}(\bm{x},t)$ is the Eulerian number density field, $K_{\bm{H}}(\bm{x},\bm{x}_d^i)$ is a kernel which provides a weighting for the contribution of the droplet $\bm{x}_d^i$ to the accumulation point $\bm{x}$, and $N$ is the number of droplets for which the range of support of the associated kernel includes the point $\bm{x}$. In Eq.~\eqref{eq:kernel-smoother}, the numerator provides the weighted sum of the Lagrangian number density contributions $n(\bm{x}_d^i,t)$ at $\bm{x}$, whilst the denominator is equal to the sum of the weighting factors given by the kernels that contribute at $\bm{x}$. Consequently, in the accumulation procedure for ${{n}}(\bm{x},t)$ it is seen that the contributions from $n(\bm{x}_d^i,t)$ are effectively weighted by weights $w^i$ defined by
\begin{equation} \label{eq:kernel-weights}
w^i = \frac{K_{\bm{H}}(\bm{x},\bm{x}_d^i)}{\sum_{j=1}^{N} K_{\bm{H}}(\bm{x},\bm{x}_d^j)} \, .
\end{equation}
In this sense, the Nadaraya-Watson estimator \eqref{eq:kernel-smoother} can be considered as an expansion in renormalised radial basis functions $w^i$ \cite{hastie2009elements}, and it is the appearance of the denominator in Eq.~\eqref{eq:kernel-weights} which enables a representative value for ${{n}}(\bm{x},t)$ to be obtained even in the case when $N = 1$. This circumvents the requirement for a specific number of droplets to be within the kernel in order to satisfy its normalisation condition, albeit at the introduction of a statistical error arising due to the low sample size. Nonetheless, for areas of extreme sparsity in inertial droplet fields, this provides an appropriate means by which to infer the behaviour of ${{n}}(\bm{x},t)$ without having to resort to using an exhaustive number of droplets in simulations.

For specification of the kernel, a range of functions including Gaussian, Epanechnikov, and tricube kernels are commonly used. In the present work, a Gaussian form is chosen due to its possession of $C^{\infty}$ continuity and convenient geometrical interpretation. In the one-dimensional case, this is specified by
\begin{equation} \label{eq:kernel-spherical}
K_{h}(\bm{x},\bm{x}_d^i) = \exp \left[ - \frac{1}{2} \frac{\lvert\lvert \bm{x} - \bm{x}_d^i \rvert\rvert^2}{h^2} \right] \, ,
\end{equation}
where $h$ is the smoothing length of the kernel. In this form only the Euclidean distance between the droplet position $\bm{x}_d^i$ and accumulation point $\bm{x}$ is used to determine the relative contribution of droplets, resulting in a spherical kernel. A more generalised approach can be taken by using the concept of structured kernels, in which distinct smoothing lengths along arbitrary axes of alignment can be defined \cite{hastie2009elements}. For the specific case of a multivariate Gaussian form, such a kernel is given by
\begin{equation} \label{eq:kernel-structured}
K_{\bm{H}}(\bm{x},\bm{x}_d^i) = \exp \left[ - \frac{1}{2} \left( \bm{x} - \bm{x}_d^i \right)^\top \cdot \bm{H}^{-1} \cdot \left( \bm{x} - \bm{x}_d^i \right) \right] \, ,
\end{equation}
where $\bm{H}$ is the bandwidth matrix that contains information about the smoothing lengths in different directions. The particular choice of $\bm{H} = h^2 \bm{I}$ corresponds to a spherical kernel, in which case Eq.~\eqref{eq:kernel-structured} accordingly reduces to Eq.~\eqref{eq:kernel-spherical}.

Control over the kernel in simulations is specified solely by the choice of the smoothing length for spherical kernels, or the bandwidth matrix for structured kernels, and this is dependent on the specific context in which the method is being applied. In general, however, it is important that the smoothing length(s) are large enough to create a field representation which varies smoothly in space, yet not so large that excessive smoothing is applied, in which case detail from the original data will be lost. Therefore, a crucial aspect in the application of kernel regression is to ensure that care is taken in the selection of an appropriate smoothing length for a given situation.

\subsection{Utilisation of the Eulerian-Lagrangian transformation tensor}

The main strength of using kernel regression for accumulation of the FLA number density data lies within the ability to use information from the FLA to specify the smoothing length $h$ or bandwidth matrix $\bm{H}$ of the kernels, in order to gain a finer degree of control over the reconstruction of the number density field than is offered by a constant-sized kernel. The rationale for this procedure is in alignment with the more general framework of metric learning methods for kernel regression, in which the kernel is adaptively defined according to the local behaviour of the interpolation data \cite{Huang2013,Weinberger2007}. In the case of the standard FLA formulation this can be accomplished by using the Eulerian-Lagrangian transformation tensor $\bm{J}^{\bm{x}\bm{x}}$ to specify the kernel size and shape associated with each droplet, which enables the highly varying sparsity that is observed in the droplet number density field of unsteady flows to be accounted for. Such a procedure is consistent with the meshfree nature of kernel methods, since it respects the individuality of Lagrangian trajectories by having a distinct bandwidth matrix for each droplet $\bm{x}_d^i$ rather than each Eulerian position $\bm{x}$. Furthermore, using the Eulerian-Lagrangian transformation to determine the bandwidth matrix is also theoretically sound, as it makes use of the physical interpretation of $\bm{J}^{\bm{x}\bm{x}}$ in describing how the influence of a given droplet is exerted locally upon the surrounding domain, and accordingly specifies the range and shape of support for the kernel that is used to accumulate the number density contribution from that droplet to the associated Eulerian field representation. The physical correctness of this can be inferred directly from the FLA using Eq.~\eqref{eq:Lagrangian-COM}, by considering that if the number density associated with an individual droplet is high, the domain over which it acts must be reduced to conserve the mass contribution of that droplet along its trajectory. The necessary information about the domain size is seen to be provided by the Jacobian determinant $\lvert \det (\bm{J}^{\bm{x}\bm{x}}) \rvert$, and directly associating this with the domain of support of a kernel-based method then naturally incorporates this physical information within the numerical accumulation procedure for constructing the Eulerian number density field. Additionally, specifying the kernel using $\bm{J}^{\bm{x}\bm{x}}$ is equivalent to interpreting the smoothing length(s) as a representative lengthscale(s) for a given kernel. This is consistent with the introduction of a lengthscale within the definition of number density, a concept which has been emphasised in previous work as being necessary to provide a meaningful representation of the number density field \cite{Monchaux2012,Papoutsakis2020}.

Using the Eulerian-Lagrangian transformation tensor $\bm{J}^{\bm{x}\bm{x}}$ to determine the range and shape of support of the kernel for a droplet also provides a way of mitigating against the occurrence of singularities in the Lagrangian number density data that the FLA provides. This is because whenever a singularity in number density occurs, the corresponding value of $\det (\bm{J}^{\bm{x}\bm{x}})$ passes smoothly through zero, and this is associated with the domain over which the local Eulerian-Lagrangian transformation applies also reducing to zero. Therefore using $\bm{J}^{\bm{x}\bm{x}}$ to specify the kernel at this point will result in a kernel with a zero range of support, i.e. mathematically equivalent to the Dirac delta function. Since the Eulerian number density field is constructed over a discrete grid with finite grid spacing, once the range of support of a kernel becomes small enough in the case when $\lvert \det (\bm{J}^{\bm{x}\bm{x}}) \rvert$ is sufficiently small, the contribution from the droplet at that point will fall below the size of an Eulerian cell and no longer extend to cover any grid points, and will therefore not be included in the Eulerian field representation. Thus specification of the kernel in accordance with $\bm{J}^{\bm{x}\bm{x}}$ offers a means of effectively filtering out or limiting the propagation of the non-physical contributions associated with the high number density values that are inherent to the FLA, and thereby maintaining a more realistic description of the number density field in regions of the droplet field that contain crossing trajectories.

In practice, such a procedure can be applied in the cases of both spherical and structured kernels. For a spherical kernel, the smoothing length $h$ is determined by setting the size of the spherical domain of the kernel to be equal to the elemental volume deformation $\lvert \det (\bm{J}^{\bm{x}\bm{x}}) \rvert$ associated with a droplet along its trajectory. For a flow configuration with a spatial domain of $\mathbb{R}^d$, this determines the smoothing length for the kernel associated with a droplet at a given time $t$ as
\begin{equation} \label{eq:h-spherical-kernel}
h(\bm{x}_0,t) = h_0 \lvert \det (\bm{J}^{\bm{x}\bm{x}} (\bm{x}_0,t)) \rvert^{1/d} \, ,
\end{equation}
where $h_0$ is the initial smoothing length associated with the droplet at time $t_0$ and position $\bm{x}_0$, and $d$ is the number of spatial dimensions in the system under consideration. Using Eq.~\eqref{eq:h-spherical-kernel}, the smoothing length $h(\bm{x}_0,t)$ is automatically updated at each timestep, thereby adjusting the size of the kernel directly in accordance with the value of $\lvert \det (\bm{J}^{\bm{x}\bm{x}}) \rvert$ at that time. This methodology is consistent with the concept of a variable smoothing length in SPH, where the simplest approach to updating $h$ utilises the number density $n (\bm{x},t)$ such that \cite{fraga2019smoothed}
\begin{equation} \label{eq:SPH-variable-smoothing-length}
h (\bm{x},t) = h_0 \left( \frac{n (\bm{x}_0,t_0)}{n (\bm{x},t)} \right)^{1 / d} \, .
\end{equation}
For monodisperse droplets, it is seen that substitution of the Lagrangian continuity equation \eqref{eq:Lagrangian-COM} into Eq.~\eqref{eq:SPH-variable-smoothing-length} demonstrates the equivalence between Eqs.~\eqref{eq:h-spherical-kernel} and \eqref{eq:SPH-variable-smoothing-length} as procedures for adaptively scaling $h (\bm{x}_0,t)$, which highlights the rational physical basis for Eq.~\eqref{eq:h-spherical-kernel}.

In the case of a structured kernel, a positive semi-definite bandwidth matrix $\bm{H}$ is constructed by taking
\begin{equation} \label{eq:h-structured-kernel}
\bm{H}(\bm{x}_0,t) = h_0^2 \, \bm{J}^{\bm{x}\bm{x}} (\bm{x}_0,t) \cdot {\bm{J}^{\bm{x}\bm{x}}}^\top (\bm{x}_0,t)  \, .
\end{equation}
It is appropriate to use the initial one-dimensional smoothing length $h_0$ within the definition of this structured kernel due to the initial condition on the Eulerian-Lagrangian transformation tensor of $\bm{J}^{\bm{x}\bm{x}}(\bm{x}_0,t_0) = \bm{I}$ in Eqs.~\eqref{eq:Jacobian-initial-conditions}, from which it follows by Eq.~\eqref{eq:h-structured-kernel} that the initial bandwidth matrix $\bm{H}_0 = \bm{H}(\bm{x}_0,t_0) = h_0^2 \bm{I}$, and is therefore a spherical kernel. The bandwidth matrix $\bm{H}(\bm{x}_0,t)$ can then subsequently be updated at each timestep using Eq.~\eqref{eq:h-structured-kernel}, adjusting the size, shape, and orientation of the kernel directly in accordance with the Eulerian-Lagrangian transformation tensor $\bm{J}^{\bm{x}\bm{x}}$ along a trajectory at that time. Specification by means of Eq.~\eqref{eq:h-structured-kernel} results in an ellipsoidal kernel which represents the deformation of the initial spherical kernel at that point along a trajectory, with the lengths of the semi-axes and angles of rotation from the Cartesian axes determined by the components of $\bm{J}^{\bm{x}\bm{x}}$ at that point. Importantly, this procedure can be physically interpreted as specification of the shape and size of the range of support for the kernel associated with an individual droplet in accordance with the spatial structures of the Eulerian number density field. For instance, in regions of high droplet number density where droplets cluster in elongated structures, the ellipsoids associated with the kernels for these droplets will be aligned along these structures, and this ensures that the number density values along trajectories are projected only onto the areas of the Eulerian number density field which have similar physical behaviour in terms of droplet clustering. Therefore using the Eulerian-Lagrangian transformation tensor to specify the kernel provides a data-driven approach to the accumulation of droplet contributions in a manner which is physically consistent with the droplet number density field, and requires no further assumptions to be made beyond specification of the initial smoothing length $h_0$. A demonstration that specification of $\bm{H}$ by Eq.~\eqref{eq:h-structured-kernel} defines a kernel for which the domain size varies proportionately to $\lvert \det \left( \bm{J}^{\bm{x}\bm{x}} \right) \rvert$ is provided in Appendix \ref{sec:kernel-domain-detJ}.

\subsection{Extension of the phase space to include droplet radius} \label{sec:kernel-gfla}

The kernel regression procedure can straightforwardly be extended to include the droplet radius $r_d$ within the phase space coordinate vector, enabling the droplet size distribution to be determined for polydisperse flows using the gFLA by means of Eq.~\eqref{eq:kernel-smoother}, where now instead of $n (\bm{x},t)$ it is $p (\bm{x},r,t)$ that is being reconstructed. In this case, it is inappropriate to use a spherical kernel, since the scales of the domains for which physical space $\bm{x}$ and radial space ${r}$ are relevant will be different. For the same reason, and also due to the higher dimensionality meaning that the same number of droplets will be more sparsely distributed in $(\bm{x},{r})$ space, consideration of the structures in the droplet probability density field $p (\bm{x},r,t)$ is no longer as important, and it is sufficient to vary the kernel size based on just the magnitude of the phase space Jacobian determinant $\det (\bm{J}(\bm{x}_0,\bm{r}_0,t))$. In this case a structured kernel of the same form as in Eq.~\eqref{eq:kernel-structured} is used, however it is now written in terms of the phase space coordinate $\boldsymbol{\xi} = (\bm{x},{r})$ and phase space trajectories $\bm{z}_d = (\bm{x}_d,r_d)$ as
\begin{equation} \label{eq:H-kernel-structured-gFLA}
K_{\bm{H}}(\boldsymbol{\xi},\bm{z}_d^i) = \frac{1}{\sqrt{\det (\bm{H})}} \exp \left[ - \frac{1}{2} \left( \boldsymbol{\xi} - \bm{z}_d^i \right)^\top \cdot \bm{H}^{-1} \cdot \left( \boldsymbol{\xi} - \bm{z}_d^i \right) \right] \, ,
\end{equation}
with the bandwidth matrix $\bm{H}$ now being defined by
\begin{equation} \label{eq:H-kernel-gFLA}
\bm{H}(\bm{x}_0,{r}_0,t) =
\left[
\begin{array}{cc}
h_{\bm{x}0}^2 \lvert \det( {J}^{\bm{x}\bm{x}} (\bm{x}_0,t) ) \rvert^{2/d} \bm{I} & \bm{0} \\
\bm{0} & h_{{r}0}^2 \lvert {J}^{r{r}} ({r}_0,t) \rvert^{2}
\end{array}
\right] \, ,
\end{equation}
where $h_{\bm{x}0}$ and $h_{{r}0}$ are the initial smoothing lengths in physical and radial space respectively. Whilst the resulting kernel is still ellipsoidal, it remains aligned with the Cartesian axes rather than rotating to align with any structures in the droplet number density field. Furthermore, the kernel remains spherical across all spatial dimensions $\bm{x}$, with separate control of $h_{\bm{x}0}$ and $h_{{r}0}$ then allowing for the range of support for the kernel in physical and radial space to be determined independently. In this manner the contributions from droplets can be appropriately accumulated to produce the Eulerian probability density field in $(\bm{x},{r})$ space, and thereby a description of the droplet size distribution at each point $\bm{x}$ in space.

\subsection{Numerical implementation}

In practice, a Gaussian kernel has an infinite range of support with non-zero contributions at every Eulerian grid point in the domain, although these will quickly become negligible at a certain distance away from a droplet. Since kernel regression normalises the resultant Eulerian number density field using only those droplets which contribute at a given point, it is computationally favourable to limit the extent of the range of support for droplets by imposing an artificial range of compact support. This can be realised in the general case of the structured multivariate Gaussian kernel \eqref{eq:H-kernel-structured-gFLA} by using the Mahalanobis distance associated with a given kernel to determine which Eulerian grid points will receive non-negligible contributions from that kernel. Specifically, employing the often used 3-sigma rule of the Gaussian distribution, which states that 99.8\% of the contributions from a given droplet lie within three smoothing lengths of its position, the condition for Eulerian grid points to receive a contribution from that droplet is
\begin{equation} \label{eq:Mahalanobis-distance}
\sqrt{\left( \bm{x} - \bm{x}_d^i \right)^\top \cdot \bm{H}^{-1} \cdot \left( \bm{x} - \bm{x}_d^i \right)} \le 3 \, .
\end{equation}
This procedure then automatically accounts for the shape and orientation of the kernel, and cuts off contributions beyond the isocontour of the kernel at a distance of three smoothing lengths from the droplet position. The number of gridpoints that a given kernel contributes to is then drastically reduced from the full domain, significantly enhancing the computational economy of the kernel regression process. The same principle can also be applied for the gFLA by replacing $\bm{x}$ and $\bm{x}_d$ in Eq.~\eqref{eq:Mahalanobis-distance} with $\boldsymbol{\xi}$ and $\bm{z}_d$ respectively.

Having a compact support for a kernel is advantageous when it comes to accumulating contributions from different droplets onto the Eulerian field. Since kernel regression is a meshfree method, and the kernel is specified uniquely for each individual droplet, it is convenient to consider the extent of all the contributions made by a single droplet onto the Eulerian gridpoints. As the kernel possesses a compact support given by Eq.~\eqref{eq:Mahalanobis-distance}, the subset of gridpoints which receive a non-zero contribution from a given droplet can then be determined. In the case of a uniform Cartesian grid, it is judicious to use the minimum bounding box which encloses the region defined by the Mahalanobis distance using Eq.~\eqref{eq:Mahalanobis-distance}. This can be realised for the multivariate Gaussian kernel \eqref{eq:H-kernel-structured-gFLA}, since the maximum extent of the range of compact support in the $k^{\text{th}}$ coordinate direction is given  by
\begin{equation} \label{eq:grid-point-range-selection}
{s}_k^{\text{max}} = 3\sqrt{{H}_{kk}} \, ,
\end{equation}
in which summation over $k$ is not implied, and the coefficient of $3$ arises due to invoking the 3-sigma rule in defining the compact support. In this manner, the subset of the domain to which a single droplet makes contributions at a given point in time can be easily determined from the uniquely defined kernel associated with that particular droplet, enabling efficient accumulation of these values onto the Eulerian field. This is particularly simple as the use of a droplet search algorithm is not required, and takes advantage of the Lagrangian nature of kernel regression along with the fact that the Eulerian gridpoints are uniformly spaced. For more general or dynamically generated grids a further procedure is needed to determine the gridpoints which lie within the Mahalanobis distance of a droplet, adding to the computational expense of the interpolation procedure.

Kernel regression is able to construct a representative Eulerian number density field that respects conservation of droplet phase properties due to the manner in which the accumulation from individual droplets is performed. Specifically, the weights $w^i$ for the Nadaraya-Watson estimator defined by Eq.~\eqref{eq:kernel-weights} normalise the droplet contributions at a given gridpoint regardless of the number of droplets which contain that gridpoint within their compact support, meaning that the influence of a given droplet on the accumulated number density field varies. This is exemplified for droplets with a sufficiently high number density, in which case the domain of the associated kernel will become correspondingly small and not extend to any of the neighbouring gridpoints, thus effectively filtering the contribution of that droplet from the Eulerian field at that instant in time. Despite this, kernel regression is still able to produce a physically consistent Eulerian field, and in particular is well suited to pairing with FLA number density data as it provides an estimator for the Eulerian number density field that is independent of the number of sampled trajectories. This is in keeping with the ethos of the FLA, in that by representing the dispersed phase as a continuum, only a subset of the trajectories required by a conventional accumulation procedure are needed to build an accurate description of the number density field.

For the case when the full shape and size of the kernel are specified using $\bm{J}^{\bm{x}\bm{x}}$ by means of Eq.~\eqref{eq:h-structured-kernel}, an additional consideration must be made when droplets are in the vicinity of the envelope of folds in the droplet phase and the number density is almost singular. Due to $\lvert \det (\bm{J}^{\bm{x}\bm{x}}) \rvert$ becoming zero when a trajectory crosses the envelope of folds, the shape of the local deformation associated with $\bm{J}^{\bm{x}\bm{x}}$ becomes elongated as the droplet approaches the envelope, and eventually collapses in the dimension perpendicular to the envelope so that it is fully aligned at the point where the droplet crosses the envelope. The consequence of defining the kernel using Eq.~\eqref{eq:h-structured-kernel} is then that as the droplet approaches and moves away from the fold envelope, the kernel becomes excessively elongated far beyond the extent of local structures in the droplet phase, and makes non-physical contributions to the regression procedure at distant gridpoints. It is therefore necessary to impose a restriction on the maximum extent to which a kernel can elongate in order to ensure that the accumulation procedure maintains stability and produces a representative reconstruction of the number density field. This is performed in the present work by simply restricting the kernel shape such that each of the semi-axes of the ellipsoidal domain upon which the kernel is defined cannot be greater than three times the value of the smoothing length $h$ associated with the equivalent spherical kernel at that point, as determined by Eq.~\eqref{eq:h-spherical-kernel}. In practice, this limit on the elongation factor has been found to represent an acceptable trade off between capturing the anisotropic structures that are present in the droplet number density field, whilst preserving an accurate physical description that does not contain spurious numerical effects. In the event that it is necessary to limit the length of one of these semi-axes, the others are accordingly rescaled so that the value of $\lvert \det (\bm{J}^{\bm{x}\bm{x}}) \rvert$ at that point on the given trajectory remains unchanged, thereby enabling droplets in the vicinity of the fold envelope to contribute to the kernel regression procedure in a manner which remains representative of the local structures in the droplet phase.

The important issue of selecting a suitable value for $h_0$ must be made at the beginning of a simulation, and in general is best specified depending upon the initial spatial distribution of droplets. For a seeding of droplets across an interval over which the Eulerian number density field can be considered constant, it is convenient to take $h_0$ as the same for all droplets, and the simplest way of setting this value is then to define it as the average inter-droplet spacing across the interval, denoted $\Delta x_{d0}$. Since $h_0$ is equivalent to the standard deviation in the one-dimensional Gaussian kernel \eqref{eq:kernel-spherical}, this then ensures that the compact support defined using the 3-sigma rule provides sufficient coverage of the computational domain for the given droplet seeding. It should be noted that the specification of $h_0 = \Delta x_{d0}$ is only a guideline however, and whilst this is used across most of the cases in Section \ref{sec:results}, some variation may be required in order to obtain either higher fidelity or smoothness of the Eulerian number density field as desired. In practice the range of permissible values for the initial smoothing length of a Gaussian kernel with respect to the average inter-droplet spacing is $\Delta x_{d0} / 3 \ll h_0 \ll 2 \Delta x_{d0}$, where the lower bound is the minimum size at which kernels with compact support are able to effectively provide coverage of the computational domain, and the upper bound represents the highest degree of smoothing which might reasonably be needed. Furthermore, in the case of an expanding spray in which droplets are initially injected through a narrow inlet but then spread to span a far wider region, the initial spatial distribution of droplets may no longer be an adequate metric for basing $h_0$ upon, and a larger value will be necessary in order to achieve sufficient coverage of the computational domain with the given droplet seeding.

In terms of the gFLA, $h_{\bm{x}0}$ is defined in the same way as $h_0$, whilst the radial space smoothing length $h_{r0}$ is specified in a similar manner in terms of the initial inter-droplet size spacing, denoted $\Delta r_{d0}$. The range of validity for the radial space smoothing length differs compared to $h_{\bm{x}0}$ in practice due to the less pronounced occurrence of coherent structures in the Eulerian droplet probability density field $p$, and an appropriate set of values is given by $0.5 \, \Delta r_{d0} \ll h_{r0} \ll \Delta r_{d0}$ for a Gaussian kernel with compact support. Analogously to $h_{\bm{x}0}$, the lower bound provides greater resolution of the droplet size distribution in the Eulerian probability density field, whilst the upper bound offers a higher degree of smoothing across the droplet sizes.

\section{Numerical simulations} \label{sec:results}

\subsection{Monodisperse droplets} \label{sec:results-mono}

It is instructive to first focus upon application of the kernel regression procedure outlined in Section \ref{sec:interpolation} to monodisperse droplets for which evaporation effects are not considered, and the extension of the droplet distribution into radial space to account for the droplet size therefore does not have to be included. In the following, a range of flow configurations of varying dimensionality and complexity are used to illustrate the performance of the methodology.

\subsubsection{One-dimensional unsteady droplet motion} \label{sec:fla1d}

\begin{figure*}[!ht]
    \begin{subfigure}[c]{0.495\textwidth}
        \includegraphics[width=\textwidth,trim={0 0 0 0}]{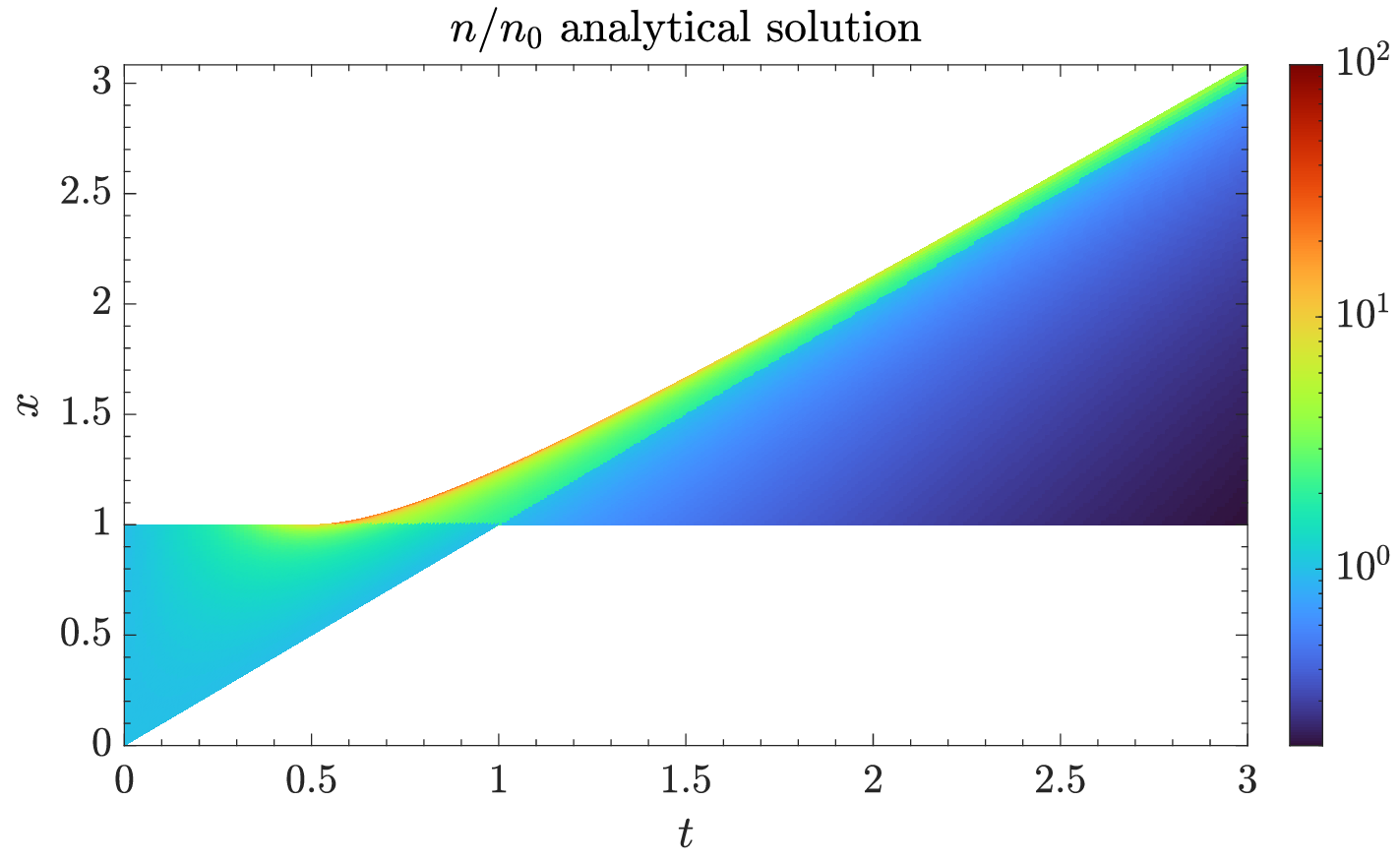}
        \caption{}
        \label{fig:1a}
    \end{subfigure}
    \begin{subfigure}[c]{0.495\textwidth}
        \includegraphics[width=\textwidth,trim={0 0 0 0}]{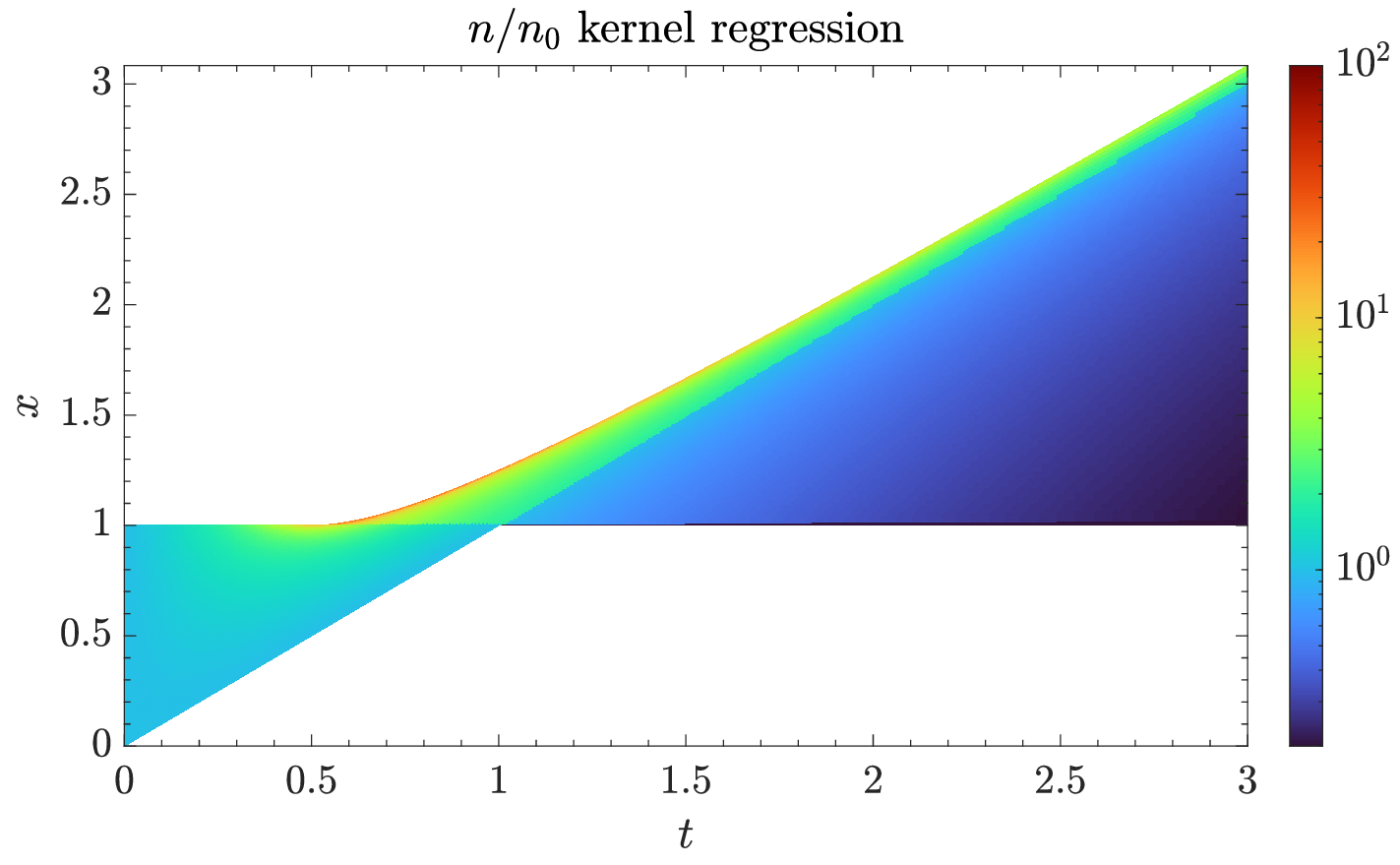}
        \caption{}
        \label{fig:1b}
    \end{subfigure}
    \begin{subfigure}[c]{0.495\textwidth}
        \includegraphics[width=\textwidth,trim={0 0 0 0}]{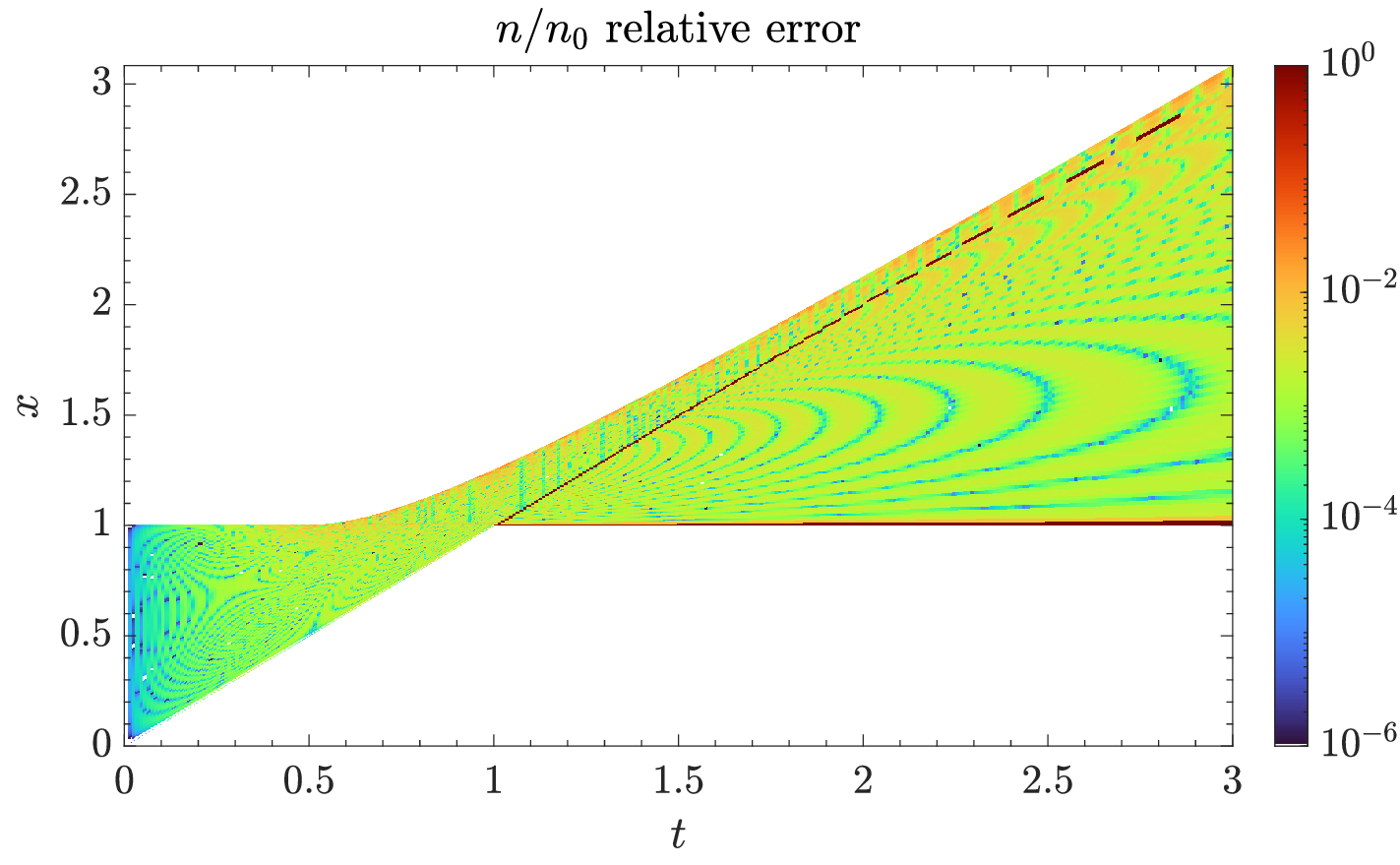}
        \caption{}
        \label{fig:1c}
    \end{subfigure}
    \begin{subfigure}[c]{0.495\textwidth}
        \includegraphics[width=\textwidth,trim={0 0 0 0}]{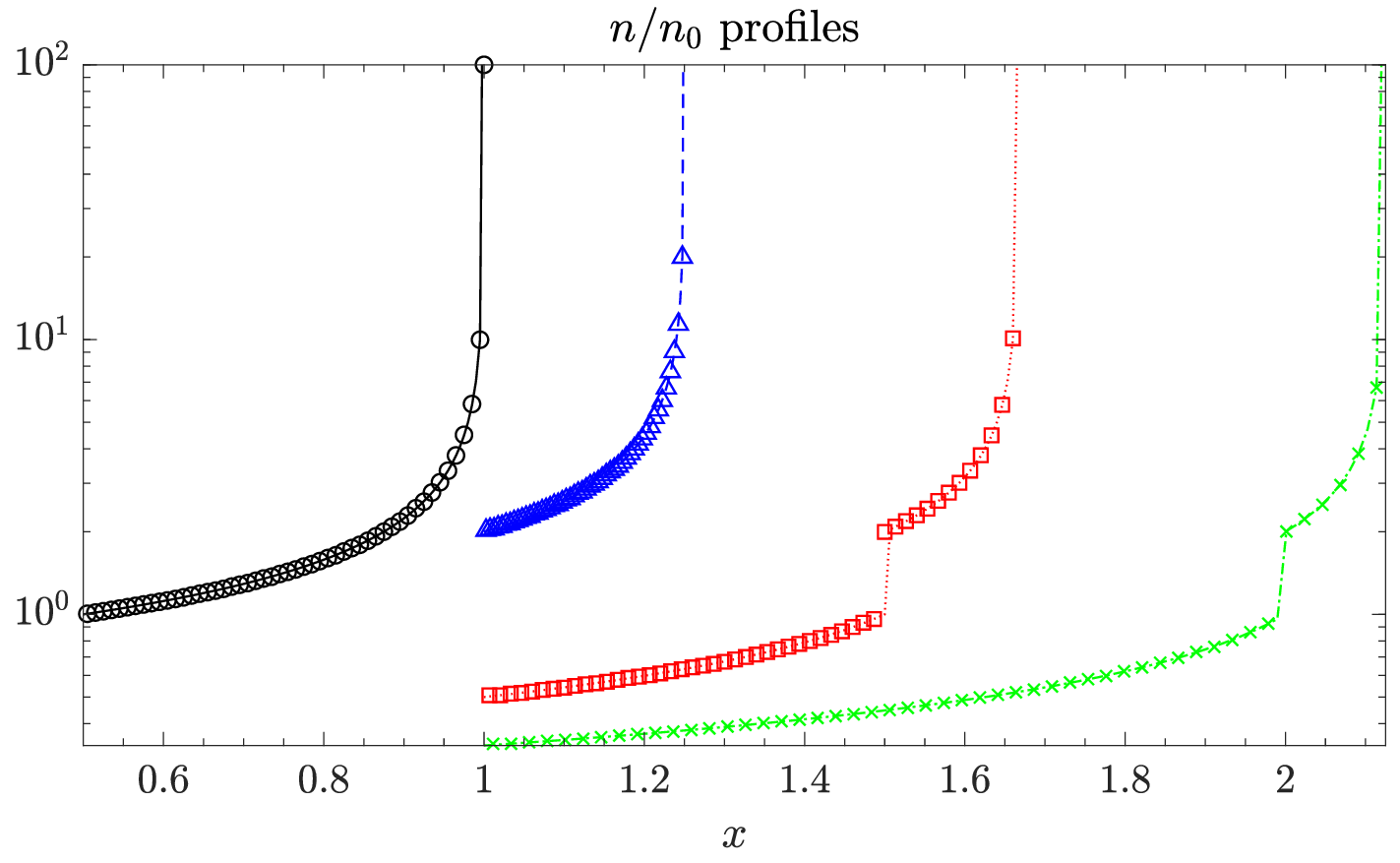}
        \caption{}
        \label{fig:1d}
    \end{subfigure}
    \caption{\eqref{fig:1a} Analytical solution for ${{n}} / n_0$ as given in Eq.~\eqref{eq:fla1d_Eulerian-nd-sol}; \eqref{fig:1b} ${{n}} / n_0$ obtained from kernel regression; \eqref{fig:1c} Relative error in ${{n}} / n_0$ between Figures \eqref{fig:1a} and \eqref{fig:1b}; \eqref{fig:1d} Comparison of kernel regression (symbols) with analytical solution (lines) for profiles of ${{n}} / n_0$ at selected instants in time:
        {\color{black}\textbf{--}$\boldsymbol{\bigcirc}$\textbf{--}} $t = 0.5$, {\color{blue}\textbf{-}\textbf{-}$\boldsymbol{\bigtriangleup}$\textbf{-}\textbf{-}} $t = 1.0$,
        {\color{red}$\cdots$$\boldsymbol{\Box}$$\cdots$} $t = 1.5$,
        {\color{green}$\cdot$\textbf{-}$\boldsymbol{\times}$$\cdot$\textbf{-}} $t = 2.0$.
    }
    \label{fig:1}
\end{figure*}

To begin with, a one-dimensional case of an unsteady droplet velocity field is considered \cite{Osiptsov1984}, which operates under the assumption that droplets are characterised by a large inertia ($St \gg 1$). Droplets are initially released with a non-uniform velocity distribution, and their motion and Jacobian evolve according to the equations
\begin{subequations}
    \label{eq:fla1d}
    \begin{align}
        \ddot{x}_d & = 0 \, ,
        & x_d(t_0) \in [0,1] \, , \, \dot{x}_d(t_0) = 1 - x_d^2(t_0) \, ,
        \label{eq:fla1d-part-eom} \\
        \ddot{{J}}^{x{x}} & = 0 \, ,
        & {J}^{x{x}}(t_0) = 1 \, , \, \dot{{J}}^{x{x}}(t_0) = -2 x_d(t_0) \, .
        \label{eq:fla1d-Jacobian-eqn}
    \end{align}
\end{subequations}
Eqs.~\eqref{eq:fla1d} admit the analytical solution
\begin{subequations}
    \label{eq:fla1d-sol}
    \begin{align}
        x_d(t) & = {x}_0 + (1 - {x}_0^2)t \, ,
        \label{eq:fla1d-part-sol} \\
        {J}^{x{x}}(t) & = 1 - 2 {x}_0 t \, .
        \label{eq:fla1d-Jacobian-sol}
    \end{align}
\end{subequations}
The associated number density $n (x_d,t)$ evolves according to Eq.~\eqref{eq:Lagrangian-COM} with initial condition $n ({x}_0,t) = n_{0}$, and the corresponding solution in this case is given by
\begin{equation} \label{eq:fla1d-nd-sol}
n (x_d,t) = \frac{n_{0}}{\lvert 1 - 2 {x}_0 t \rvert} \, .
\end{equation}
From Eq.~\eqref{eq:fla1d-part-sol} it follows that a fold is formed at $t = 0.5$, with the envelope of droplet trajectories along this being
\begin{equation} \label{eq:fla1d-envelope}
{x} = t + \frac{1}{4t} \, , \qquad t \ge 0.5 \, .
\end{equation}
In this case the Eulerian number density field can also be obtained as
\begin{align} \label{eq:fla1d_Eulerian-nd-sol}
    {{n}} ({x},t) =
    \left\{
    \begin{array}{cc}
        \left. {{n}}_0 \middle/ \sqrt{1 - 4tx + 4t^2} \right. & \qquad t < x < 1 \, \text{ or } \, 1 < x < t \, , \\
        \left. 2 {{n}}_0 \middle/ \sqrt{1 - 4tx + 4t^2} \right. & \qquad 0.5 < t < x \, \text{ and } \, x > 1 \, ,
    \end{array}
    \right.
\end{align}
where ${{n}}_0 = {{n}}({x}_0,t_0)$, and in which it is assumed that droplets moving towards the envelope of the trajectories form a separate layer of the fold to those droplets moving away from the envelope, and that these layers belong to different non-interacting continua. The total number density in the region is simply double that in the remainder of the ${x}$--$t$ domain which is occupied by droplets, resulting in the piecewise nature of Eq.~\eqref{eq:fla1d_Eulerian-nd-sol}. Along the fold envelope \eqref{eq:fla1d-envelope} the droplet number density becomes infinitely high, however this singularity remains integrable \cite{Osiptsov1984}. The analytical solution \eqref{eq:fla1d_Eulerian-nd-sol} is graphically depicted in ${x}$--$t$ space in Figure \eqref{fig:1a}, and features a discontinuity along the internal boundary of the multi-valued region given by the second sub-function of Eq.~\eqref{eq:fla1d_Eulerian-nd-sol}. The region of the domain occupied by the droplet phase is reconstructed on a grid at time intervals of $\Delta t = 0.01$ with 100 uniformly spaced points of separation $\Delta {x}$ at each time, and uses an initial seeding of 101 droplets positioned uniformly on the interval ${x}_0 \in [0,1]$, corresponding to an initial average inter-droplet spacing of $\Delta x_{d0} = 0.01$.

To assess the efficacy of the kernel regression procedure in Section \ref{sec:interpolation}, the Eulerian number density field is reconstructed from the Lagrangian number density \eqref{eq:fla1d-nd-sol} using the Nadaraya-Watson estimator in Eq.~\eqref{eq:kernel-smoother}. Since this case is one-dimensional, the kernel given in Eq.~\eqref{eq:kernel-spherical} is utilised, with the smoothing length $h$ specified by Eq.~\eqref{eq:h-spherical-kernel}. This produces the result shown in Figure \eqref{fig:1b}. For comparison purposes, it is appropriate to consider the relative error between kernel regression and the analytical solution, and this is presented in Figure \eqref{fig:1c}. It is seen that the relative error in the kernel regression is generally below $10^{-2}$ throughout the ${x}$--$t$ domain, except at the fold envelope \eqref{eq:fla1d-envelope} and along the boundary of the multi-valued region. This more marked error arises due to the smoothing property of kernels, in that the influence of a given droplet is distributed over a small region of the domain, with the consequence that representation of steep gradients and discontinuities in the number density field cannot be exactly captured using this procedure. Therefore it is expected that some additional error will result from the use of kernel regression in reconstructing the number density field, however an appropriate choice of initial smoothing length $h_0$ can mitigate against this \cite{Price2012}, with a smaller kernel generally being preferable subject to ${{n}}$ remaining smooth. In this case, the smoothing length is set proportional to the variable grid spacing at each point in time, with the initial value of $h_0 = \Delta x_{d0} / 3$ being chosen in order to achieve a high degree of fidelity in the reconstructed Eulerian number density field. The overall accuracy of the kernel regression is seen to be good, as can be inferred from examination of the spatial profiles of ${{n}}$ at selected instances in time, depicted in Figure \eqref{fig:1d}. The only discernible loss of accuracy is in the vicinity of the discontinuity between the single-valued and multi-valued regions of Eq.~\eqref{eq:fla1d_Eulerian-nd-sol} for $t \ge 1$, as can be seen in the jump in the profiles for ${y} = 1.5$ and ${y} = 2$. Moreover, it is seen that whilst ${{n}}$ remains finite in the vicinity of the fold envelope \eqref{eq:fla1d-envelope}, kernel regression is able to capture the increase in number density well, with the error only becoming greater than $10^{-2}$ along the fold itself.

\subsubsection{Two-dimensional fan spray injection in cross-flow} \label{sec:fla2d-mono}

\begin{figure*}[!ht]
    \begin{subfigure}[c]{0.495\textwidth}
        \includegraphics[width=\textwidth,trim={0 0 0 0}]{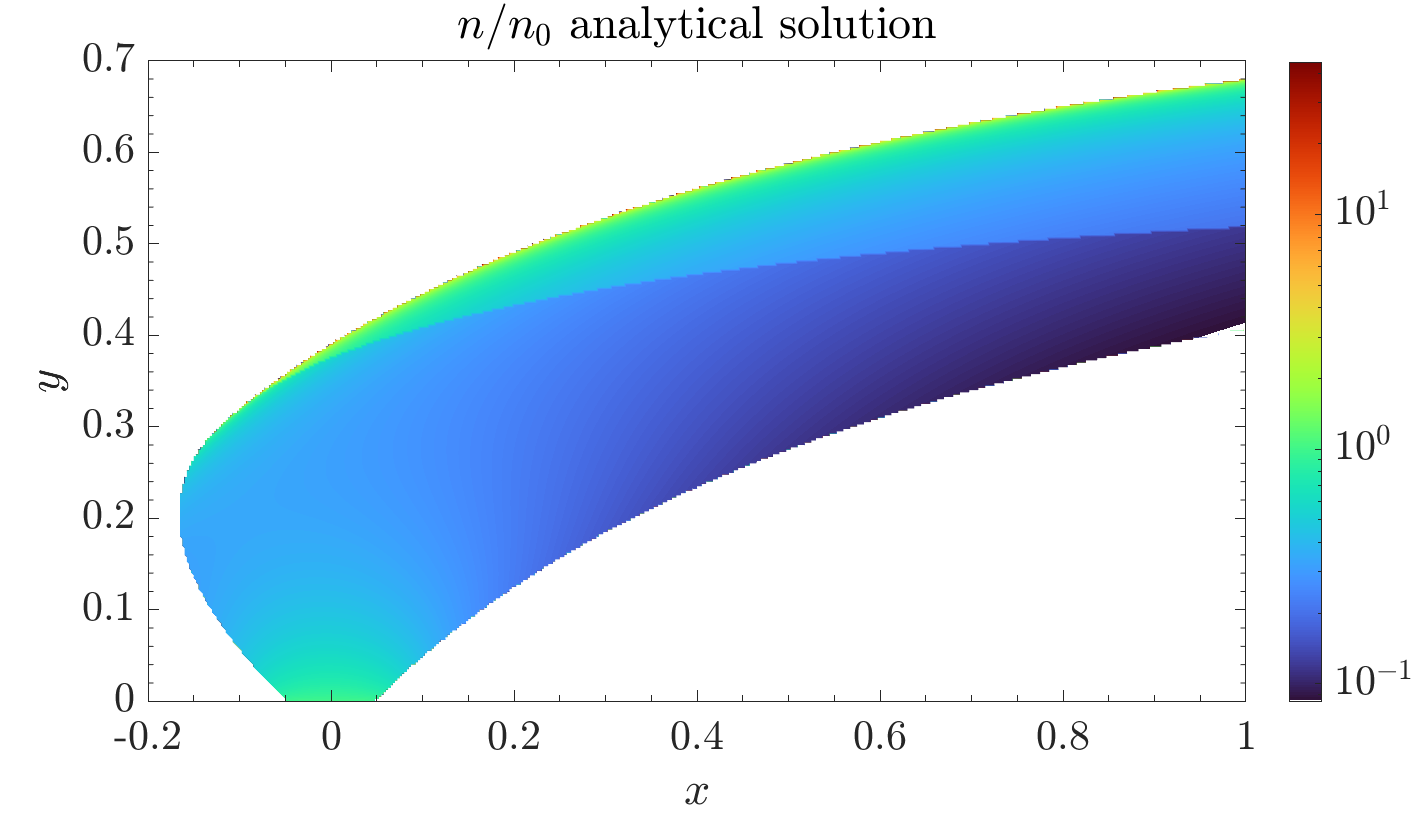}
        \caption{}
        \label{fig:2a}
    \end{subfigure}
    \begin{subfigure}[c]{0.495\textwidth}
        \includegraphics[width=\textwidth,trim={0 0 0 0}]{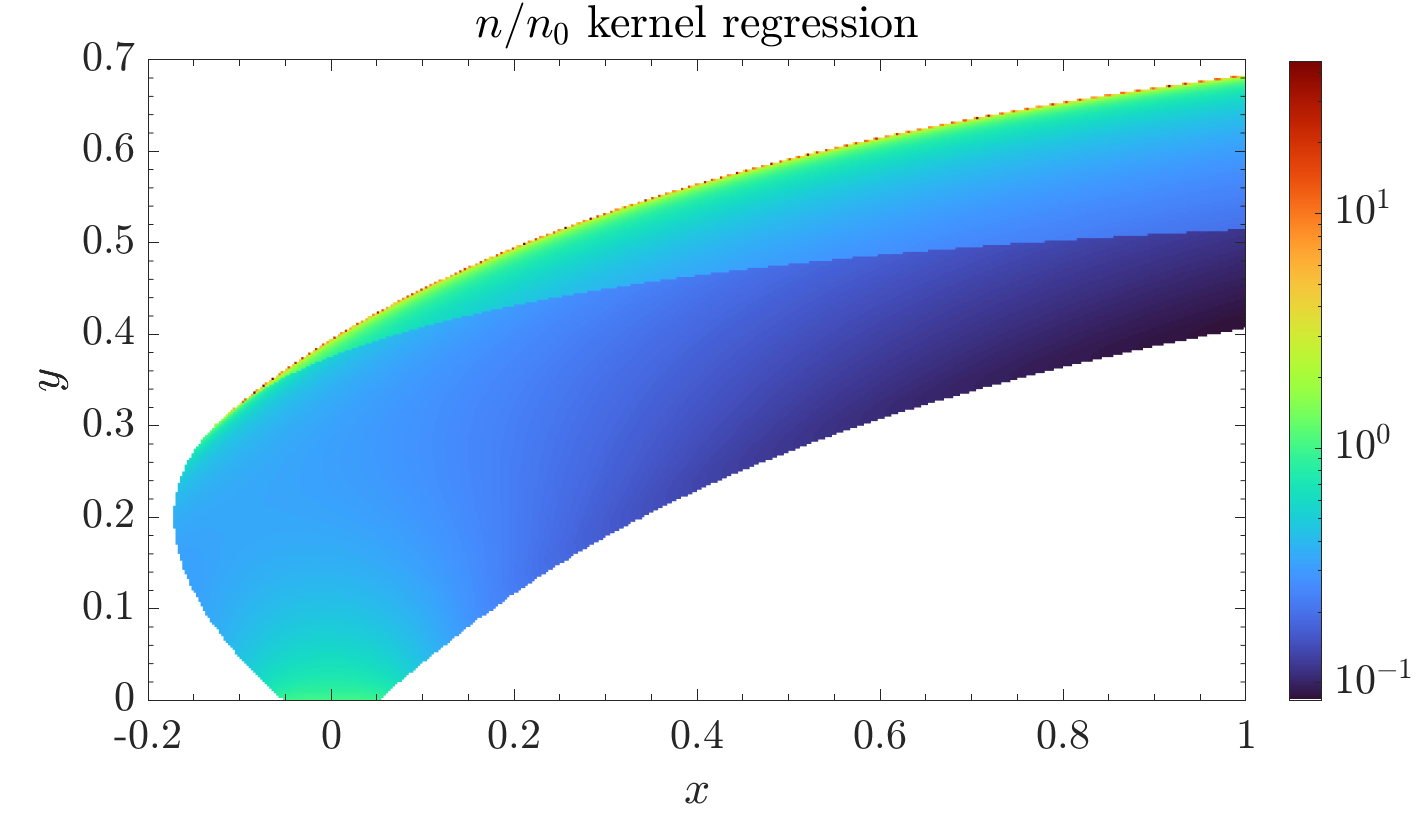}
        \caption{}
        \label{fig:2b}
    \end{subfigure}
    \begin{subfigure}[c]{0.495\textwidth}
        \includegraphics[width=\textwidth,trim={0 0 0 0}]{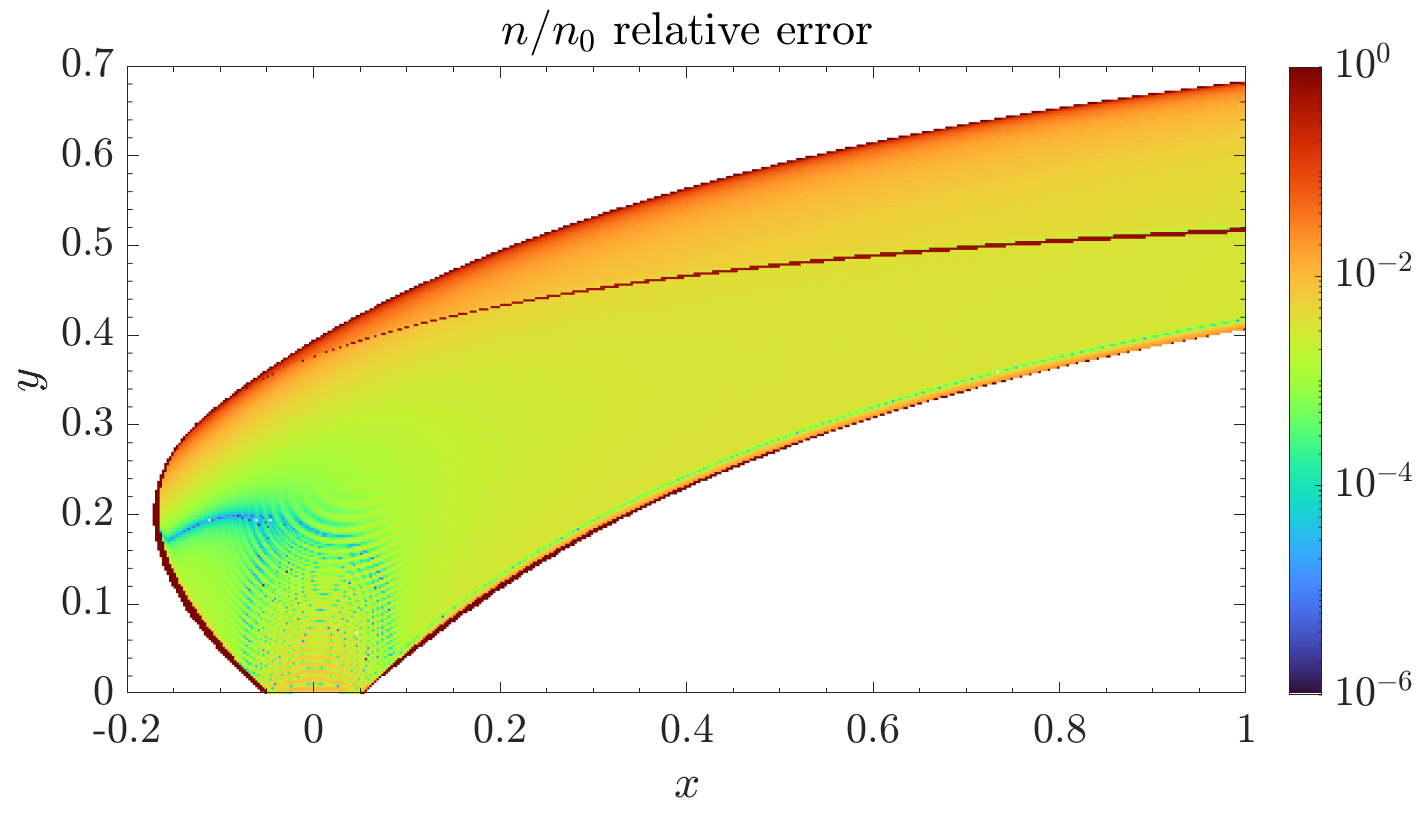}
        \caption{}
        \label{fig:2c}
    \end{subfigure}
    \begin{subfigure}[c]{0.495\textwidth}
        \includegraphics[width=\textwidth,trim={0 0 0 0}]{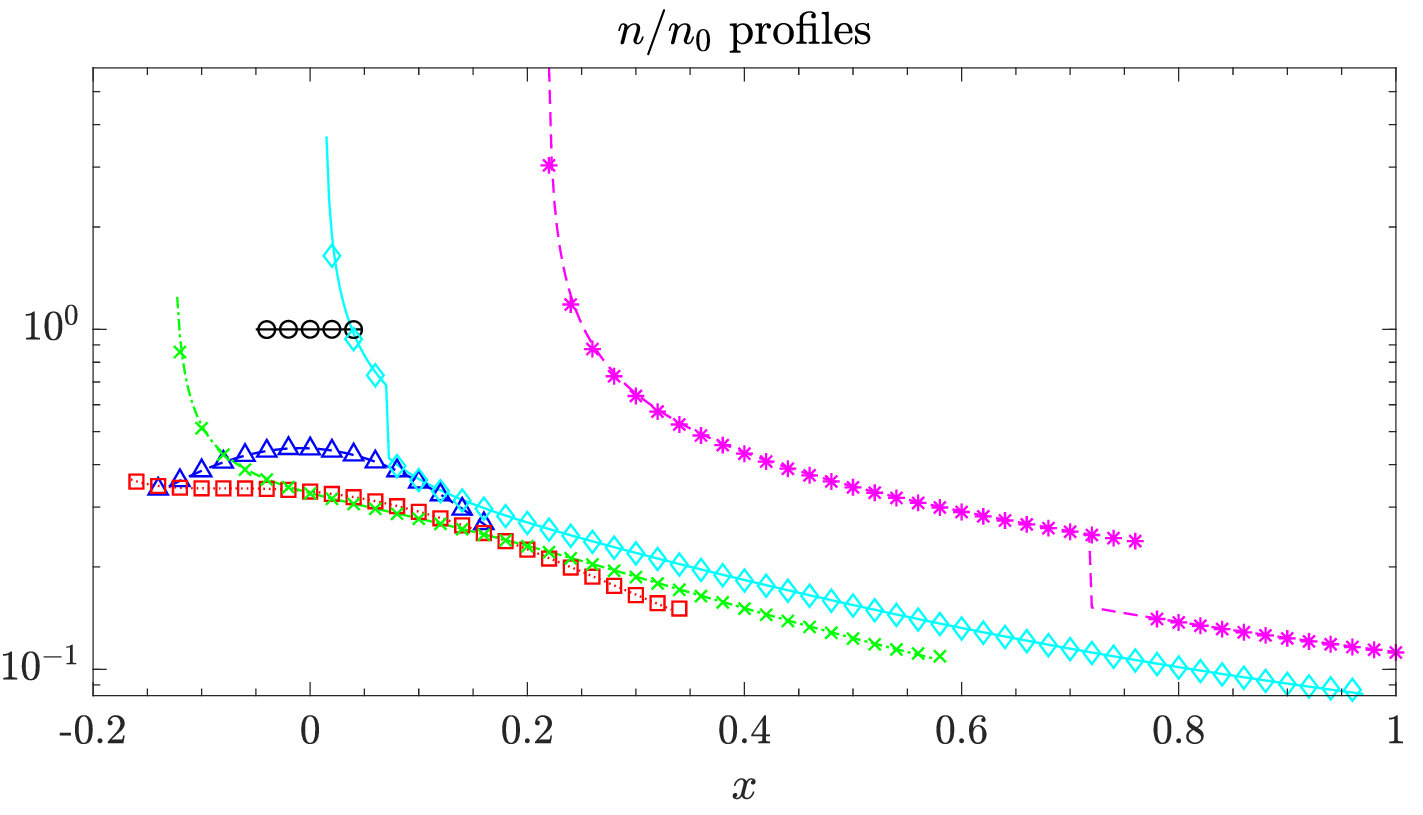}
        \caption{}
        \label{fig:2d}
    \end{subfigure}
    \caption{\eqref{fig:2a} Analytical solution for ${{n}} / n_0$ obtained from Eqs.~\eqref{eq:fla2d-analytical-solution} and \eqref{eq:fla2d-nd-sol}; \eqref{fig:2b} ${{n}} / n_0$ obtained from kernel regression; \eqref{fig:2c} Relative error in ${{n}} / n_0$ between Figures \eqref{fig:2a} and \eqref{fig:2b}; \eqref{fig:2d} Comparison of kernel regression (symbols) with analytical solution (lines) for profiles of $n / n_0$ at selected ${y}$ locations:
        {\color{black}\textbf{--}$\boldsymbol{\bigcirc}$\textbf{--}} ${y} = 0$, {\color{blue}\textbf{-}\textbf{-}$\boldsymbol{\bigtriangleup}$\textbf{-}\textbf{-}} ${y} = 0.1$,
        {\color{red}$\cdots$$\boldsymbol{\Box}$$\cdots$} ${y} = 0.2$,
        {\color{green}$\cdot$\textbf{-}$\boldsymbol{\times}$$\cdot$\textbf{-}} ${y} = 0.3$,
        {\color{cyan}\textbf{--}$\boldsymbol{\diamond}$\textbf{--}} ${y} = 0.4$,
        {\color{magenta}\textbf{-}\textbf{-}$\boldsymbol{\ast}$\textbf{-}\textbf{-}} ${y} = 0.5$.
    }
    \label{fig:2}
\end{figure*}

Consider a vertical injection of droplets with velocity magnitude $v^*$ into a horizontal cross flow with constant velocity $\bm{u} = (1,0)$. Taking the case of $St = 1$, the droplet and Jacobian equations of evolution, given by Eqs.~\eqref{eq:part-eom} and \eqref{eq:Jacobian-evolution-physical-space} respectively, are
\begin{subequations}
    \label{eq:fla2d}
    \begin{align}
        \ddot{\bm{x}}_d & = \bm{u} - \dot{\bm{x}}_d \, ,
        \label{eq:fla2d-part-eom} \\
        \ddot{\bm{J}}^{\bm{x}\bm{x}} & = -\beta \dot{\bm{J}}^{\bm{x}\bm{x}} \, ,
        \label{eq:fla2d-Jacobian-eqn}
    \end{align}
\end{subequations}
where the initial conditions for $\bm{x}_d$ in Eq.~\eqref{eq:fla2d-part-eom} are given by
\begin{align}
    x_{d,1}(t_0) & \in [-\epsilon,\epsilon] \, , &
    \dot{x}_{d,1}(t_0) & = v^* \sin \left(\frac{\pi}{4} \frac{x_{d,1}(t_0)}{\epsilon}\right) \, , \nonumber \\
    x_{d,2}(t_0) & = 0 \, , &
    \dot{x}_{d,2}(t_0) & = v^* \cos \left(\frac{\pi}{4} \frac{x_{d,1}(t_0)}{\epsilon}\right) \, , \label{eq:fla2d-x-ic}
\end{align}
and the initial conditions for $\bm{J}^{\bm{x}\bm{x}}$ in Eq.~\eqref{eq:fla2d-Jacobian-eqn} are given by
\begin{align}
    \bm{J}^{\bm{x}\bm{x}}(t_0) & = \bm{I} \, , \nonumber \\
    \dot{{J}}^{\bm{x}\bm{x}}_{11}(t_0) & = \frac{1}{\epsilon} \frac{\pi}{4} v^* \cos \left(\frac{\pi}{4} \frac{x_{d,1}(t_0)}{\epsilon}\right)  \, ,\nonumber \\
    \dot{{J}}^{\bm{x}\bm{x}}_{12}(t_0) & = -\frac{1}{\epsilon} \frac{\pi}{4} v^* \sin \left(\frac{\pi}{4} \frac{x_{d,1}(t_0)}{\epsilon}\right) + \frac{1 - \dot{x}_{d,1}(t_0)}{\dot{x}_{d,2}(t_0)} \, , \nonumber \\
    \dot{{J}}^{\bm{x}\bm{x}}_{21}(t_0) & = -\frac{1}{\epsilon} \frac{\pi}{4} v^* \sin \left(\frac{\pi}{4} \frac{x_{d,1}(t_0)}{\epsilon}\right) \, , \nonumber \\
    \dot{{J}}^{\bm{x}\bm{x}}_{22}(t_0) & = \left. \frac{1}{\epsilon} \frac{\pi}{4} v^* \sin^2 \left(\frac{\pi}{4} \frac{x_{d,1}(t_0)}{\epsilon}\right) \middle/ \cos \left(\frac{\pi}{4} \frac{x_{d,1}(t_0)}{\epsilon}\right) \right. - 1 \, , \label{eq:fla2d-j-ic}
\end{align}

\noindent in which the expressions for $\dot{{J}}^{\bm{x}\bm{x}}_{12}(t_0)$ and $\dot{{J}}^{\bm{x}\bm{x}}_{22}(t_0)$ are derived from \cite{Healy2005}
\begin{equation}
\frac{\partial}{\partial {x}_{0,2}} = - \frac{\dot{x}_{d,1}(t_0)}{\dot{x}_{d,2}(t_0)} \frac{\partial}{\partial {x}_{0,1}} + \frac{1}{\dot{x}_{d,2}(t_0)} \frac{\partial}{\partial t} \, .
\end{equation}
The system \eqref{eq:fla2d} with initial conditions \eqref{eq:fla2d-x-ic} and \eqref{eq:fla2d-j-ic} admits the analytical solution
\begin{align}
    x_{d,1}(t) & = x_{d,1}(t_0) + (1 - \dot{x}_{d,1}(t_0)) \exp(-t) \, , \nonumber \\
    x_{d,2}(t) & = x_{d,2}(t_0) (1 - \exp(-t)) \, , \nonumber \\
    \dot{x}_{d,1}(t) & = 1 - (1 - \dot{x}_{d,1}(t_0))\exp(-t) \, , \nonumber \\
    \dot{x}_{d,2}(t) & = \dot{x}_{d,2}(t_0) \exp(-t) \, , \nonumber \\
    \bm{J}^{\bm{x}\bm{x}}(t) & = \bm{I} + \dot{\bm{J}}^{\bm{x}\bm{x}}(t_0) (1 - \exp(-t)) \, , \nonumber \\
    \dot{\bm{J}}^{\bm{x}\bm{x}}(t) & = \dot{\bm{J}}^{\bm{x}\bm{x}}(t_0) \exp(-t) \, . \label{eq:fla2d-analytical-solution}
\end{align}
From Eq.~\eqref{eq:fla2d-analytical-solution}, it follows that the associated number density $n (\bm{x}_d,t)$ with initial condition $n (\bm{x}_0,t) = n_{0}$ evolving according to Eq.~\eqref{eq:Lagrangian-COM} is given by
\begin{equation} \label{eq:fla2d-nd-sol}
n (\bm{x}_d,t) = \frac{n_{0}}{\lvert 1 + (1 - \exp(-t))(\dot{{J}}^{\bm{x}\bm{x}}_{11}(t_0) + \dot{{J}}^{\bm{x}\bm{x}}_{22}(t_0)) + (1 - \exp(-t))^2 (\dot{{J}}^{\bm{x}\bm{x}}_{11}(t_0) \dot{{J}}^{\bm{x}\bm{x}}_{22}(t_0) - \dot{{J}}^{\bm{x}\bm{x}}_{12}(t_0) \dot{{J}}^{\bm{x}\bm{x}}_{21}(t_0)) \rvert} \, .
\end{equation}
It is then possible to numerically evaluate the Eulerian number density field in this case using Eq.~\eqref{eq:fla2d-nd-sol} and the parametric solutions given in Eq.~\eqref{eq:fla2d-analytical-solution}. The droplet field includes a multi-valued region containing separate layers of a fold, with the envelope of this fold occurring along the top edge of the spray, and the other boundary of the multi-valued region presenting an internal discontinuity within the spray. The analytical solution from Eqs.~\eqref{eq:fla2d-analytical-solution} and \eqref{eq:fla2d-nd-sol} is graphically depicted as a steady-state distribution in ${x}$--${y}$ space in Figure \eqref{fig:2a}, where the droplet number density field is reconstructed on a uniform Cartesian grid with spacing of $\Delta {x} = \Delta {y} = 0.0025$. In this example the value of the parameter $\epsilon$ used to define the interval in which droplets are injected in Eq.~\eqref{eq:fla2d-x-ic} is taken to be $\epsilon = 0.05$, with a total of 101 droplets being injected uniformly over $\bm{x} \in ([-\epsilon,\epsilon],0)$ at the start of the simulation, giving a value of $\Delta x_{d0} = 0.001$.

Numerical reconstruction of the steady-state droplet field from the Lagrangian number density \eqref{eq:fla2d-nd-sol} using the Nadaraya-Watson estimator \eqref{eq:kernel-smoother} is shown in Figure \eqref{fig:2b}. The structured kernel in Eq.~\eqref{eq:kernel-structured} is used for this multidimensional case, with the bandwidth matrix $\bm{H}$ being specified by Eq.~\eqref{eq:h-structured-kernel}. The relative error of kernel regression is shown in Figure \eqref{fig:2c}, and as with the one-dimensional case in Section \ref{sec:fla1d} generally remains below $10^{-2}$ except in the vicinity of the fold envelope, along the discontinuity between the multi-valued and single-valued regions of the spray, and along the edges of the spray. Again, it is the perceptible size of the kernel which causes these more marked errors to arise, and this is especially apparent along the edges of the spray where the kernel extends beyond the region occupied by droplets into the surrounding area. This effect is balanced against the need to keep the kernel large enough to produce a smooth number density field from the sample of trajectories used in the simulation, once again requiring careful choice of the initial smoothing length $h_0$. In this instance, the default value of $h_0 = \Delta x_{d0}$ is used to achieve this. Notwithstanding this, the overall accuracy of kernel regression can be observed from the profiles of ${{n}}$ at selected ${y}$ values in Figure \eqref{fig:2d}, and is seen to generally be good with the exception of the discontinuity along the edge of the multi-valued region in the spray. This is observed in the profile for ${y} = 0.5$, where kernel regression does not capture the exact location of the discontinuity as a result. Aside from this, the physical behaviour and increases in number density near the fold envelope are well captured.

\subsubsection{Flow around a cylinder: steady-state case (Re = 20)} \label{sec:fpc-mono-steady}

To determine the capabilities of the kernel regression procedure in a context which is more representative of a general engineering flow, the case of a two-dimensional gas-droplet flow around a cylinder is considered. This is a prototypical problem, and exhibits distinct flow regimes depending upon the flow Reynolds number $Re$; specifically a symmetric steady-state flow at low $Re$, and a periodic von K\'{a}rm\'{a}n vortex street beyond the critical value of $Re$ at which the transition to unsteadiness occurs.

The evolution of the carrier flow is described using the Navier-Stokes equations for an incompressible fluid
\begin{subequations}
    \label{eq:carrier-flow-eqs}
    \begin{align}
        \boldsymbol{\nabla} \cdot \bm{u} & = \bm{0} \, ,
        \label{eq:incompressibility} \\
        \frac{\partial \bm{u}}{\partial t} + \left( \bm{u} \cdot \boldsymbol{\nabla} \right) \bm{u} & = - \boldsymbol{\nabla} p + \frac{1}{Re} \nabla^2 \bm{u} \, .
        \label{eq:navier-stokes}
    \end{align}
\end{subequations}
The cylinder is taken to have radius $R$ which is used as the representative lengthscale, and the uniform free-stream velocity at which fluid enters the inlet of the domain is chosen as the associated velocity scale $U$. The Reynolds number is then defined in this work as $Re = UR / \nu$, in contrast to much of the existing literature in which the cylinder diameter is chosen as the characteristic lengthscale. In this context, the critical Reynolds number at which the transition to unsteadiness occurs is $Re_{c} = 23.5$.

To investigate droplet behaviour in the steady-state regime, the case of $Re = 20$ is considered. Since the underlying carrier flow is symmetrical about the cylinder in the direction normal to the flow, the droplet distribution also exhibits this symmetry. Droplets are injected at ${x} / R = -5$ over the interval ${y} / R \in [-4,4]$ with the free-stream carrier flow velocity $\bm{u} / U = (1,0)$. In the steady-state cases, a total of 101 droplets are injected over this interval in a square-law profile at the start of the simulation to give an average inter-droplet spacing of $\Delta x_{d0} = 0.08$, with the droplet seeding increasing towards the centreline ${y} / R = 0$.
Interaction of the droplets with the cylinder is not included within the simulation, and any droplets which do come into contact with the cylinder are subsequently omitted from the calculation of the number density field at later times.

The droplet trajectories evolve according to a linear drag law, which for the case of monodisperse non-evaporating droplets determines the equations for motion and Jacobian evolution from Eqs.~\eqref{eq:part-eom} and \eqref{eq:Jacobian-evolution-physical-space} as
\begin{subequations}
    \label{eq:fpc-part-eqns}
    \begin{align}
        \ddot{\bm{x}}_d & = \, \frac{1}{St} \left( \bm{u}(\bm{x}_d(t),t) - \dot{\bm{x}}_d \right) \, ,
        \label{eq:fpc-part-eom} \\
        \ddot{\bm{J}}^{\bm{x}\bm{x}} & = \, \frac{1}{St} \frac{\partial \bm{u}}{\partial \bm{x}}(\bm{x}_d(t),t) \cdot \bm{J}^{\bm{x}\bm{x}}
        - \frac{1}{St} \dot{\bm{J}}^{\bm{x}\bm{x}} \, .
        \label{eq:fpc-Jacobian-eqn}
    \end{align}
\end{subequations}
The associated initial conditions are given by
\begin{subequations}
    \label{eq:fpc-part-eqns-ic}
    \begin{align}
        & \bm{x}_d(t_0) = \, (-5,[-4,4]) \, , &
        & \dot{\bm{x}}_d(t_0) = \, (1,0) \, ,
        \label{eq:fpc-part-eom-ic} \\
        & \bm{J}^{\bm{x}\bm{x}}(\bm{x}_0,t_0) = \, \bm{I} \, , &
        & \dot{\bm{J}}^{\bm{x}\bm{x}}(\bm{x}_0,t_0) = \, \bm{0} \, ,
        \label{eq:fpc-Jacobian-eqn-ic}
    \end{align}
\end{subequations}
where the initial condition on $\dot{\bm{J}}^{\bm{x}\bm{x}}$ is due to the uniformity of the droplet phase at the point of injection. The Lagrangian number density $n (x_d,t)$ associated with a droplet is then calculated using Eq.~\eqref{eq:Lagrangian-COM} without the appearance of the droplet radius $r_d$ as a parameter.

Eqs.~\eqref{eq:fpc-part-eqns} are solved using a customisation of the Lagrangian particle tracking library in OpenFOAM, and can be coupled to the required solver for the carrier flow; in this case pimpleFoam has been used to solve Eqs.~\eqref{eq:carrier-flow-eqs}. The domain size for the simulations is $-20 \le {x} / R \le 30$ and $-20 \le {y} / R \le 20$ in a Cartesian coordinate system centred on the cylinder. Further details of the numerical setup, mesh independence, and validation can be found in \cite{Li2021}.

\begin{figure*}[!ht]
    \begin{subfigure}[c]{0.495\textwidth}
        \includegraphics[width=\textwidth,trim={0 0 0 0}]{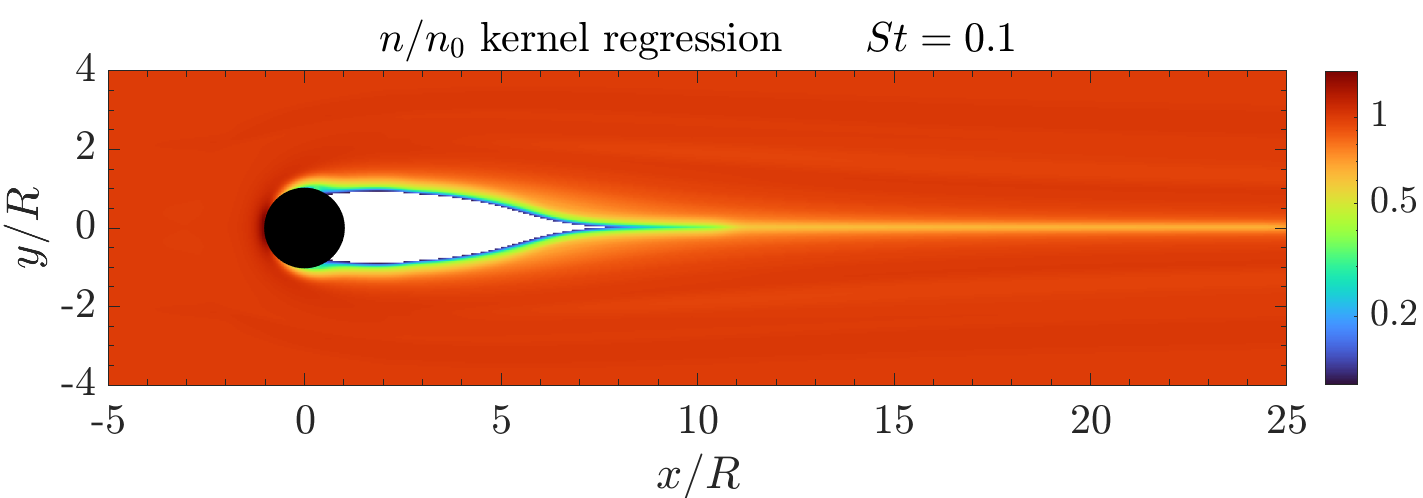}
        \caption{}
        \label{fig:3a}
    \end{subfigure}
    \begin{subfigure}[c]{0.495\textwidth}
        \includegraphics[width=\textwidth,trim={0 0 0 0}]{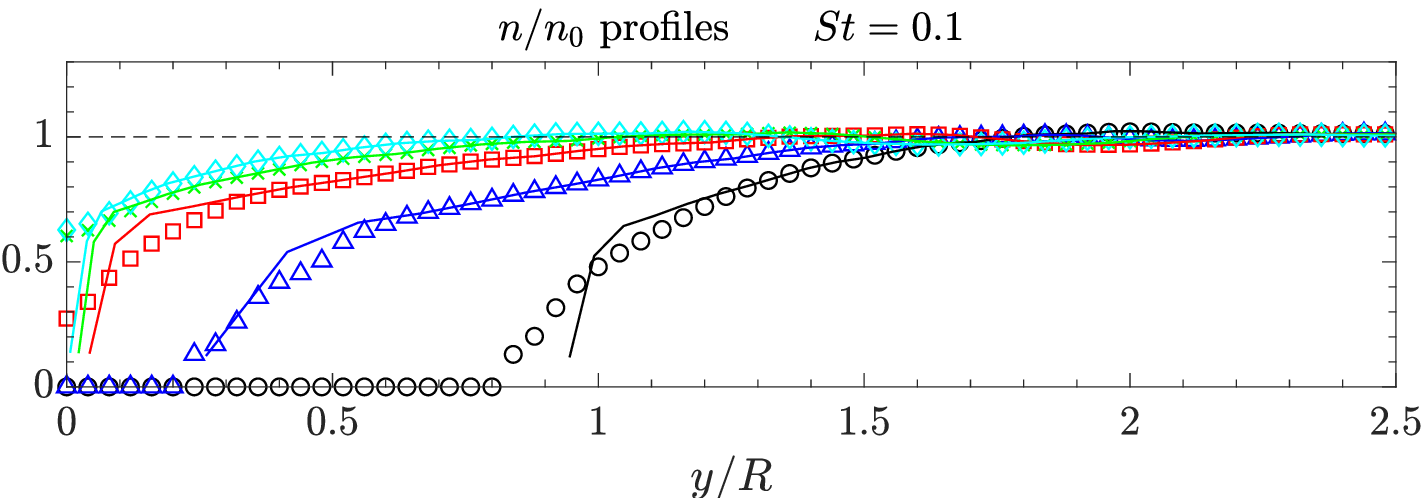}
        \caption{}
        \label{fig:3b}
    \end{subfigure}
    \begin{subfigure}[c]{0.495\textwidth}
        \includegraphics[width=\textwidth,trim={0 0 0 0}]{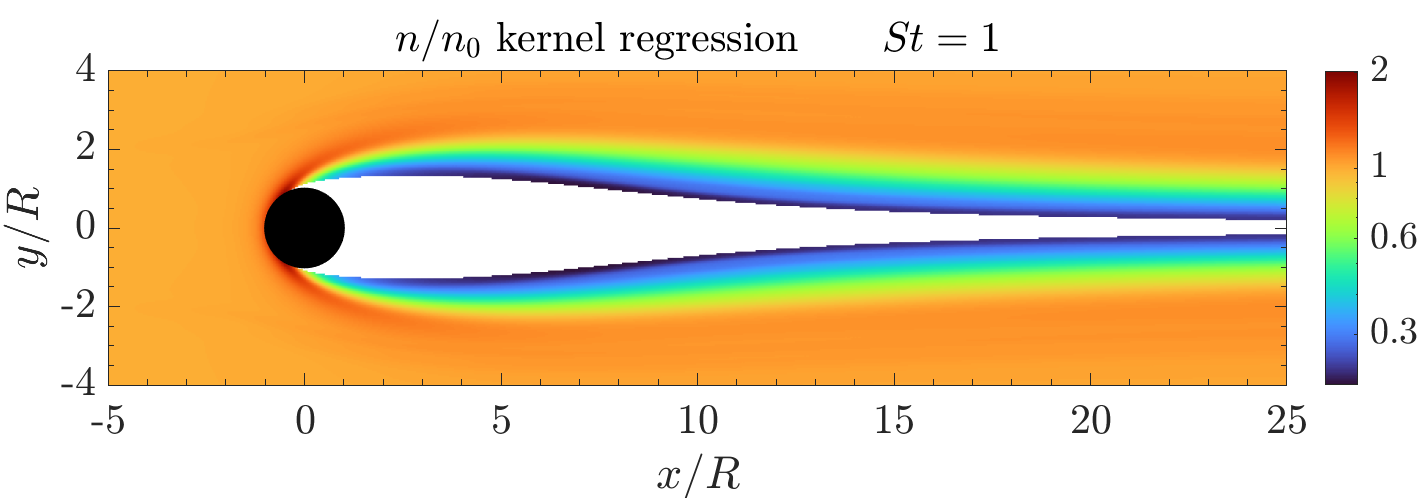}
        \caption{}
        \label{fig:3c}
    \end{subfigure}
    \begin{subfigure}[c]{0.495\textwidth}
        \includegraphics[width=\textwidth,trim={0 0 0 0}]{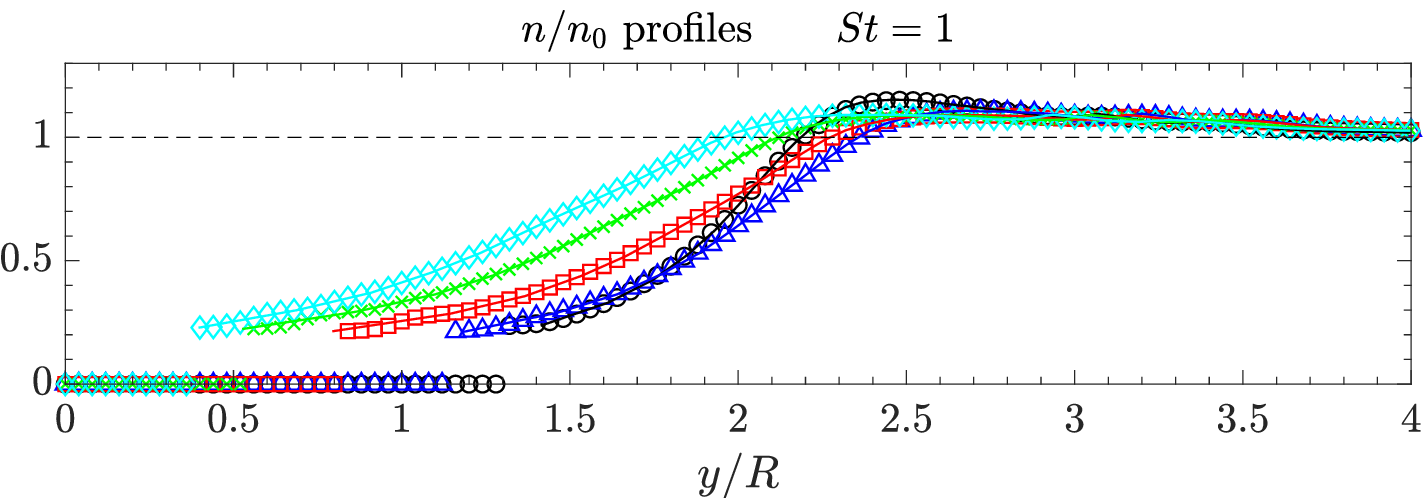}
        \caption{}
        \label{fig:3d}
    \end{subfigure}
    \begin{subfigure}[c]{0.495\textwidth}
        \includegraphics[width=\textwidth,trim={0 0 0 0}]{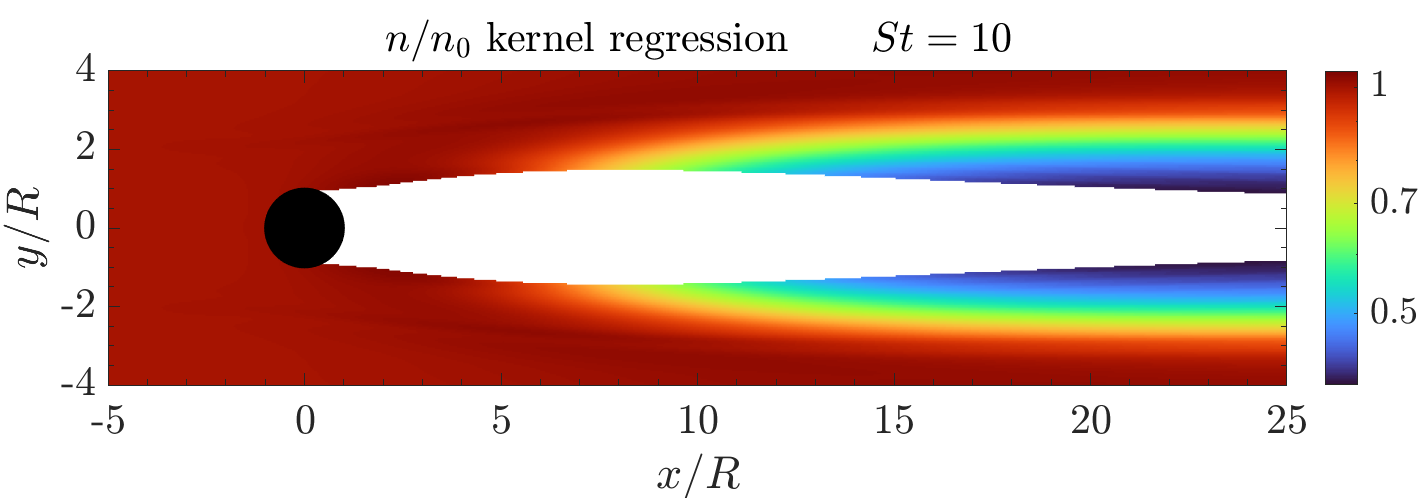}
        \caption{}
        \label{fig:3e}
    \end{subfigure}
    \begin{subfigure}[c]{0.495\textwidth}
        \includegraphics[width=\textwidth,trim={0 0 0 0}]{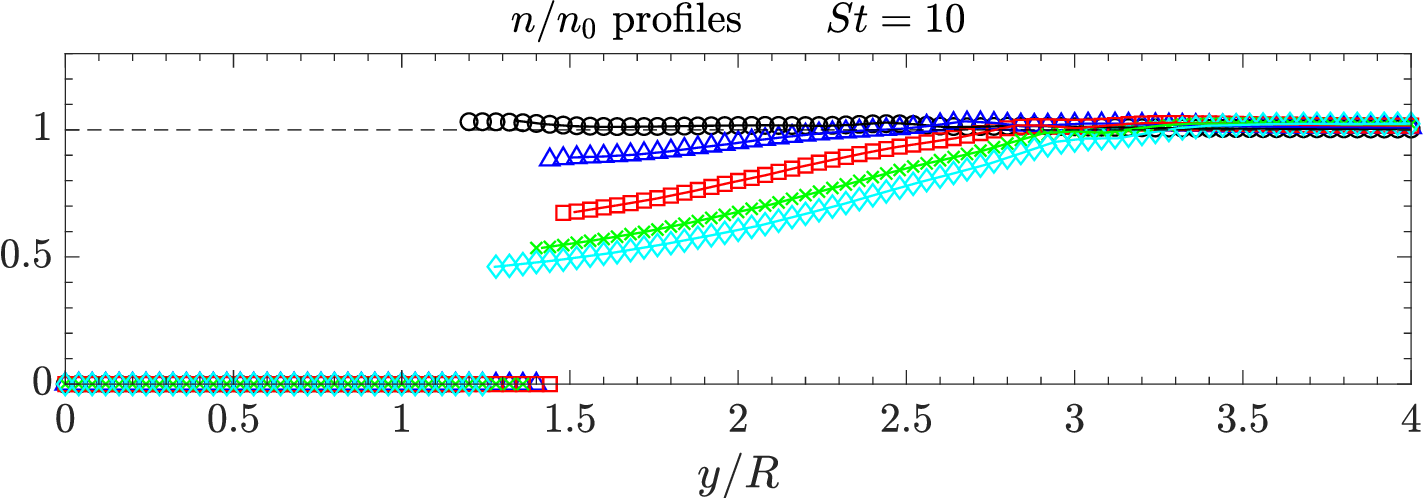}
        \caption{}
        \label{fig:3f}
    \end{subfigure}
    \caption{
        Reconstruction of ${{n}} / n_0$ using kernel regression in monodisperse steady-state flow around a cylinder at $Re = 20$ for: \eqref{fig:3a}  $St = 0.1$; \eqref{fig:3c} $St = 1$; \eqref{fig:3e} $St = 10$;
        Profiles of ${{n}} / n_0$ obtained using kernel regression (symbols) at selected locations
        {\color{black}$\boldsymbol{\bigcirc}$} ${x} / R = 3$, {\color{blue}$\boldsymbol{\bigtriangleup}$} ${x} / R = 6$,
        {\color{red}$\boldsymbol{\Box}$} ${x} / R = 9$,
        {\color{green}$\boldsymbol{\times}$} ${x} / R = 12$,
        {\color{cyan}$\boldsymbol{\diamond}$} ${x} / R = 15$
        compared against linear interpolation (lines) of FLA number density data for: \eqref{fig:3b}  $St = 0.1$; \eqref{fig:3d} $St = 1$; \eqref{fig:3f} $St = 10$.
    }
    \label{fig:3}
\end{figure*}

In order to reconstruct the Eulerian droplet number density field using the Nadaraya-Watson estimator~\eqref{eq:kernel-smoother} in the general multidimensional case, the structured kernel in Eq.~\eqref{eq:kernel-structured} is used with the bandwidth matrix $\bm{H}$ specified by Eq.~\eqref{eq:h-structured-kernel}. The reconstruction is done on a uniform Cartesian grid with a spacing of $\Delta {x} / R = \Delta {y} / R = 0.04$. This is considered for three different droplet sizes which correspond to $St = 0.1$, $1$, and $10$, and the number density field ${{n}}$ reconstructed using kernel regression is displayed in Figures \eqref{fig:3a}, \eqref{fig:3c}, and \eqref{fig:3e} respectively for each of these cases. The profiles of the number density field at selected values of ${x}$ are also displayed in Figures \eqref{fig:3b}, \eqref{fig:3d}, and \eqref{fig:3f} for these respective values of $St$.

Both the extent and number density distribution of the droplet field is seen to vary markedly with $St$, with the common feature to all cases being the wake behind the cylinder which is devoid of droplets. For $St = 0.1$ the droplets follow the flow relatively closely, and the two regions of the droplet field formed by droplets travelling above and below the cylinder meet downstream of the cylinder after a distance of $\sim 8R$. In contrast, the droplet field at the higher values of $St$ remains separated into two distinct regions beyond ${x} = 25R$. The other key difference between the cases lies within the rate of variation in number density of the droplet field. For $St = 0.1$ the droplet field is largely uniform away from the cylinder, and only experiences rapid variation along the edges of the cylinder wake, whilst $St = 1.0$ and $St = 10$ display a larger region of variation in the number density field, but with the change being progressively more gradual.

The profiles in Figures \eqref{fig:3b}, \eqref{fig:3d}, and \eqref{fig:3f} provide a means of assessing the efficacy of kernel regression against the exact values of number density obtained from the FLA. This is possible since in the case of a steady-state flow it is permissible to interpolate the values of number density along trajectories between time points \cite{Li2021}, and therefore obtain an accurate descriptor of the number density field along given cross-section profiles independently of kernel regression. Owing to the steady-state behaviour of the flow, the number density varies smoothly without the presence of folds and also largely monotonically, and these characteristics essentially mean that linear interpolation of the number density values along trajectories onto a chosen profile provides an exact solution against which kernel regression can be compared. Of particular note is the case for $St = 0.1$, in which the kernel regression performs well across a large part of the flow, but loses accuracy in the region of rapid variation along the edge of the cylinder wake, as depicted in Figure \eqref{fig:3b}. This is due to the perceptible size of the kernel causing the regression procedure to introduce a certain level of smoothing into the result, and therefore limiting the level of variation in number density which can be accurately captured. However, this is balanced against the kernel needing to be large enough to provide a smooth representation of the number density field for a given initial droplet seeding within a simulation, and as with the previous examples emphasises the importance of judicious selection of the initial smoothing length $h_0$. In these cases the default value of $h_0 = \Delta x_{d0}$ is selected to provide a reasonable coverage of the computational domain for the chosen initial seeding of 101 droplets which is used. Increasing the initial droplet seeding would enable a smaller value of $h_0$ to smoothly reconstruct the number density field, which would in turn provide greater accuracy in the regions of rapid variation at lower $St$. Nonetheless, at higher $St$ the more gradual variation in number density for the different ${x}$ profiles is successfully accounted for by kernel regression across the entire droplet phase, and the procedure provides a high level of accuracy as demonstrated in Figures \eqref{fig:3d} and \eqref{fig:3f}.
\begin{figure*}[!ht]
    \begin{subfigure}[c]{0.495\textwidth}
        \includegraphics[width=\textwidth,trim={0 0 0 0}]{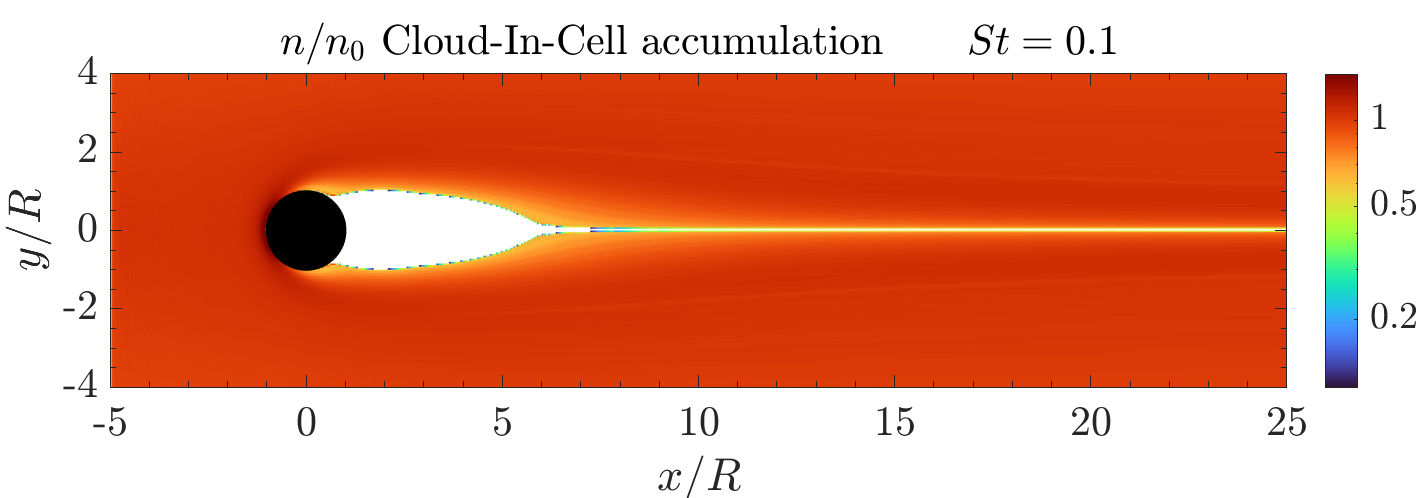}
        \caption{}
        \label{fig:cic-a}
    \end{subfigure}
    \begin{subfigure}[c]{0.495\textwidth}
        \includegraphics[width=\textwidth,trim={0 0 0 0}]{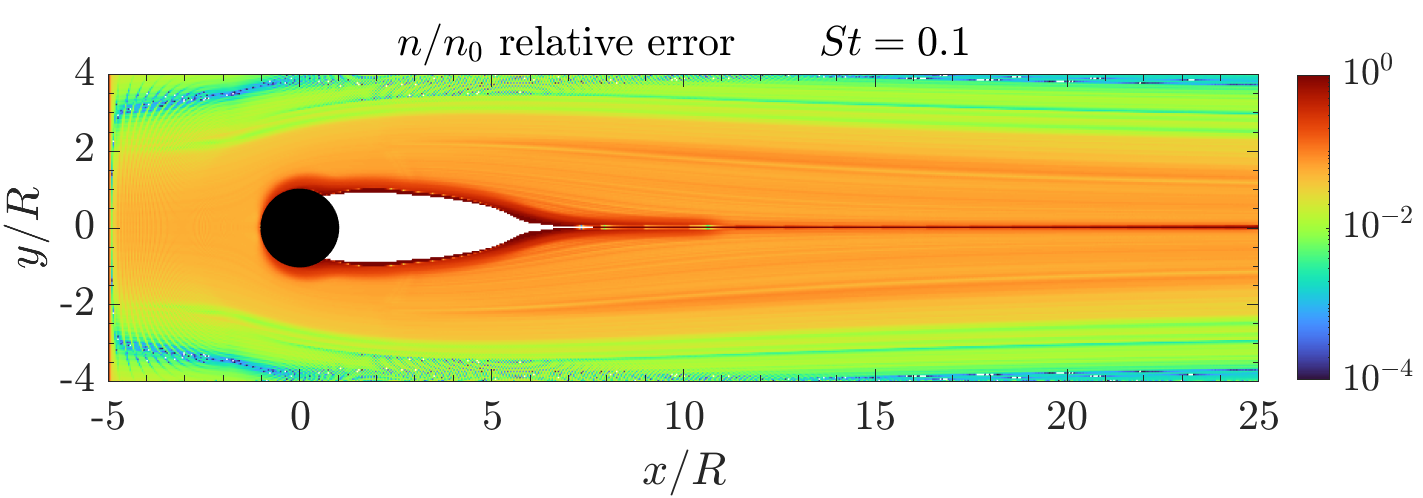}
        \caption{}
        \label{fig:cic-b}
    \end{subfigure}
    \begin{subfigure}[c]{0.495\textwidth}
        \includegraphics[width=\textwidth,trim={0 0 0 0}]{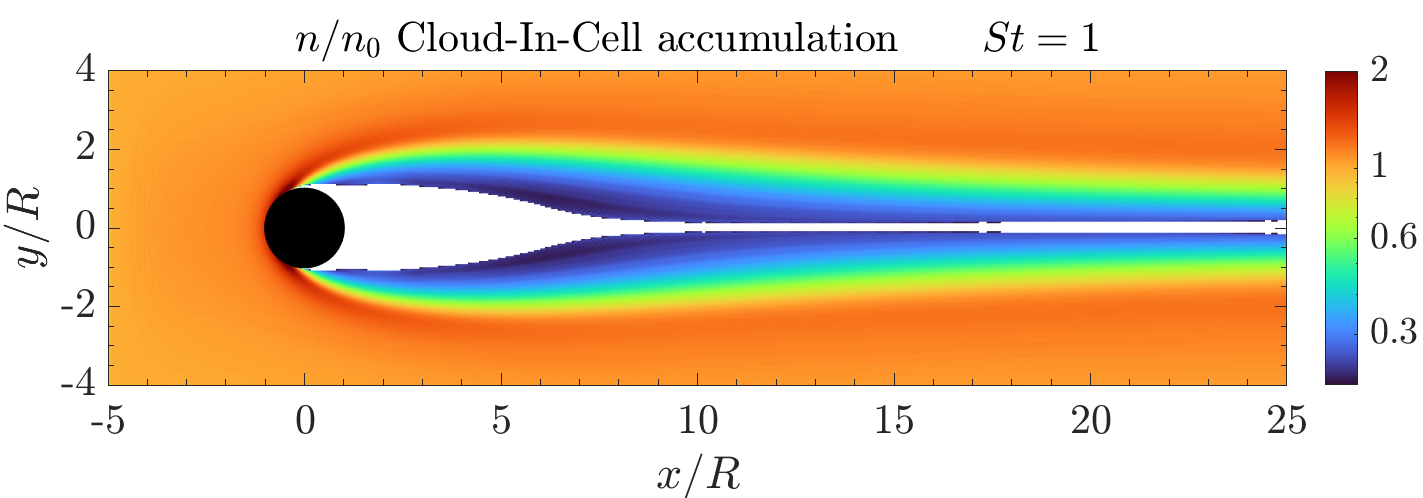}
        \caption{}
        \label{fig:cic-c}
    \end{subfigure}
    \begin{subfigure}[c]{0.495\textwidth}
        \includegraphics[width=\textwidth,trim={0 0 0 0}]{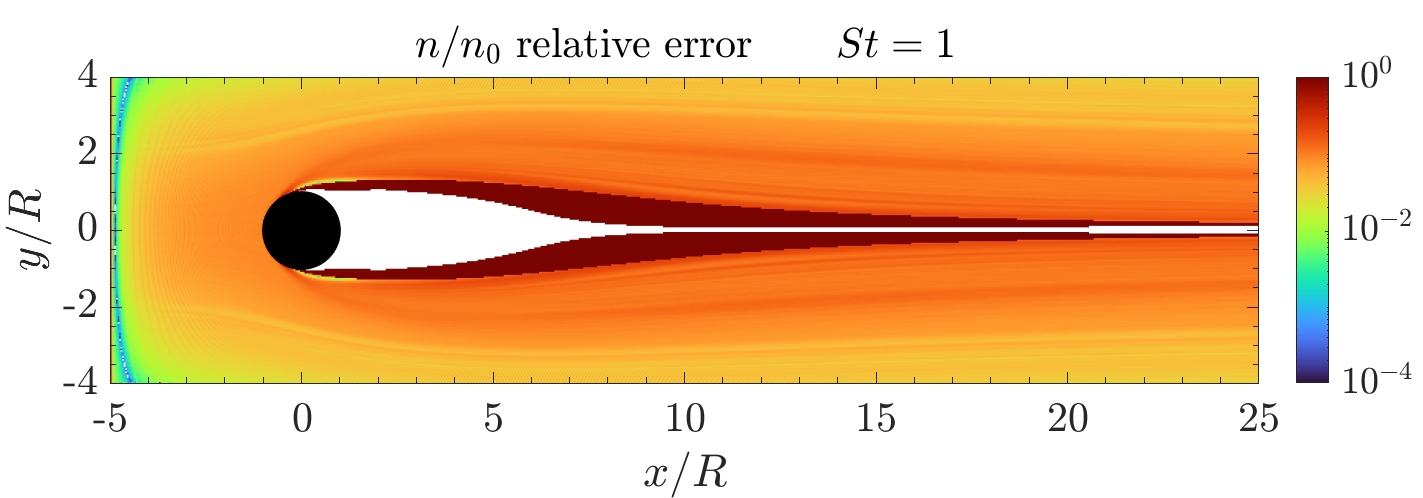}
        \caption{}
        \label{fig:cic-d}
    \end{subfigure}
    \begin{subfigure}[c]{0.495\textwidth}
        \includegraphics[width=\textwidth,trim={0 0 0 0}]{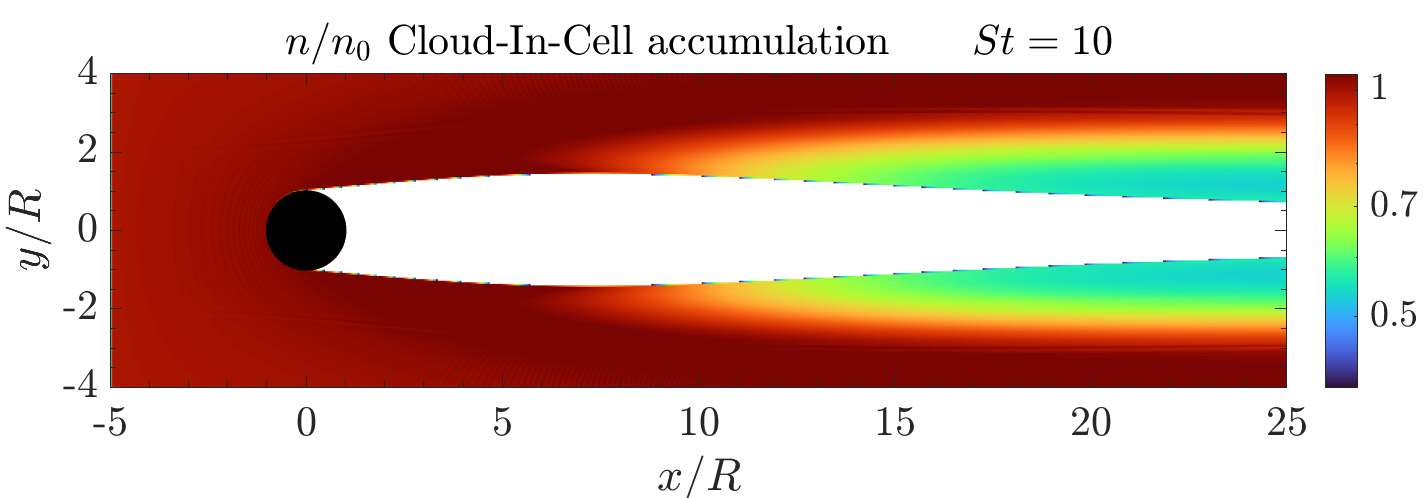}
        \caption{}
        \label{fig:cic-e}
    \end{subfigure}
    \begin{subfigure}[c]{0.495\textwidth}
        \includegraphics[width=\textwidth,trim={0 0 0 0}]{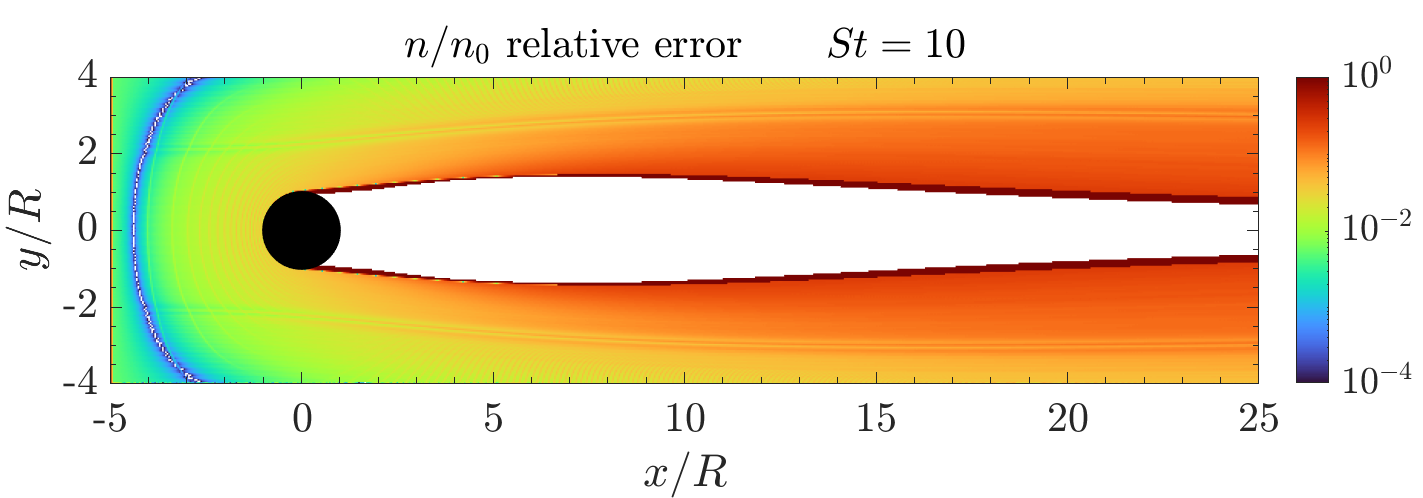}
        \caption{}
        \label{fig:cic-f}
    \end{subfigure}
    \caption{
        Reconstruction of ${{n}} / n_0$ using Cloud-In-Cell accumulation in monodisperse steady-state flow around a cylinder at $Re = 20$ for: \eqref{fig:cic-a}  $St = 0.1$; \eqref{fig:cic-c} $St = 1$; \eqref{fig:cic-e} $St = 10$;
        Relative error in ${{n}} / n_0$ between kernel regression and Cloud-In-Cell accumulation for: \eqref{fig:cic-b} $St = 0.1$ (from Figures \eqref{fig:3a} and \eqref{fig:cic-a}); \eqref{fig:cic-d} $St = 1$ (from Figures \eqref{fig:3c} and \eqref{fig:cic-c}); \eqref{fig:cic-f} $St = 10$ (from Figures \eqref{fig:3e} and \eqref{fig:cic-e}).
    }
    \label{fig:cic}
\end{figure*}

An illustration of the computational efficiency afforded by kernel regression can be made for this steady state approach by comparing to the number density field obtained using direct trajectory methods. In this instance, the Cloud-In-Cell approach that was outlined in Section \ref{sec:existing-interpolation} is used, with the accumulation made upon the same Eulerian grid as used for kernel regression. Since the number density is calculated directly in the CIC approach, a sufficient number of droplets are required in each grid cell to produce a stable result, and this is achieved by a higher initial seeding of droplets across the injection interval ${y} / R \in [-4,4]$. In contrast to the FLA which uses 101 droplets within this range, the CIC approach requires 10001 uniformly spaced droplets across the interval for the resultant Eulerian number density field to sufficiently converge, and furthermore the injection rate of the droplets also has to be increased by a factor of 10. Thus overall the number of droplet realisations required by the CIC approach is found to be $10^3$ times more than that by the FLA, consistent with previous works. The number density fields produced by this procedure are displayed in Figures \eqref{fig:cic-a}, \eqref{fig:cic-c}, and \eqref{fig:cic-e} for each of $St = 0.1$, $1$, and $10$ respectively, and it can be observed that the number density distribution is largely similar to that obtained from the FLA in the corresponding cases shown in Figures \eqref{fig:3a}, \eqref{fig:3c}, and \eqref{fig:3e}. The much higher seeding of trajectories needed for the CIC approach does, however, result in some differences to the profile of the wake behind the cylinder, and this is best seen through a direct comparison of the two procedures as illustrated in Figures \eqref{fig:cic-b}, \eqref{fig:cic-d}, and \eqref{fig:cic-f}. This depicts the relative error between the number density fields produced by the kernel regression and CIC approaches, and it is seen that this is largest along the edge of the wake, where the error can exceed $10^{-1}$. Away from the wake however, the error generally varies between $10^{-2}$ and $10^{-1}$, which is indicative of the level of accuracy that kernel regression of FLA data is able to achieve using $10^3$ times fewer droplet realisations, and the associated computational speedup that comes with this.

\subsubsection{Flow around a cylinder: transient case (Re = 100)} \label{sec:fpc-mono-transient}

\begin{figure*}[!ht]
    \begin{subfigure}[c]{0.495\textwidth}
        \includegraphics[width=\textwidth,trim={0 0 0 0}]{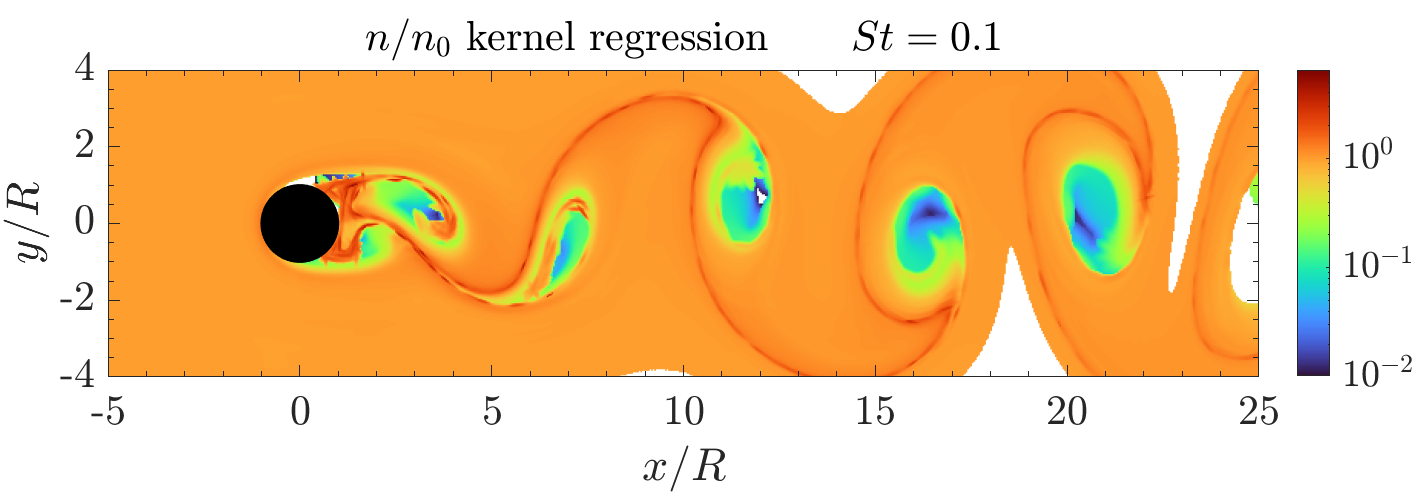}
        \caption{}
        \label{fig:4a}
    \end{subfigure}
    \begin{subfigure}[c]{0.495\textwidth}
        \includegraphics[width=\textwidth,trim={0 0 0 0}]{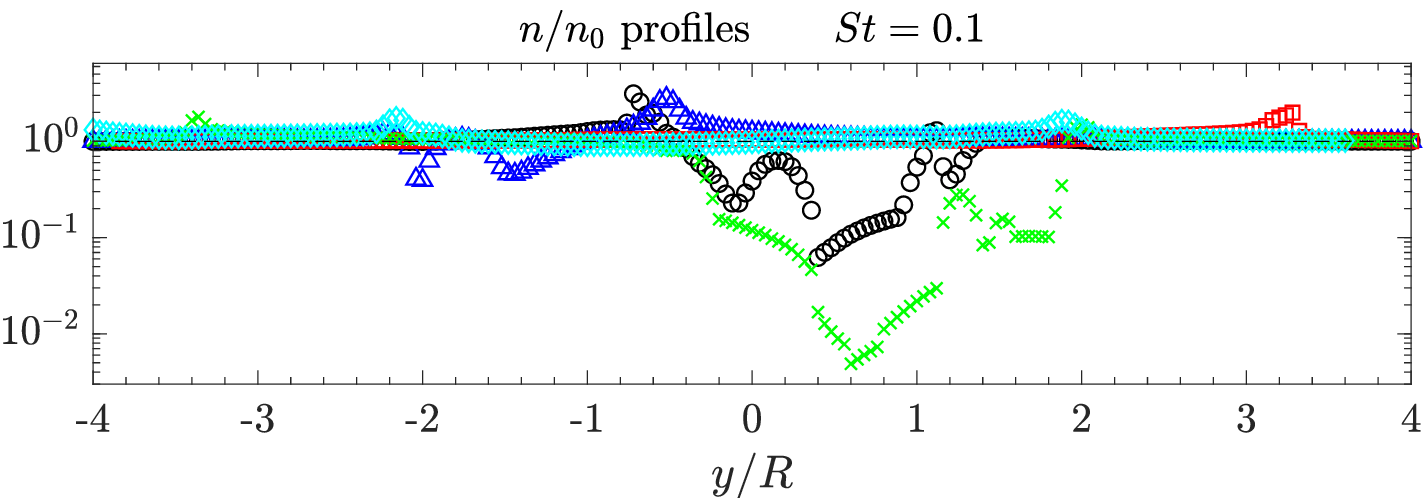}
        \caption{}
        \label{fig:4b}
    \end{subfigure}
    \begin{subfigure}[c]{0.495\textwidth}
        \includegraphics[width=\textwidth,trim={0 0 0 0}]{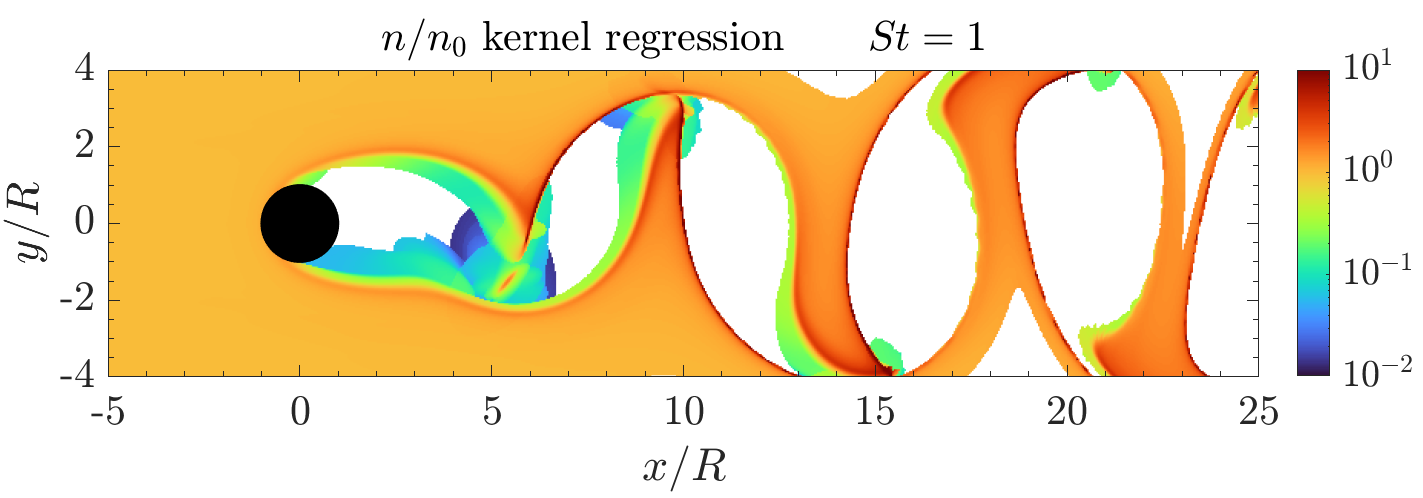}
        \caption{}
        \label{fig:4c}
    \end{subfigure}
    \begin{subfigure}[c]{0.495\textwidth}
        \includegraphics[width=\textwidth,trim={0 0 0 0}]{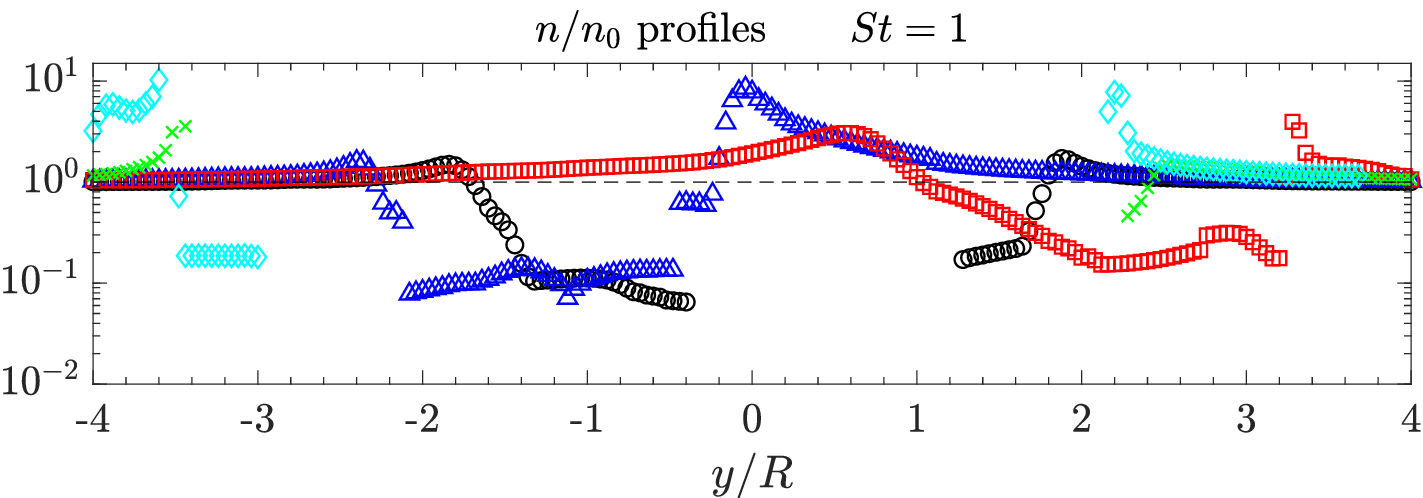}
        \caption{}
        \label{fig:4d}
    \end{subfigure}
    \begin{subfigure}[c]{0.495\textwidth}
        \includegraphics[width=\textwidth,trim={0 0 0 0}]{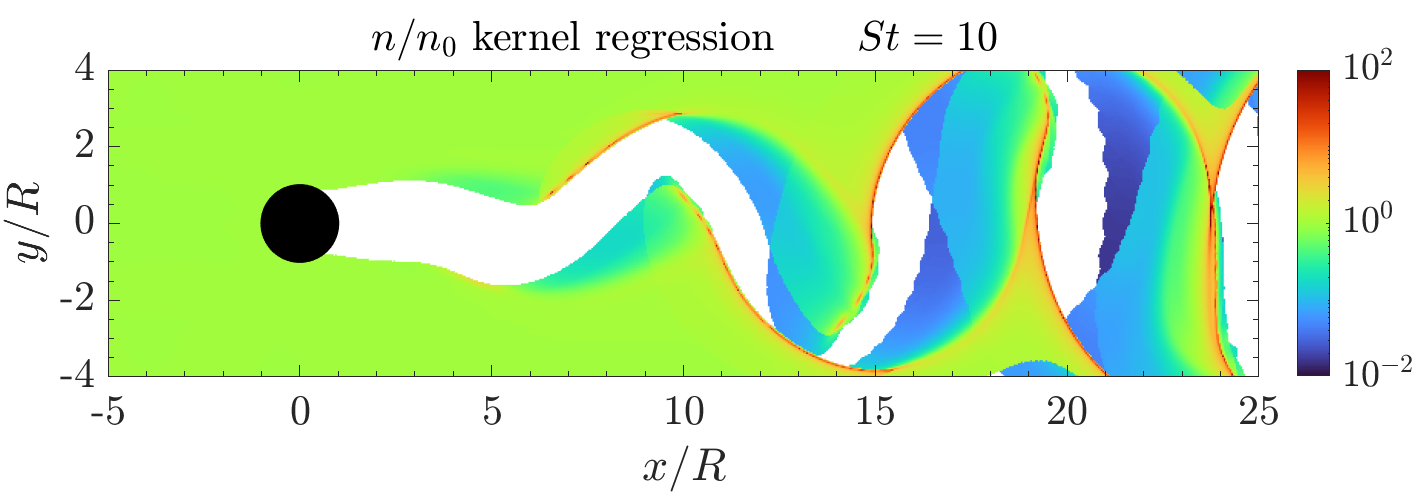}
        \caption{}
        \label{fig:4e}
    \end{subfigure}
    \begin{subfigure}[c]{0.495\textwidth}
        \includegraphics[width=\textwidth,trim={0 0 0 0}]{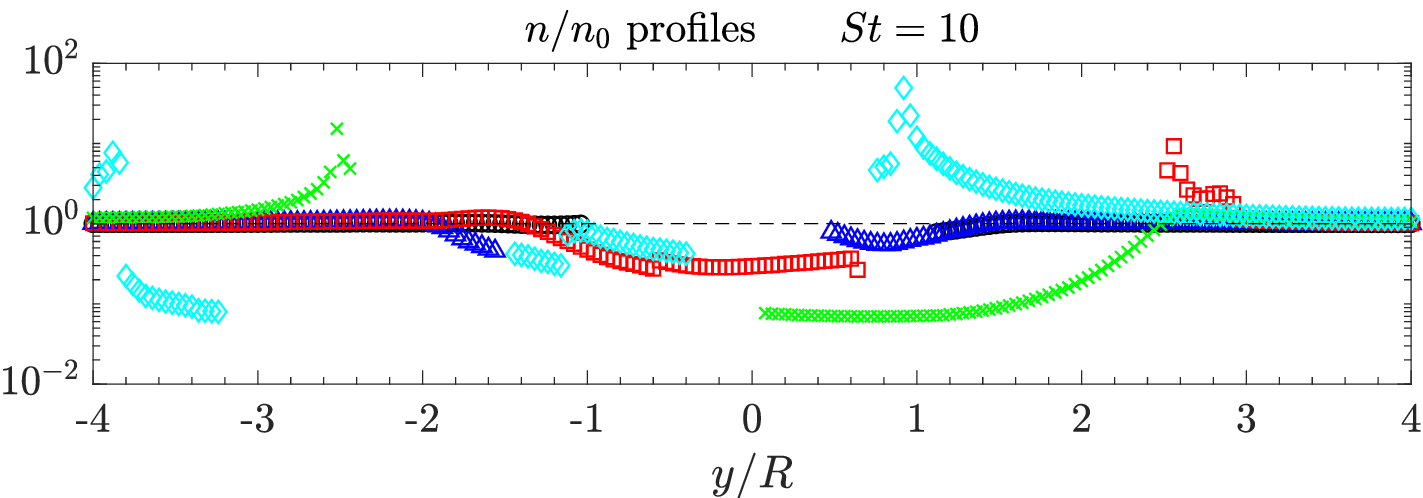}
        \caption{}
        \label{fig:4f}
    \end{subfigure}
    \caption{
        Reconstruction of ${{n}} / n_0$ using kernel regression in monodisperse transient flow around a cylinder at $Re = 100$ and time $t = 40$ for: \eqref{fig:4a} $St = 0.1$; \eqref{fig:4c} $St = 1$; \eqref{fig:4e} $St = 10$;
        Profiles of ${{n}} / n_0$ obtained using kernel regression at selected locations
        {\color{black}$\boldsymbol{\bigcirc}$} ${x} / R = 3$, {\color{blue}$\boldsymbol{\bigtriangleup}$} ${x} / R = 6$,
        {\color{red}$\boldsymbol{\Box}$} ${x} / R = 9$,
        {\color{green}$\boldsymbol{\times}$} ${x} / R = 12$,
        {\color{cyan}$\boldsymbol{\diamond}$} ${x} / R = 15$
        for: \eqref{fig:4b}  $St = 0.1$; \eqref{fig:4d} $St = 1$; \eqref{fig:4f} $St = 10$.
    }
    \label{fig:4}
\end{figure*}

For consideration of droplet behaviour in the transient regime, the case of $Re = 100$ is used. The underlying carrier flow becomes periodic at this level of unsteadiness, and forms the well-known phenomenon of a von K\'{a}rm\'{a}n vortex street. The configuration is identical to that in Section \ref{sec:fpc-mono-steady} except that the value of $Re$ is higher and the initial droplet seeding consists of 81 droplets that are injected over the interval ${y} / R \in [-4,4]$ with uniform spacing at each timestep. This corresponds to an initial average inter-droplet spacing of $\Delta x_{d0} = 0.1$, however the initial smoothing length is kept as $h_0 = 0.08$ as for the steady-state case, in order to achieve better resolution of the transient structures in the droplet number density field. In practice, this value of $h_0 = 0.8 \, \Delta x_{d0}$ also represents the lower size limit of initial smoothing length that will result in a smooth number density field for transient configurations, and as such has been found to be the optimum value for achieving this trade-off in these cases. As before, the carrier flow evolves according to Eqs.~\eqref{eq:carrier-flow-eqs} and the droplets and Jacobian are governed by Eqs.~\eqref{eq:fpc-part-eqns}. The three distinct droplet sizes corresponding to $St = 0.1$, $1$, and $10$ are again considered, with the number density field ${{n}}$ reconstructed using kernel regression at $t = 40$ displayed in Figures \eqref{fig:4a}, \eqref{fig:4c}, and \eqref{fig:4e} respectively for each case, and the associated profiles of the number density field at selected values of ${x}$ given in Figures \eqref{fig:4b}, \eqref{fig:4d}, and \eqref{fig:4f}.

The droplet behaviour is seen to reflect that of the carrier flow, with clear build-ups and voids in the number density field that are directly influenced by the structure of the vortices that form in the wake of the cylinder. The crucial feature of the dispersed phase behaviour is the effect of different levels of droplet inertia on the number density field at the different values of $St$. For $St = 0.1$ the voids in the droplet number density field are closely aligned with the location of vortices in the carrier flow due to the relatively low droplet inertia, as observed in Figure \eqref{fig:4a}. This is consistent with Maxey's centrifuging mechanism \cite{Maxey1987a}, which argues that low inertia droplets are ejected from areas of high vorticity. Of greater interest are the areas of higher droplet number density, which form along distinct curves between the vortices. The representation of this in Figure \eqref{fig:4a} is a demonstration that kernel regression is able to capture this level of detail accurately in the reconstructed number density field, and successfully account for the complex behaviour of the droplet field in transient flows. At higher $St$, the voids of low droplet number density that occur around vortices become much larger, in addition to the appearance of a wake downstream of the cylinder in which no droplets are present, as seen in Figures \eqref{fig:4c} and \eqref{fig:4e}. For $St = 1$ in \eqref{fig:4c}, the droplet field evolves as alternating regions in which droplets occur followed by voids that contain no droplets, with the region in which droplets are present characterised by a fold line of high number density along the leading edge and an area of lower number density along the trailing edge. For $St = 10$ this behaviour is also observed, with the fold lines exhibiting a higher number density than for $St = 1$, and the trailing edge of the droplet field containing both regions of lower number density and distinct layers of folds as observed from the discontinuities in the droplet field in Figure \eqref{fig:4e}. The successful capturing of these fold layers shows that kernel regression is able to retain this level of detail from the FLA number density data, and account for this phenomenon at a far lower computational cost than direct trajectory methods \cite{Mishchenko2019}.

The ${x}$ profiles of the number density field in Figures \eqref{fig:4b}, \eqref{fig:4d}, and \eqref{fig:4f} exhibit the different levels of variation in number density that the kernel regression procedure is able to reproduce. For intervals where there is no data for a given profile, the number density is zero as there are no droplets present at that point. It can be seen that as $St$ increases, the maximum number density within the flow becomes higher, and the maximum spatial gradient of the number density also increases. In particular, for $St = 1$ and $St = 10$ the profiles in the vicinity of fold lines display rapid variation in number density, however kernel regression is capable of providing a smooth representation of this behaviour, which serves to illustrate the flexibility of this procedure for use in complex flow configurations.

\subsection{Polydisperse droplets} \label{sec:results-poly}

Generalisation to the case of polydisperse droplets enables investigation of the droplet size distribution within a flow. Extension of the kernel regression framework into radial space is outlined in Section \ref{sec:kernel-gfla}, and in the following some of the flow configurations from the monodisperse cases in Section \ref{sec:results-mono} are used to illustrate the efficacy of this procedure.

To specify the different droplet sizes within simulations, an appropriate probability distribution is defined at the outset as a function of the initial droplet radius $r_0 = r_d (t_0)$. Following previous work \cite{Li2021}, a log-normal distribution that is the same at all initial locations $\bm{x}_0 = \bm{x}_d (t_0)$ is assumed,
\begin{equation} \label{eq:init-size-distribution}
p (\bm{x}_0,r_0,t_0) = \frac{1}{r_0}\frac{1}{\sqrt{2\pi}\sigma} \exp \left[ - \frac{(\ln (r_0) - \mu)^2}{2\sigma^2} \right] \, ,
\end{equation}
in which the mean and standard deviation parameters are chosen to be $\mu = 0.16$ and $\sigma = 0.4$ respectively. The droplet size $r_0$ in Eq.~\eqref{eq:init-size-distribution} is nondimensionalised by $r_{d0}^*$, which is a reference droplet radius corresponding to the peak of the distribution. Whilst a log-normal distribution is representative of the spread of droplet sizes in some applications \cite{Kooij2018}, an alternative distribution, for instance Rosin-Rammler, can easily be specified if it is more physically appropriate.

Of interest in polydisperse droplet flows is not only the distribution of droplet sizes, but also the
spatial distribution and average size statistics across all sizes of droplet. These can be determined by considering the moments of ${{p}}$, which utilises its interpretation as a probability density field. Specifically, the number density ${{n}}$, average radius $\overline{r}$, and radius variance $\overline{r^{\prime}r^{\prime}}$ can be obtained using the definitions
\begin{subequations}
    \label{eq:pdf-moments}
    \begin{align}
        {{n}} (\bm{x},t) & = \int_{r} {{p}} (\bm{x},{r},t) \, d{r} \, ,
        \label{eq:number-density} \\
        \overline{r} (\bm{x},t) & = \frac{1}{n (\bm{x},t)} \int_{r} r {{p}} (\bm{x},{r},t) \, d{r} \, ,
        \label{eq:average-radius} \\
        \overline{r^{\prime}r^{\prime}} (\bm{x},t) & = \frac{1}{n (\bm{x},t)} \int_{r} (r - \overline{r}) (r - \overline{r}) {{p}} (\bm{x},{r},t) \, d{r} \, .
        \label{eq:radius-variance}
    \end{align}
\end{subequations}
The averaged field variables in Eqs.~\eqref{eq:pdf-moments} are evaluated within simulations by numerically integrating the probability density ${{p}}$ obtained from the kernel regression procedure across all droplet sizes in accordance with the various radius weightings used. Here attention is restricted to just the moments of ${{p}}$ given in Eqs.~\eqref{eq:pdf-moments}, however it is possible to define further higher-order statistics of the droplet size distribution as required by the case under consideration, for instance the skewness and kurtosis may be of interest in transient flows. It is also possible to obtain relevant statistics for industrial spray systems, for instance the Sauter mean diameter, from ${{p}}$ using a similar procedure.

\subsubsection{One-dimensional quiescent flow with evaporating droplets}

To illustrate the behaviour of evaporating droplets in a simple case, a one-dimensional flow of droplets in quiescent air with uniform temperature is considered, as previously used in \cite{Li2021}. Droplets are initially located at ${x}_0 = 0$, with velocity ${v}_0 = 1$, radius ${r}_0 \in [0,4]$ and the probability density $p (\bm{x}_d (t_0),r_d (t_0),t_0)$ as specified by Eq.~\eqref{eq:init-size-distribution}. For the case of $St_0^* = 1$, the droplet motion and evaporation governed by Eqs.~\eqref{eq:part-rad-eom} become
\begin{subequations}
    \label{eq:gfla1d}
    \begin{align}
        \ddot{x}_d & = - \frac{1}{r_d^2} \dot{x}_d \, ,
        & x_d(t_0) = 0 \, , \, \dot{x}_d(t_0) = 1 \, ,
        \label{eq:gfla1d-part-eom} \\
        \dot{r}_d & = - \frac{\delta}{2 r_d} \, ,
        & {r}_d (t_0) \in [0,4] \, ,
        \label{eq:gfla1d-rad-eom}
    \end{align}
\end{subequations}
and the corresponding Jacobian evolution given in Eqs.~\eqref{eq:Jacobian-evolution} becomes
\begin{subequations}
    \label{eq:gfla1d-Jacobian-evolution}
    \begin{align}
        \ddot{{J}}^{x{x}} & = - \frac{1}{r_d^2} \dot{{J}}^{x{x}}
        + \frac{2}{r_d^3} \dot{x}_d {J}^{r{x}} \, ,
        \\
        \ddot{{J}}^{x{r}} & = - \frac{1}{r_d^2} \dot{{J}}^{x{r}}
        + \frac{2}{r_d^3} \dot{x}_d {J}^{r{r}} \, ,
        \\
        \dot{{J}}^{r{x}} & = \frac{\delta}{2 r_d^2} {J}^{r{x}} \, ,
        \\
        \dot{{J}}^{r{r}} & = \frac{\delta}{2 r_d^2} {J}^{r{r}} \, ,
    \end{align}
\end{subequations}
with the initial conditions as given in Eqs.~\eqref{eq:Jacobian-initial-conditions} along with $\dot{{J}}^{x{x}}({x}_0,{r}_0,t_0) = 0$. The systems \eqref{eq:gfla1d} and \eqref{eq:gfla1d-Jacobian-evolution} admit analytical solutions along trajectories as detailed in \cite{Li2021}. The probability density along trajectories is calculated using Eq.~\eqref{eq:Lagrangian-COM}, however to reconstruct the probability density field it is necessary to use multidimensional kernel regression as outlined in Section \ref{sec:kernel-gfla}, with the corresponding phase space for this case being $\boldsymbol{\xi} = ({x},{r})$. For this case, the droplet evaporation rate is specified by $\delta = 1$, and a total of 100 droplet sizes defined uniformly over the range $r_0 \in [0,4]$ are injected at the start of the simulation, giving an initial inter-droplet size spacing of $\Delta r_{d0} / r_{d0}^* = 0.04$. The initial radial space smoothing length is set as $h_{{r}0} = 0.5 \, \Delta r_{d0}$ to achieve a good resolution over the droplet size distribution, whilst as all droplets are started from the same location at $x = 0$ and $\Delta x_{d0}$ is consequently not defined in this case, the initial physical space smoothing length is also taken to be $h_{x0} = h_{{r}0} / r_{d0}^* = 0.5 \, \Delta r_{d0} / r_{d0}^*$ for consistency. The instantaneous phase space probability density field ${{p}}$ is reconstructed on a uniform Cartesian grid with spacing $\Delta {x} = 0.04$ and $\Delta {r} / r_{d0}^* = 0.04$
\begin{figure*}[!ht]
    \begin{subfigure}[c]{0.495\textwidth}
        \includegraphics[width=\textwidth,trim={0 0 0 0}]{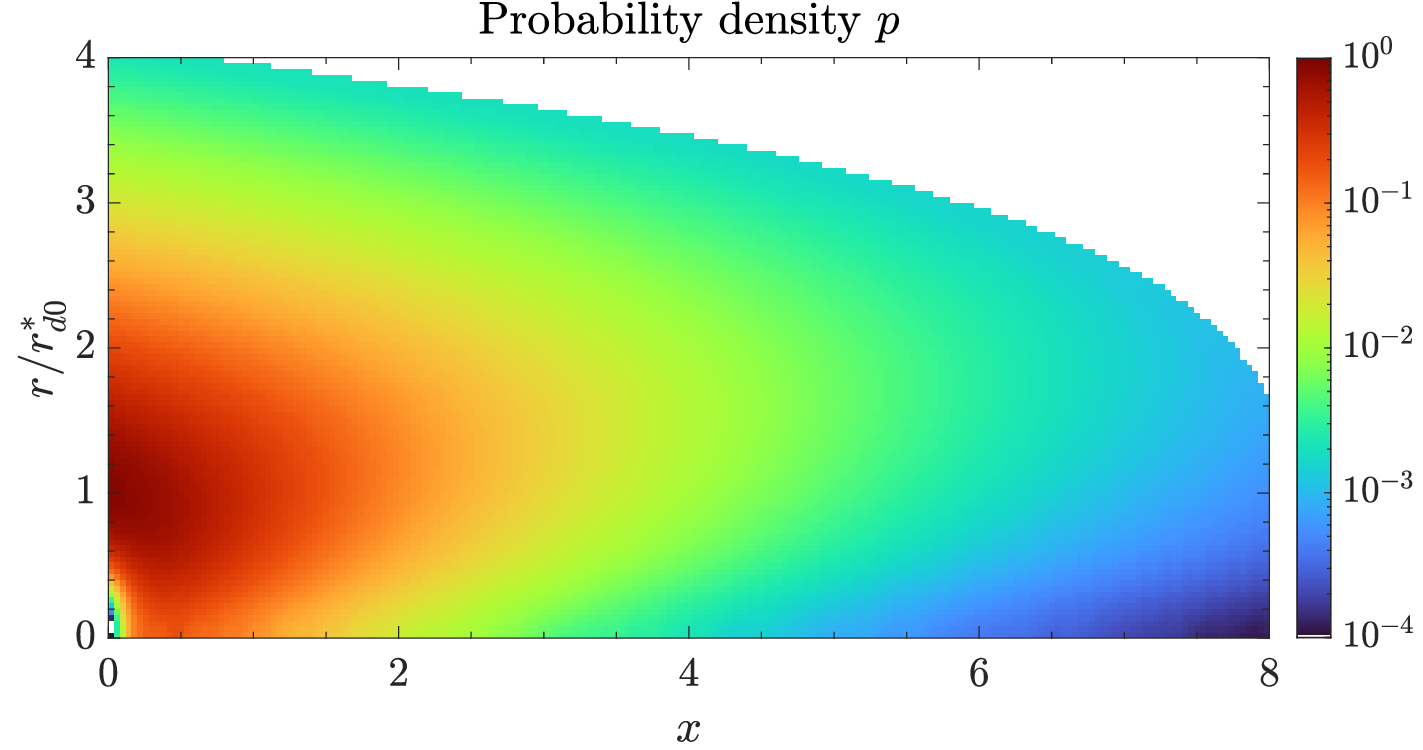}
        \caption{}
        \label{fig:5a}
    \end{subfigure}
    \begin{subfigure}[c]{0.495\textwidth}
        \includegraphics[width=\textwidth,trim={0 0 0 0}]{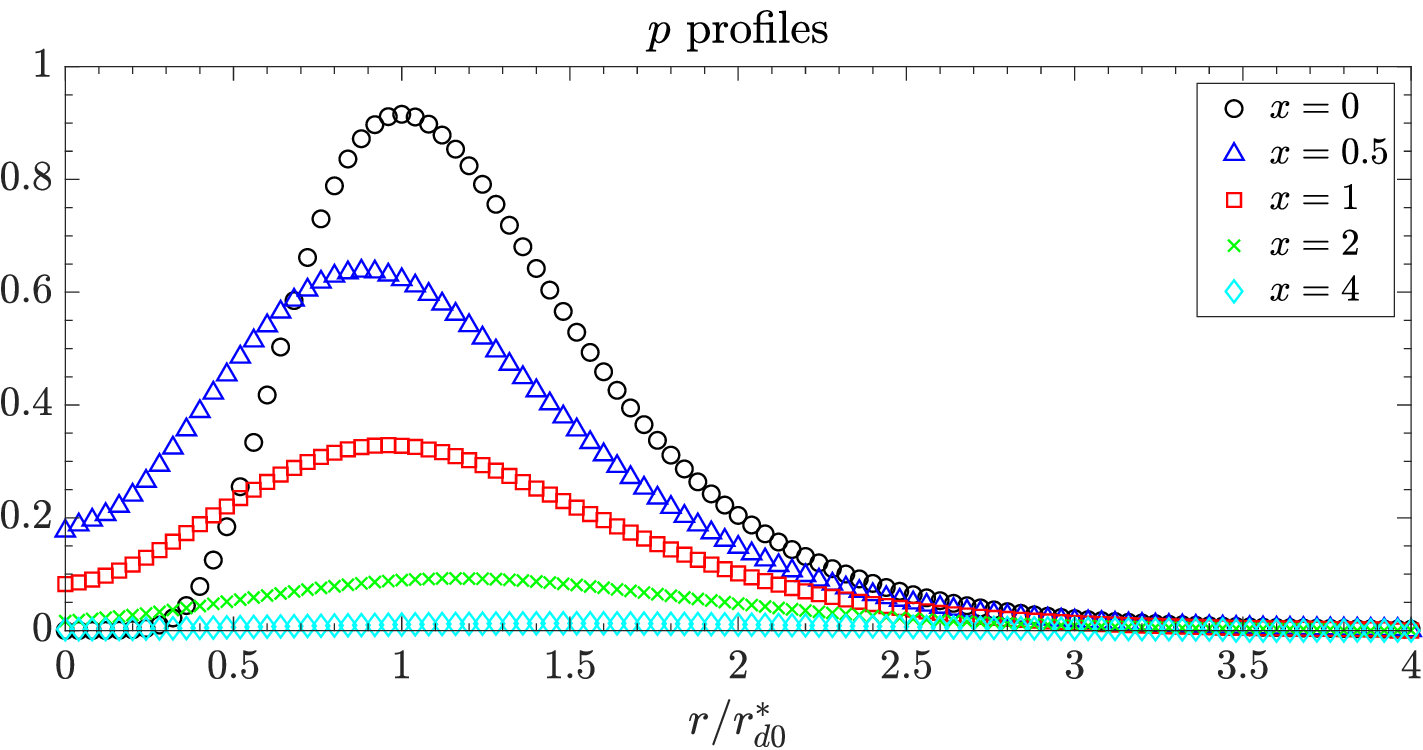}
        \caption{}
        \label{fig:5b}
    \end{subfigure}
    \begin{subfigure}[c]{0.495\textwidth}
        \includegraphics[width=\textwidth,trim={0 0 0 0}]{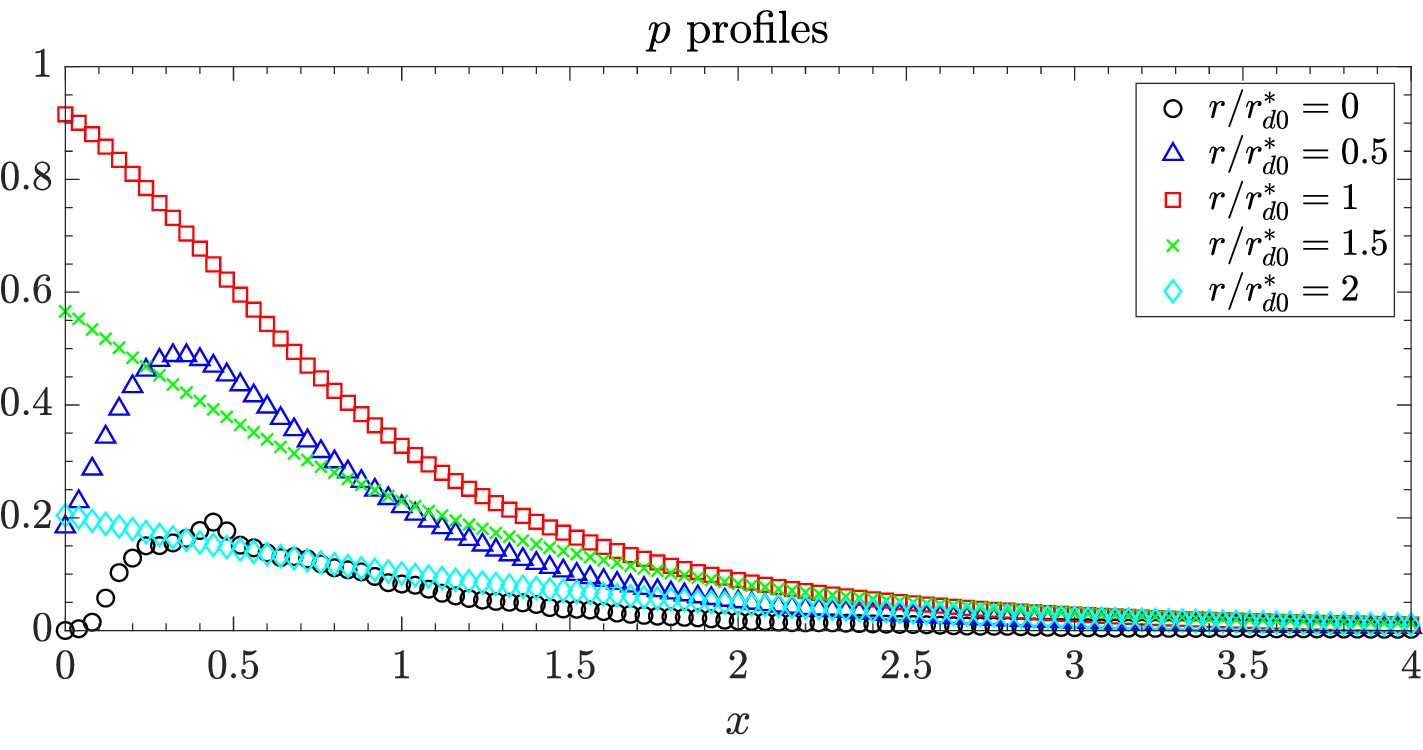}
        \caption{}
        \label{fig:5c}
    \end{subfigure}
    \begin{subfigure}[c]{0.495\textwidth}
        \includegraphics[width=\textwidth,trim={0 0 0 0}]{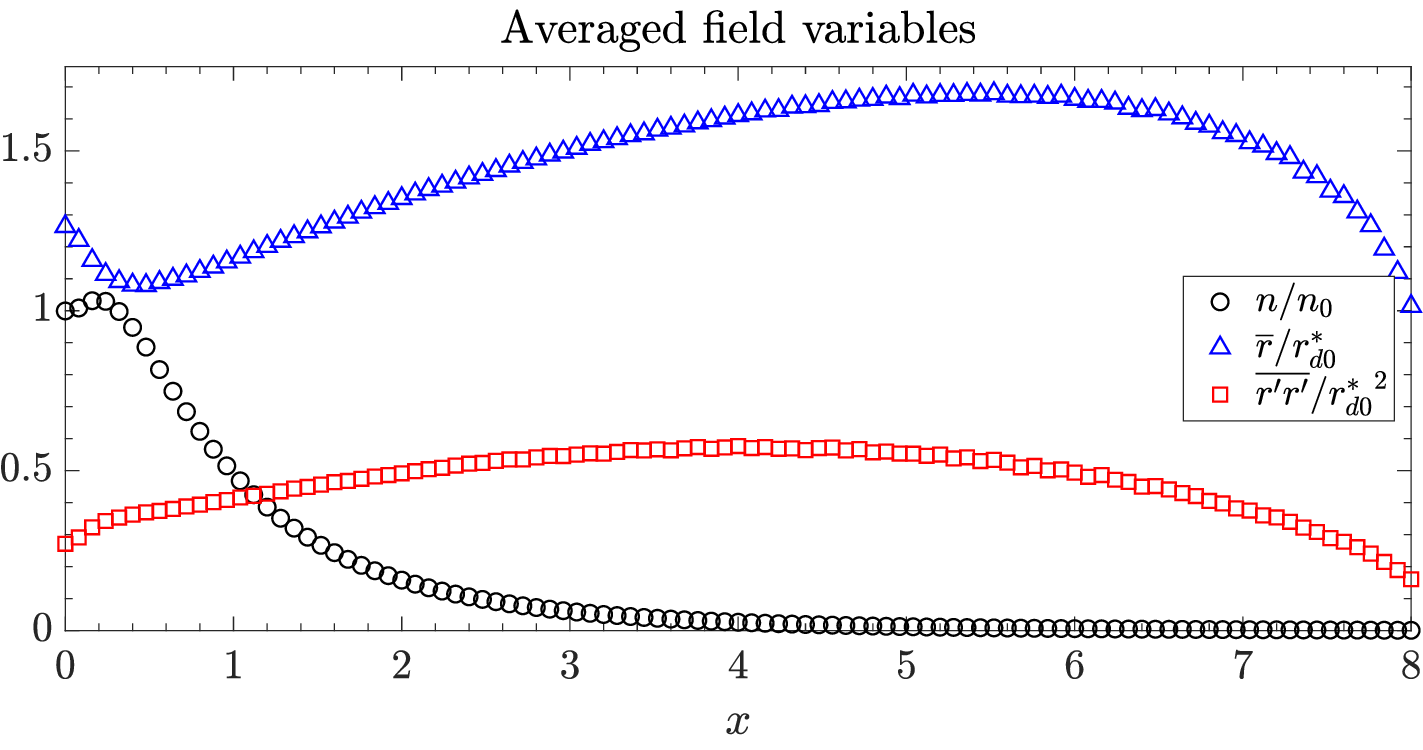}
        \caption{}
        \label{fig:5d}
    \end{subfigure}
    \caption{
        \eqref{fig:5a} Reconstruction of the probability density field ${{p}}$ using kernel regression for a one-dimensional quiescent flow with evaporating droplets; \eqref{fig:5b} Profiles of ${{p}}$ at selected $x$ locations;
        \eqref{fig:5c} Profiles of ${{p}}$ at selected values of $r / r_{d0}^*$;
        \eqref{fig:5d} Averaged field variables:
        {\color{black}$\boldsymbol{\bigcirc}$} $n / n_0$,
        {\color{blue}$\boldsymbol{\bigtriangleup}$} $\overline{r} / r_{d0}^*$,
        {\color{red}$\boldsymbol{\Box}$} $\overline{r^{\prime}r^{\prime}} / {r_{d0}^*}^2$.
    }
    \label{fig:5}
\end{figure*}

The complete Eulerian probability density field in $({x},{r})$ space is depicted in Figure \eqref{fig:5a}, and it is seen that kernel regression is able to reproduce a smoothly varying distribution across both physical space and radial space, enabling the full behaviour of the evaporation process to be examined. In Figure \eqref{fig:5b}, the droplet size distribution profiles at selected ${x}$ locations are shown. The profile at ${x} = 0$ is the initial probability density distribution specified in Eq.~\eqref{eq:init-size-distribution}, with the profiles at subsequent locations reflecting the decrease in probability density as droplets evaporate whilst they travel through the domain. Figure \eqref{fig:5c} depicts the droplet spatial distribution profiles at selected droplet sizes, with the peak probability occurring for $r = 1$ as also determined by Eq.~\eqref{eq:init-size-distribution}. The averaged field variables obtained from the moments of ${{p}}$ using Eqs.~\eqref{eq:pdf-moments} are given in Figure \eqref{fig:5d}. The number density $n$ initially shows a minor increase at small $x$, before droplets begin to fully evaporate and $n$ subsequently decreases exponentially. In contrast, the average radius $\overline{r}$ initially decreases at small $x$, before gradually growing and reaching its peak value at $x \approx 5.5$. This is indicative of the small droplets which form the peak of the distribution \eqref{eq:init-size-distribution} evaporating more quickly than larger droplets, causing the average size to increase during this period. Following this, $\overline{r}$ decreases as the larger droplets evaporate. Finally, the variance in droplet radius $\overline{r^{\prime}r^{\prime}}$ slowly increases as droplets travel across the domain, reaching a peak value at $x \approx 4$. This reflects the increased spread of the profiles displayed in Figure \eqref{fig:5b} at successive locations, before $\overline{r^{\prime}r^{\prime}}$ slowly decreases across the remainder of the domain. Although a simple example, this case serves to demonstrate the wealth of information that kernel regression is able to efficiently extract from the trajectory data in polydisperse flows.

\subsubsection{Two-dimensional fan spray injection in cross-flow} \label{sec:fla2d-poly}

This case is an extension of the configuration that was considered in Section \ref{sec:fla2d-mono} to polydisperse evaporating droplets, and has also been previously studied in \cite{Li2021}. For the case of $St_0^* = 1$, droplet motion and evaporation are governed by
\begin{subequations}
    \label{eq:gfla2d}
    \begin{align}
        \ddot{\bm{x}}_d & = \frac{1}{r_d^2} \left( \bm{u} - \dot{\bm{x}}_d \right) \, ,
        \label{eq:gfla2d-part-eom} \\
        \dot{r}_d & = - \frac{\delta}{2 r_d} \, ,
        & {r}_d (t_0) \in [0,4] \, ,
        \label{eq:gfla2d-rad-eom}
    \end{align}
\end{subequations}
where $\bm{u} = (1,0)$ as before, and the initial conditions for $\bm{x}_d$ and $\dot{\bm{x}}_d$ are identical to those in Eq.~\eqref{eq:fla2d-x-ic}. The corresponding Jacobian components evolve according to
\begin{subequations}
    \label{eq:gfla2d-Jacobian-evolution}
    \begin{align}
        \ddot{\bm{J}}^{\bm{x}\bm{x}} & = - \frac{1}{r_d^2} \dot{\bm{J}}^{\bm{x}\bm{x}}
        - \frac{2}{r_d^3} \left( \bm{u} - \dot{\bm{x}}_d \right) \cdot \bm{J}^{r\bm{x}} \, ,
        \\
        \ddot{\bm{J}}^{\bm{x}{r}} & = - \frac{1}{r_d^2} \dot{\bm{J}}^{\bm{x}{r}}
        - \frac{2}{r_d^3} \left( \bm{u} - \dot{\bm{x}}_d \right) {J}^{{r}{r}} \, ,
        \\
        \dot{\bm{J}}^{r\bm{x}} & = \frac{\delta}{2 r_d^2} \bm{J}^{r\bm{x}} \, ,
        \\
        \dot{{J}}^{r{r}} & = \frac{\delta}{2 r_d^2} {J}^{r{r}} \, ,
    \end{align}
\end{subequations}
where the initial conditions are as in Eqs.~\eqref{eq:Jacobian-initial-conditions}, and those for $\dot{\bm{J}}^{\bm{x}\bm{x}}$ being the same as in Eq.~\eqref{eq:fla2d-j-ic}. As in the monodisperse case, the systems \eqref{eq:gfla2d} and \eqref{eq:gfla2d-Jacobian-evolution} admit analytical solutions along trajectories, and these are given in \cite{Li2021}. Eq.~\eqref{eq:Lagrangian-COM} is used to calculate the probability density along trajectories, with multidimensional kernel regression subsequently used to reconstruct the probability density field over the phase space $\boldsymbol{\xi} = ({x},{y},{r})$ through the use of Eqs.~\eqref{eq:H-kernel-structured-gFLA} and \eqref{eq:H-kernel-gFLA}. In this configuration, the phase space probability density field ${{p}}$ is reconstructed on a uniform Cartesian grid with spacing $\Delta {x} = \Delta {y} = 0.01$ and $\Delta {r} / r_{d0}^* = 0.04$, with the droplet evaporation rate being determined by $\delta = 1$.  As in the monodisperse case, $\epsilon = 0.05$ is used to define the interval for droplet injection, and in this case at the start of the simulation 100 droplet sizes defined uniformly over the range ${r}_0 \in [0,4]$ are injected at each of 101 locations uniformly spread over the interval $\bm{x}_0 \in ([-\epsilon,\epsilon],0)$. This yields initial inter-droplet spacings of $\Delta x_{d0} = 0.001$ and $\Delta r_{d0} / r_{d0}^* = 0.04$ in position and size respectively, however due to the expanding nature of the spray in the spatial domain a larger initial physical space smoothing length of $h_{\bm{x}0} = 5 \Delta x_{d0}$ is required to achieve adequate coverage of the droplet field, whilst the default initial radial space smoothing length of $h_{{r}0} = \Delta r_{d0}$ is used.
\begin{figure*}[!ht]
    \begin{subfigure}[c]{0.495\textwidth}
        \includegraphics[width=\textwidth,trim={0 15 0 0}]{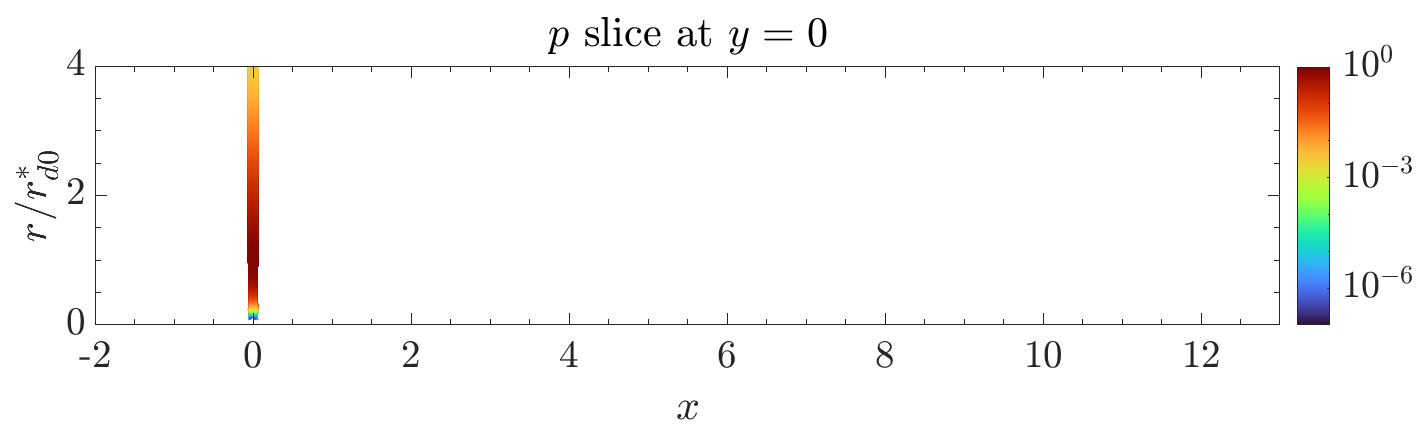}
        \caption{}
        \label{fig:6a}
    \end{subfigure}
    \begin{subfigure}[c]{0.495\textwidth}
        \includegraphics[width=\textwidth,trim={0 15 0 0}]{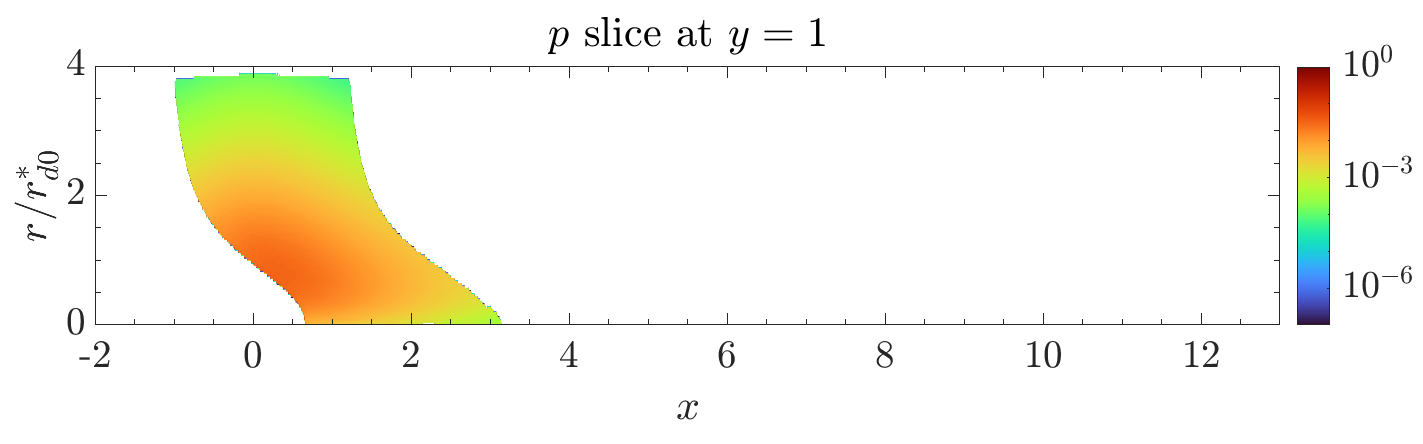}
        \caption{}
        \label{fig:6b}
    \end{subfigure}
    \begin{subfigure}[c]{0.495\textwidth}
        \includegraphics[width=\textwidth,trim={0 15 0 0}]{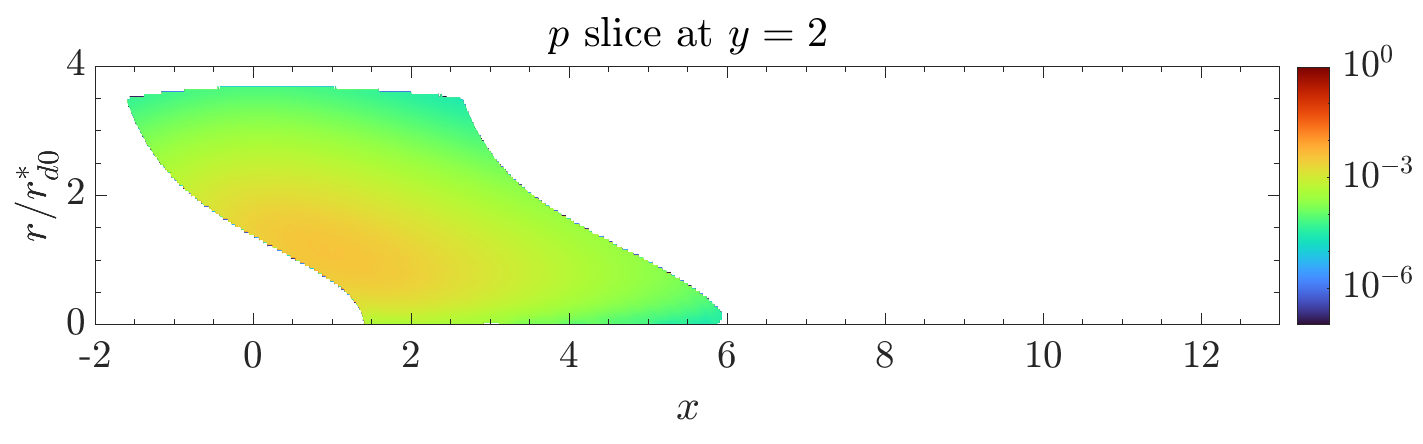}
        \caption{}
        \label{fig:6c}
    \end{subfigure}
    \begin{subfigure}[c]{0.495\textwidth}
        \includegraphics[width=\textwidth,trim={0 15 0 0}]{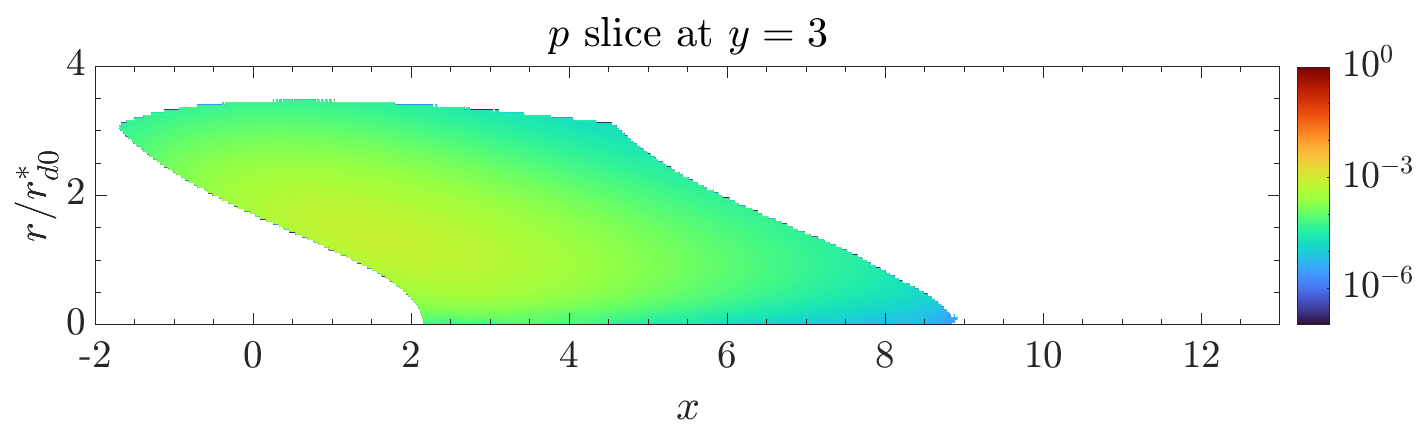}
        \caption{}
        \label{fig:6d}
    \end{subfigure}
    \begin{subfigure}[c]{0.495\textwidth}
        \includegraphics[width=\textwidth,trim={0 15 0 0}]{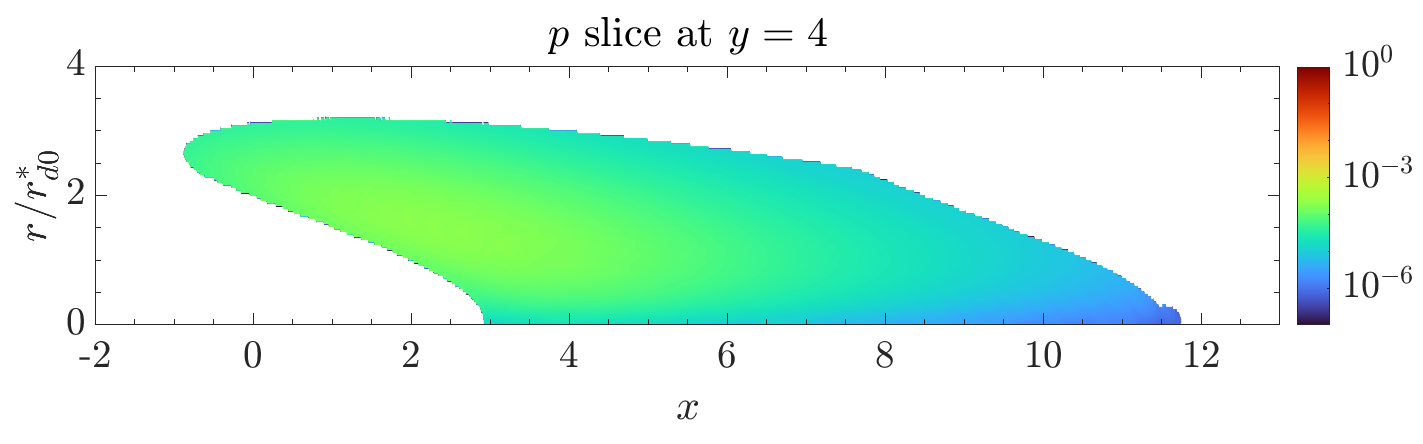}
        \caption{}
        \label{fig:6e}
    \end{subfigure}
    \begin{subfigure}[c]{0.495\textwidth}
        \includegraphics[width=\textwidth,trim={0 15 0 0}]{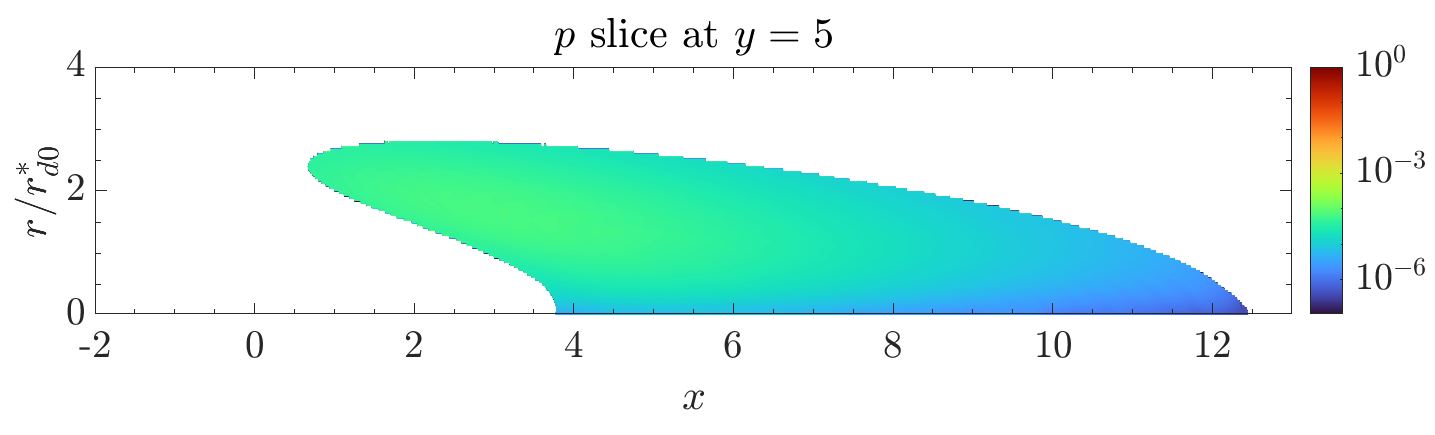}
        \caption{}
        \label{fig:6f}
    \end{subfigure}
    \caption{
        Slices of the probability density field ${{p}}$ obtained using kernel regression for a two-dimensional fan spray injection in cross-flow at selected ${y}$ locations: \eqref{fig:6a} ${y} = 0$; \eqref{fig:6b} ${y} = 1$; \eqref{fig:6c} ${y} = 2$; \eqref{fig:6d} ${y} = 3$; \eqref{fig:6e} ${y} = 4$; \eqref{fig:6f} ${y} = 5$.
    }
    \label{fig:6}
\end{figure*}

\begin{figure*}[!ht]
    \begin{subfigure}[c]{0.495\textwidth}
        \includegraphics[width=\textwidth,trim={0 15 0 0}]{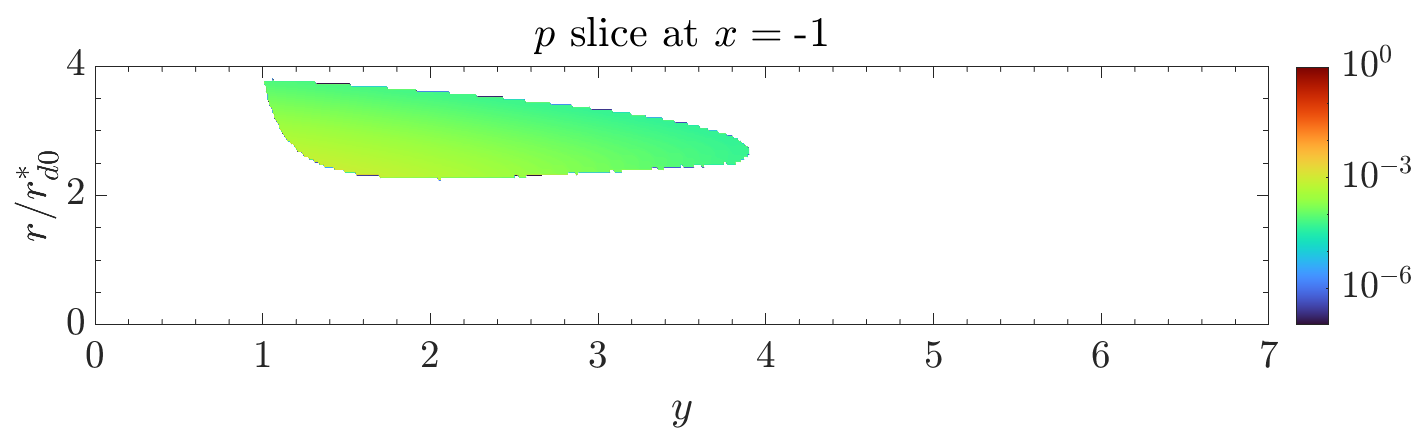}
        \caption{}
        \label{fig:7a}
    \end{subfigure}
    \begin{subfigure}[c]{0.495\textwidth}
        \includegraphics[width=\textwidth,trim={0 15 0 0}]{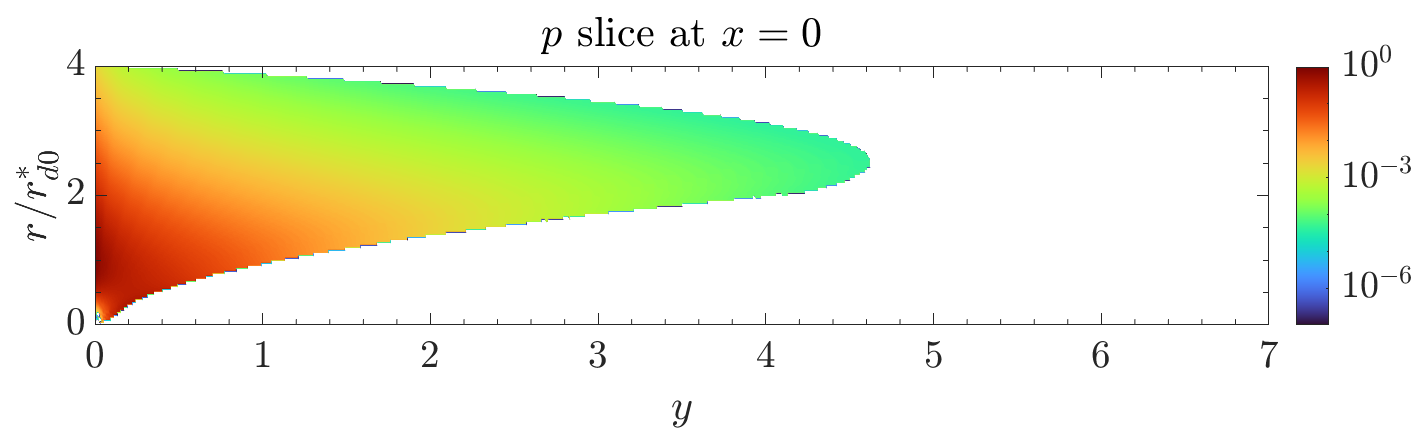}
        \caption{}
        \label{fig:7b}
    \end{subfigure}
    \begin{subfigure}[c]{0.495\textwidth}
        \includegraphics[width=\textwidth,trim={0 15 0 0}]{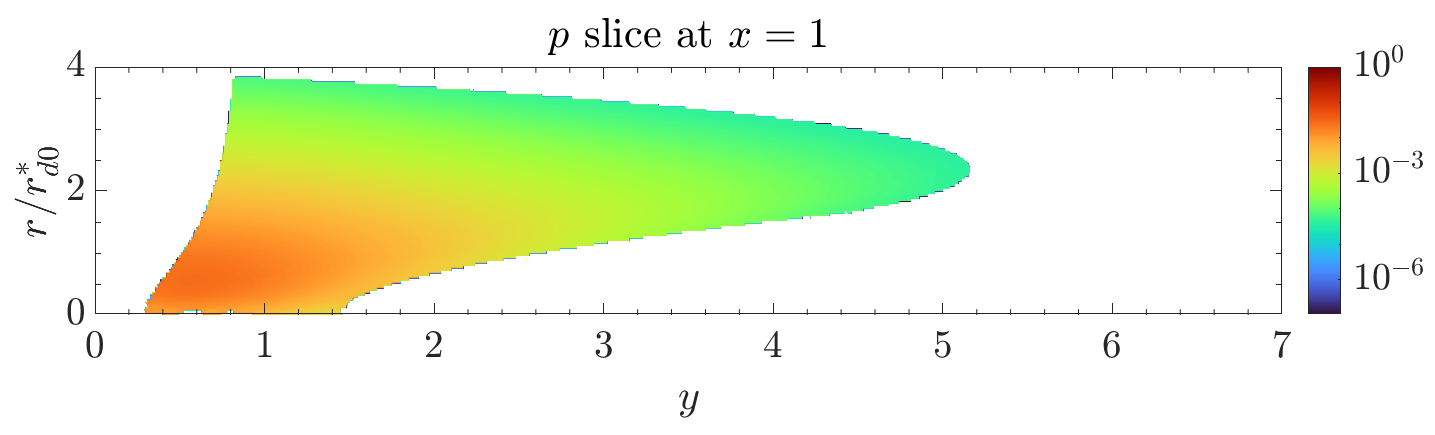}
        \caption{}
        \label{fig:7c}
    \end{subfigure}
    \begin{subfigure}[c]{0.495\textwidth}
        \includegraphics[width=\textwidth,trim={0 15 0 0}]{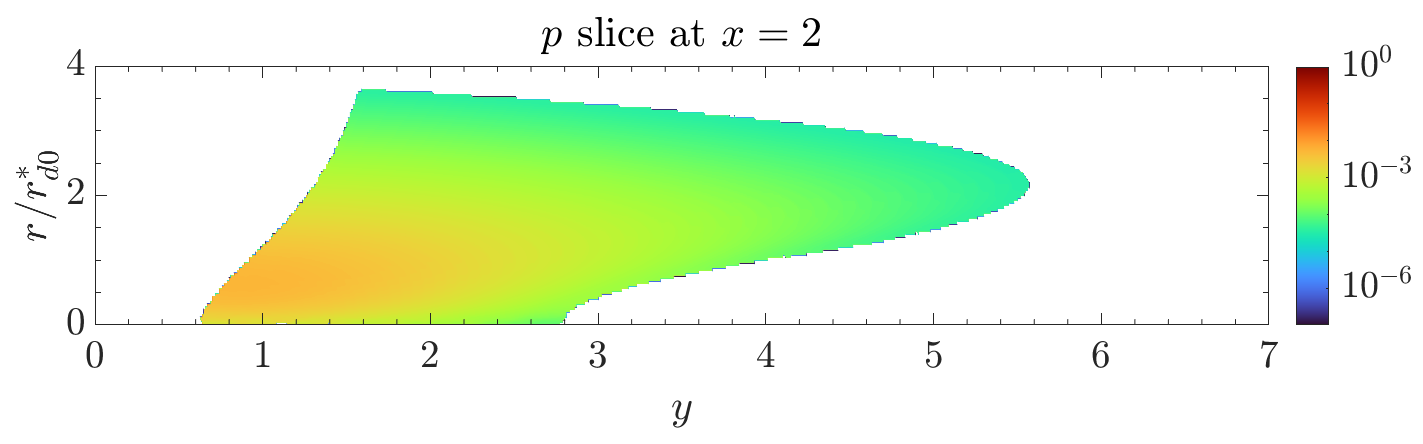}
        \caption{}
        \label{fig:7d}
    \end{subfigure}
    \begin{subfigure}[c]{0.495\textwidth}
        \includegraphics[width=\textwidth,trim={0 15 0 0}]{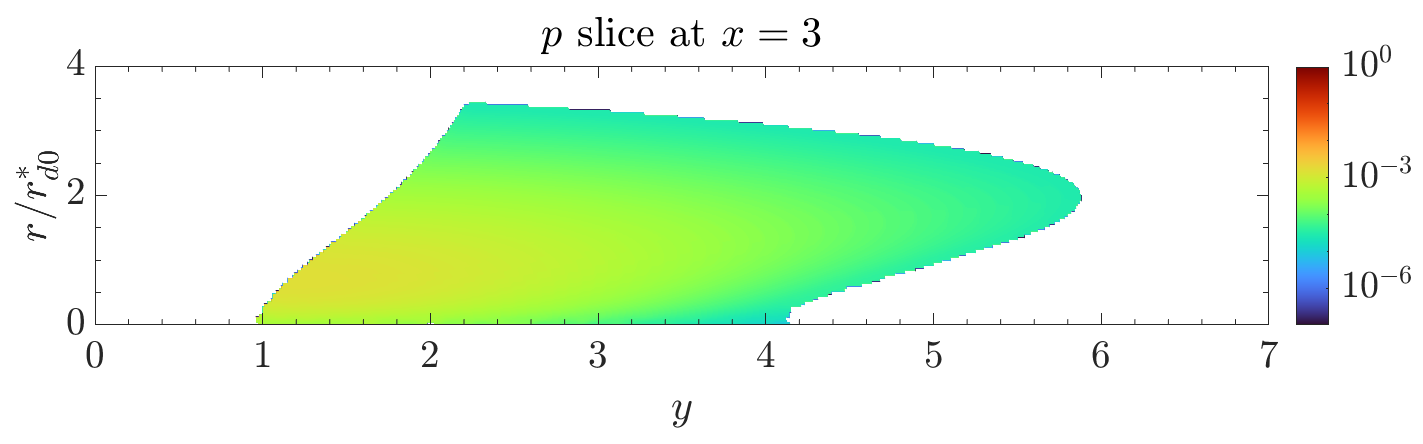}
        \caption{}
        \label{fig:7e}
    \end{subfigure}
    \begin{subfigure}[c]{0.495\textwidth}
        \includegraphics[width=\textwidth,trim={0 15 0 0}]{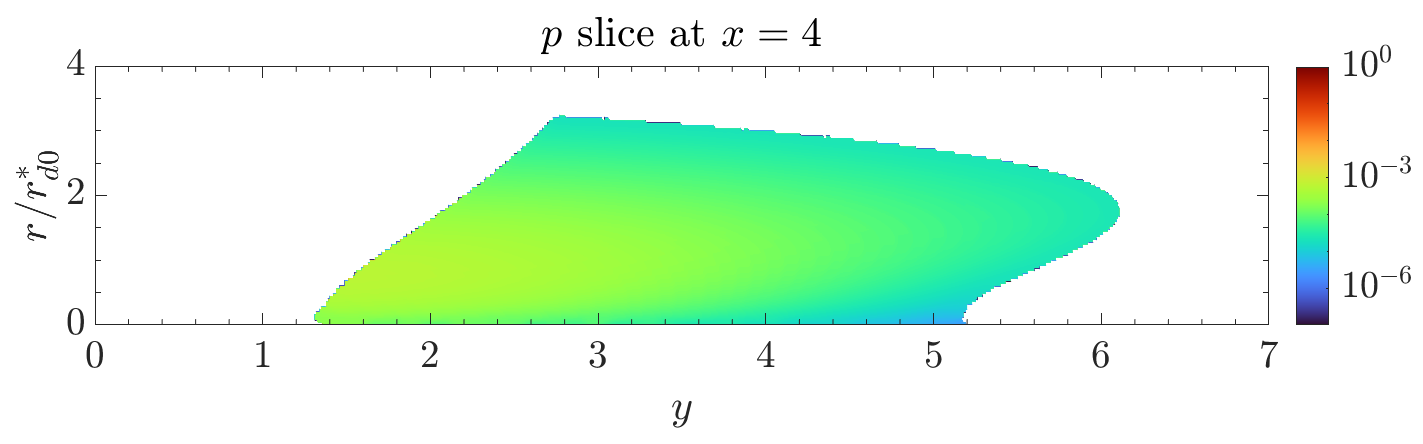}
        \caption{}
        \label{fig:7f}
    \end{subfigure}
    \caption{
        Slices of the probability density field ${{p}}$ obtained using kernel regression for a two-dimensional fan spray injection in cross-flow at selected ${x}$ locations: \eqref{fig:7a} ${x} = -1$; \eqref{fig:7b} ${x} = 0$; \eqref{fig:7c} ${x} = 1$; \eqref{fig:7d} ${x} = 2$; \eqref{fig:7e} ${x} = 3$; \eqref{fig:7f} ${x} = 4$.
    }
    \label{fig:7}
\end{figure*}

For the case of a multidimensional polydisperse droplet flow this enables the collection of a wealth of information about the steady-state droplet distribution across the spatial dimensions and range of droplet sizes, and Figures \ref{fig:6} and \ref{fig:7} are used to illustrate this, with the variation in droplet size distribution in one spatial direction (${x}$ and ${y}$) depicted as slices in the other spatial direction (${y}$ and ${x}$ respectively). This shows both how the spatial distribution of droplets varies as they travel away from the injection interval, and the range of droplet sizes which are present at a given location, but also importantly the distribution of droplet sizes within that range. It is seen that as droplets travel away from the injection interval in both the $x$ and $y$ directions, the range of droplet sizes decreases as the larger droplets evaporate, and the droplets become more dispersed in an asymmetrical profile which is reflective of the cross-flow configuration for the case. Of note is Figure \eqref{fig:7a} for the profile at $x = -1$, which illustrates the droplets that initially travel against the flow in the $x$ direction from the injection interval. It is observed that only sufficiently large droplets appear at this location, which is a consequence of them having enough inertia to maintain their initial momentum, in contrast to smaller droplets which are more responsive to the flow and do not reach $x = -1$. This highlights the ability of the kernel regression procedure to provide detailed insight into the evolution of polydisperse droplet flows.
\begin{figure*}[!ht]
    \begin{subfigure}[c]{0.495\textwidth}
        \includegraphics[width=\textwidth,trim={0 0 0 0}]{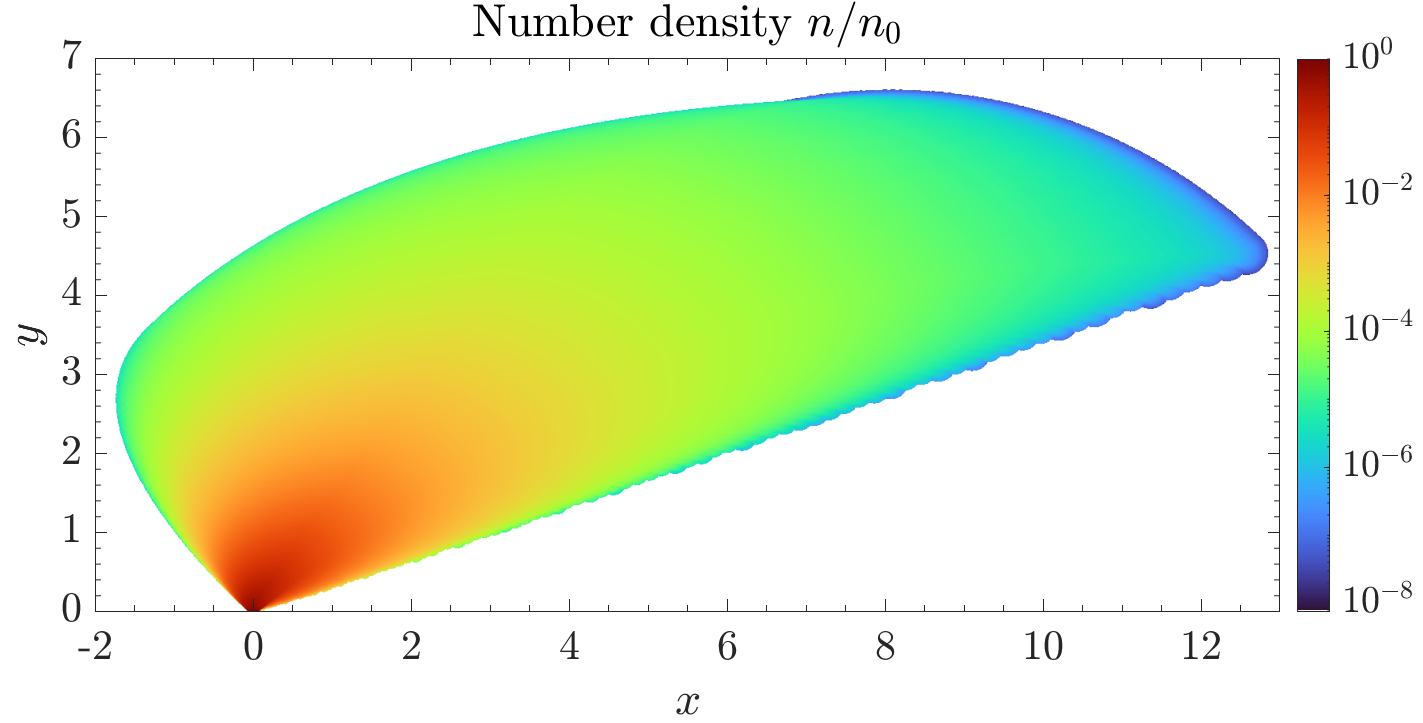}
        \caption{}
        \label{fig:8a}
    \end{subfigure}
    \begin{subfigure}[c]{0.495\textwidth}
        \includegraphics[width=\textwidth,trim={0 0 0 0}]{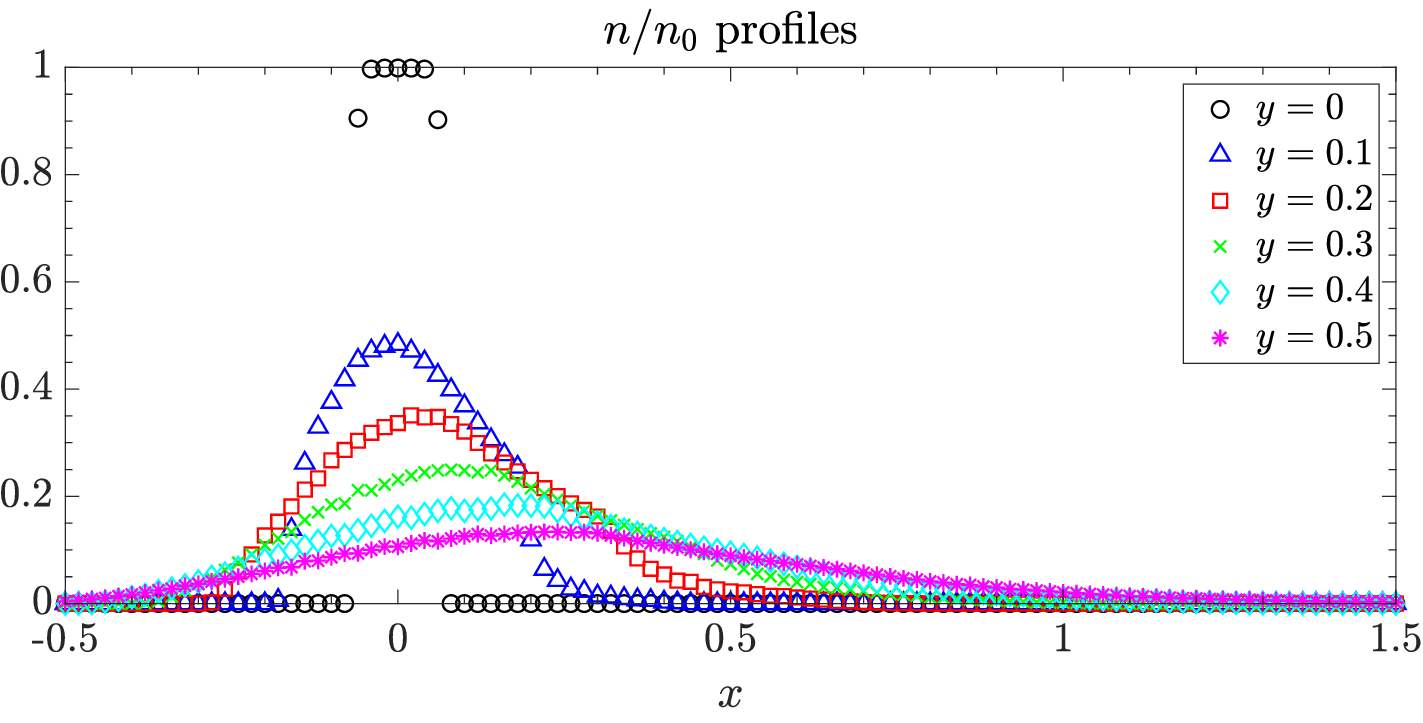}
        \caption{}
        \label{fig:8b}
    \end{subfigure}
    \begin{subfigure}[c]{0.495\textwidth}
        \includegraphics[width=\textwidth,trim={0 0 0 0}]{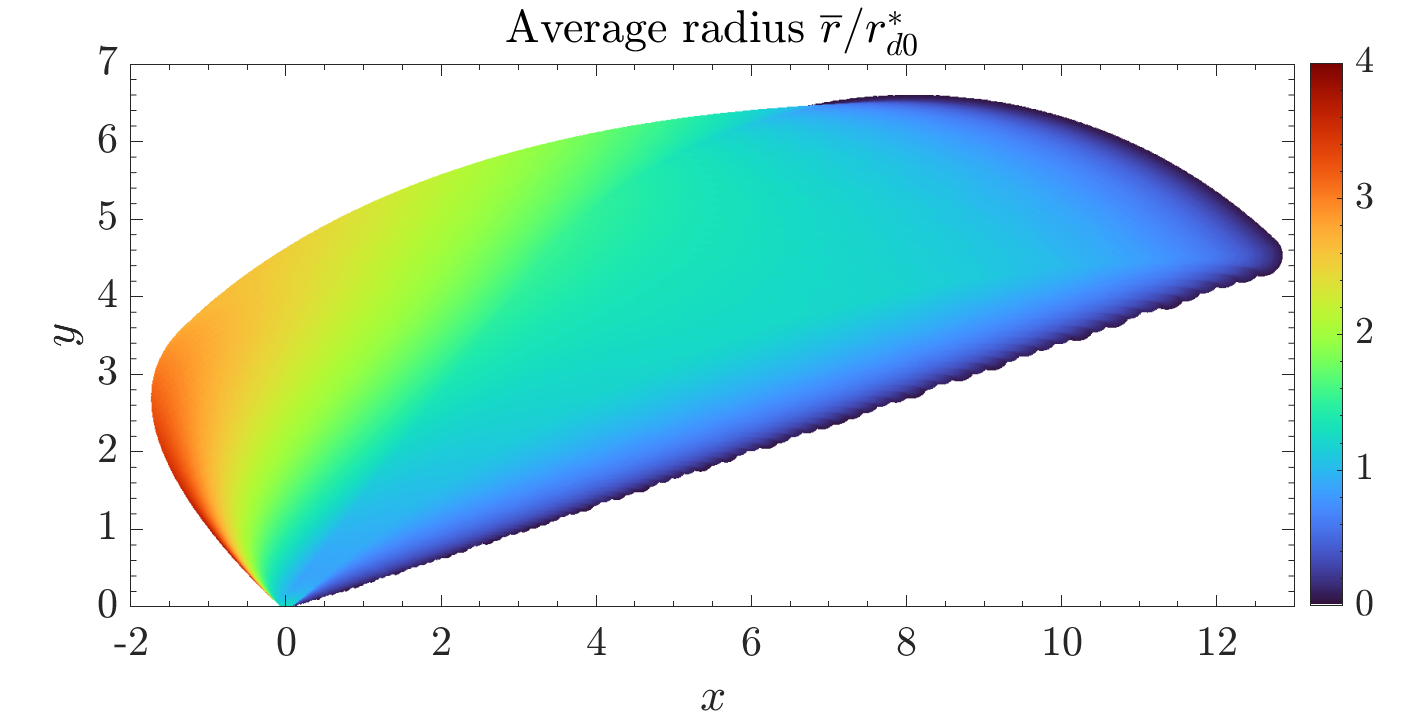}
        \caption{}
        \label{fig:8c}
    \end{subfigure}
    \begin{subfigure}[c]{0.495\textwidth}
        \includegraphics[width=\textwidth,trim={0 0 0 0}]{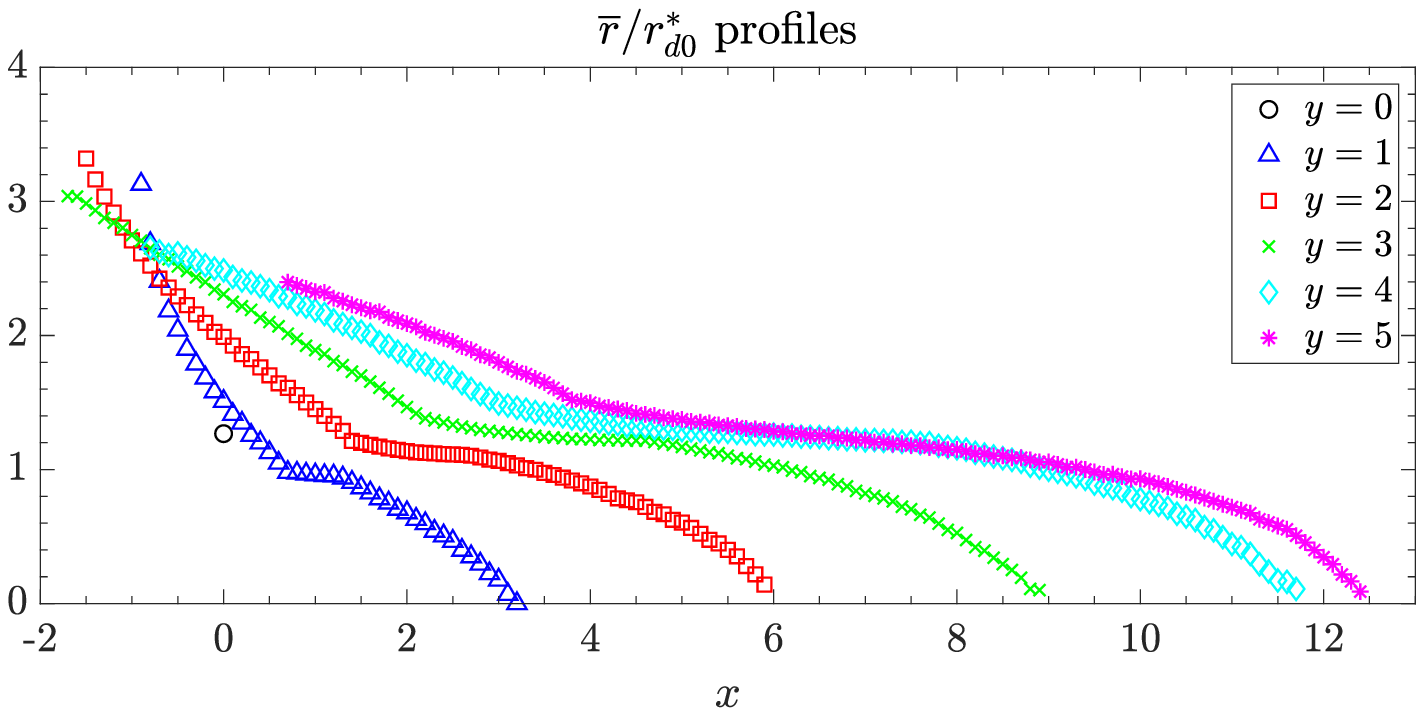}
        \caption{}
        \label{fig:8d}
    \end{subfigure}
    \begin{subfigure}[c]{0.495\textwidth}
        \includegraphics[width=\textwidth,trim={0 0 0 0}]{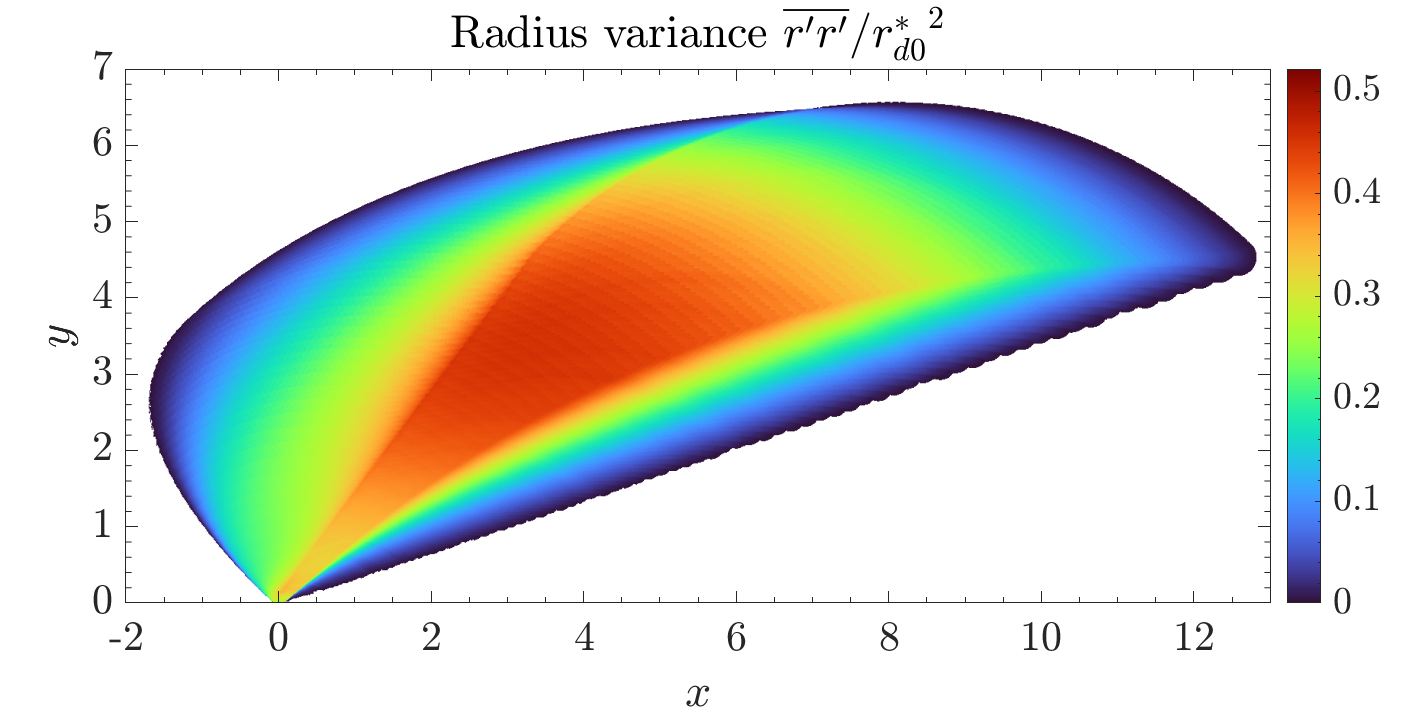}
        \caption{}
        \label{fig:8e}
    \end{subfigure}
    \begin{subfigure}[c]{0.495\textwidth}
        \includegraphics[width=\textwidth,trim={0 0 0 0}]{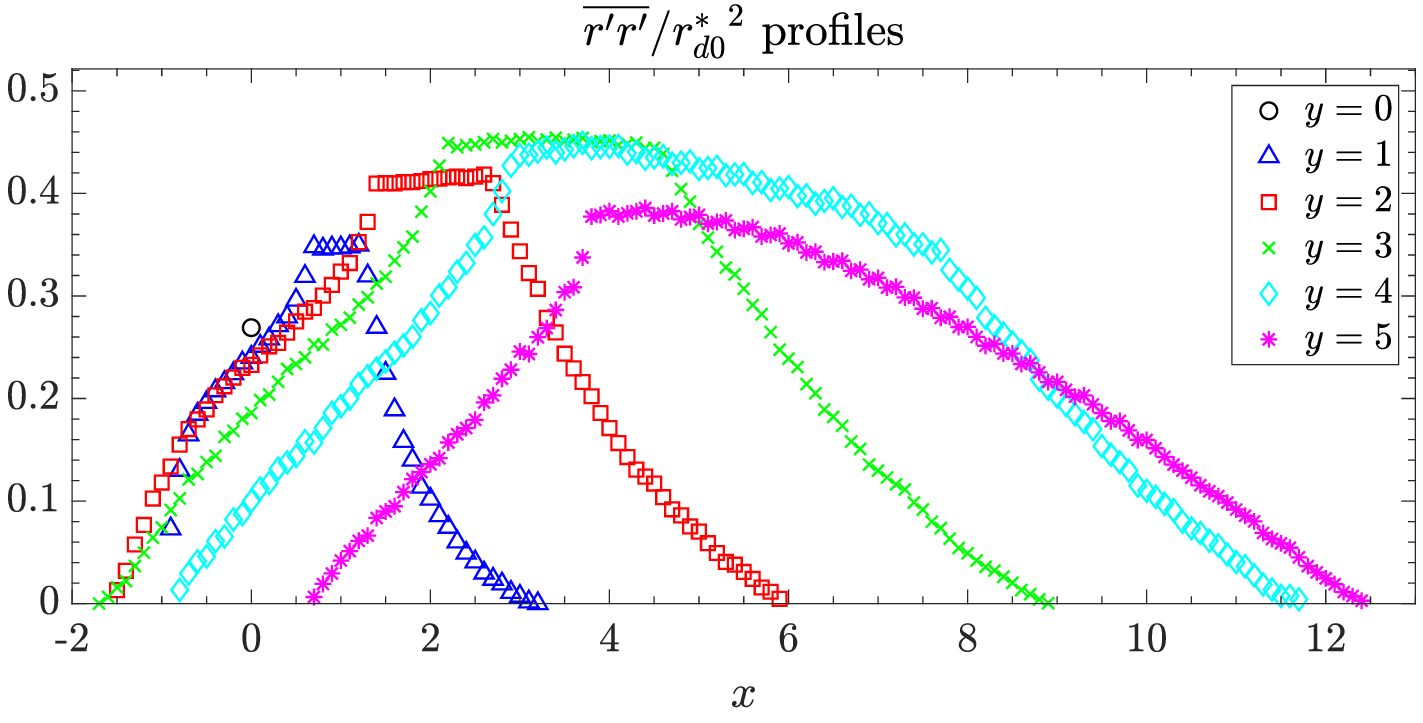}
        \caption{}
        \label{fig:8f}
    \end{subfigure}
    \caption{
        Reconstruction of the averaged field variables using kernel regression for a two-dimensional fan spray injection in cross-flow: \eqref{fig:8a} Number density ${{n}} / n_0$; \eqref{fig:8c} Average radius $\overline{r} / r_{d0}^*$; \eqref{fig:8e} Radius variance $\overline{r^{\prime}r^{\prime}} / {r_{d0}^*}^2$; Profiles of the averaged field variables at selected ${y}$ locations:
        \eqref{fig:8b} ${{n}} / n_0$; \eqref{fig:8d} $\overline{r} / r_{d0}^*$; \eqref{fig:8f} $\overline{r^{\prime}r^{\prime}} / {r_{d0}^*}^2$.
    }
    \label{fig:8}
\end{figure*}

Of particular interest within polydisperse droplet flows is the number density ${{n}}$ obtained using Eq.~\eqref{eq:number-density}. This is depicted for this case in Figure \eqref{fig:8a}, and shows the full extent of the droplet spatial distribution across all droplet sizes, with the number density decreasing as droplets move away from the injection interval and evaporate. Profiles of ${{n}}$ at selected ${y}$ locations are shown in Figure \eqref{fig:8b}, and illustrate the rapid decrease in number density close to the injection interval.
The distribution for the average droplet radius $\overline{r}$ calculated using Eq.~\eqref{eq:average-radius} is shown in Figure \eqref{fig:8c}, and it is observed that this decreases across the flow with increasing $x$ position. This is consistent with the observation from Figure \eqref{fig:7a}, in which only the largest droplets are able to travel against the carrier flow from the injection interval to reach the profile at $x = -1$. The profiles of $\overline{r}$ in Figure \eqref{fig:8d} reflect the fact that at increasing $y$ locations, the droplet field fills a wider part of the domain in the $x$ direction, whilst both the peak value and rate of decrease in $\overline{r}$ are reduced. The droplet radius variance distribution $\overline{r^{\prime}r^{\prime}}$ in Figure \eqref{fig:8e} is obtained from  Eq.~\eqref{eq:radius-variance} and illustrates that the highest spread in droplet size is found in the centre region of the droplet field, with less variation found along the lower and upper edges of the spray, where in accordance with Figure \eqref{fig:8c} smaller and larger droplets are respectively more dominant. The profiles of $\overline{r^{\prime}r^{\prime}}$ in Figure \eqref{fig:8f} show that the highest droplet variance occurs at $y \approx 3$, whilst the spread of variance in the $x$ direction increases with greater values of $y$ position. Together this collection of information exemplifies how kernel regression is able to produce stable distributions for the averaged field variables from trajectory data provided by the gFLA, and highlights its suitability for application to more general gas-droplet flows.

\subsubsection{Flow around a cylinder: steady-state case (Re = 20)} \label{sec:fpc-poly-steady}

For investigation into the behaviour of polydisperse droplets in a more general flow, the example of two-dimensional gas-droplet flow around a cylinder that was described in Sections \ref{sec:fpc-mono-steady} and \ref{sec:fpc-mono-transient} is returned to. The carrier flow is governed by Eqs.~\eqref{eq:carrier-flow-eqs} as previously, and now 41 different droplet sizes within the range $St \in [0.1,10]$ are introduced at each of 101 injection points across the interval ${y} / R \in [-4,4]$. The probability density is initialised in accordance with the distribution \eqref{eq:init-size-distribution}. In the following the focus is on the ability of kernel regression to reconstruct the droplet size distribution within the flow, and therefore the use of non-evaporating droplets is maintained. To achieve this, the droplet motion and Jacobian evolution are described as in the monodisperse case by Eqs.~\eqref{eq:fpc-part-eqns}, and supplemented with the droplet evaporation rate $\varphi = 0$ and equations and associated initial conditions for the other blocks in Eqs.~\eqref{eq:Jacobian-evolution} and \eqref{eq:Jacobian-initial-conditions} respectively to yield $\bm{J}^{\bm{x}r} = \bm{0}$, $\bm{J}^{r\bm{x}} = \bm{0}$ and $J^{rr} = 1$.

As with the previous polydisperse droplet examples, reconstruction of the droplet probability density field ${{p}}$ follows the procedure outlined in Section \ref{sec:kernel-gfla}, with the structured kernel defined as in Eq.~\eqref{eq:H-kernel-structured-gFLA} and the bandwidth matrix $\bm{H}$ given by Eq.~\eqref{eq:H-kernel-gFLA}. The probability density field is accumulated onto a uniform Cartesian grid with a spatial spacing of $\Delta {x} / R = \Delta {y} / R = 0.04$ as in Section \ref{sec:fpc-mono-steady}, and radial spacing $\Delta {r} / r_{d0}^* = 0.133$. The initial inter-droplet spacings in location and size are given by $\Delta x_{d0} / R = 0.08$ and $\Delta r_{d0} / r_{d0}^* = 0.1$ respectively, and the initial smoothing lengths set as $h_{\bm{x}0} = \Delta x_{d0}$ and $h_{r0} = 0.667 \, \Delta r_{d0}$ accordingly.

\begin{figure*}[!ht]
    \begin{subfigure}[c]{0.495\textwidth}
        \includegraphics[width=\textwidth,trim={0 15 0 0}]{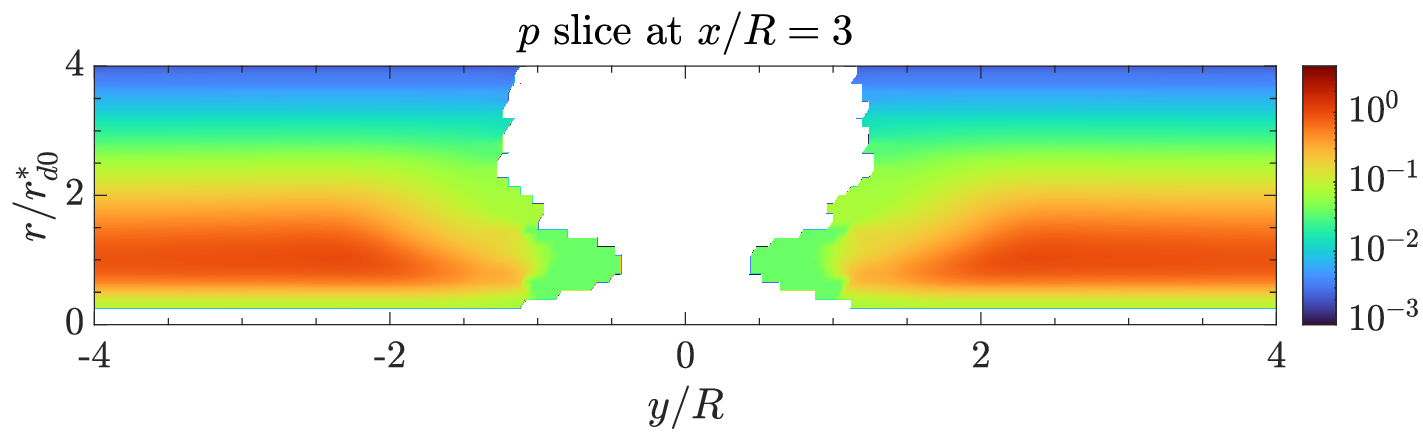}
        \caption{}
        \label{fig:9a}
    \end{subfigure}
    \begin{subfigure}[c]{0.495\textwidth}
        \includegraphics[width=\textwidth,trim={0 15 0 0}]{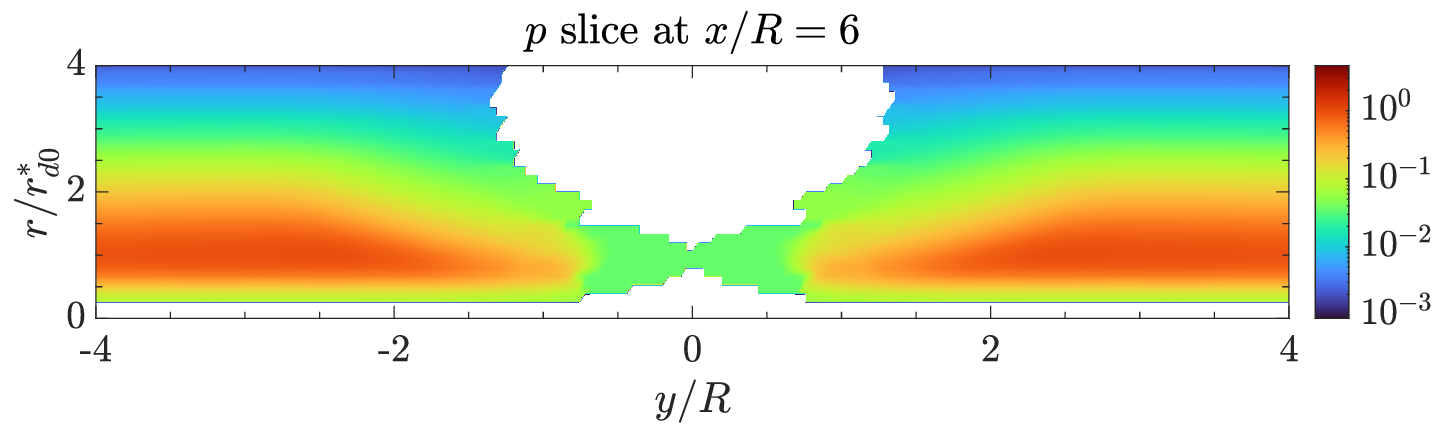}
        \caption{}
        \label{fig:9b}
    \end{subfigure}
    \begin{subfigure}[c]{0.495\textwidth}
        \includegraphics[width=\textwidth,trim={0 15 0 0}]{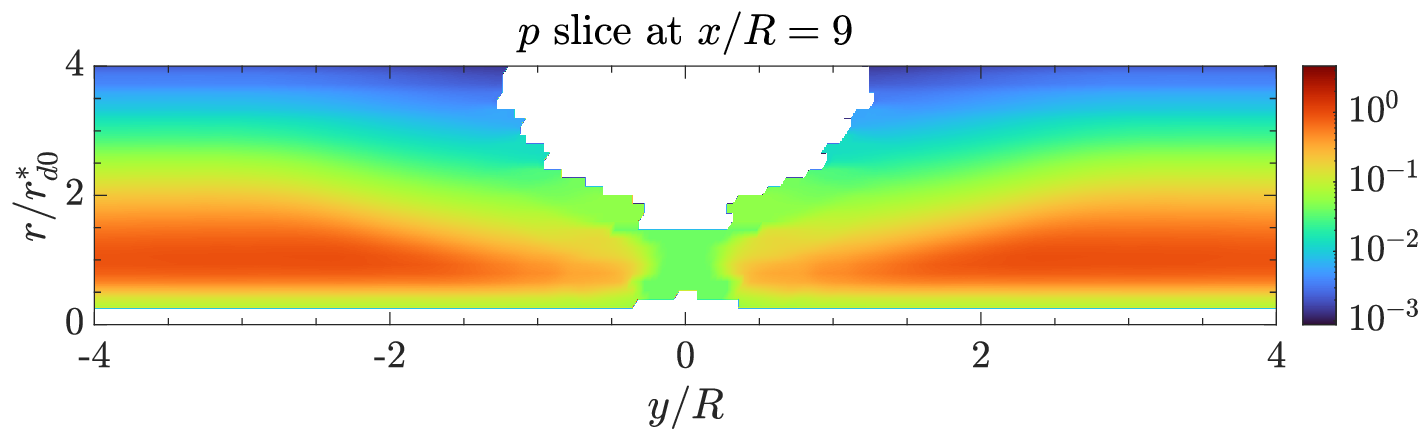}
        \caption{}
        \label{fig:9c}
    \end{subfigure}
    \begin{subfigure}[c]{0.495\textwidth}
        \includegraphics[width=\textwidth,trim={0 15 0 0}]{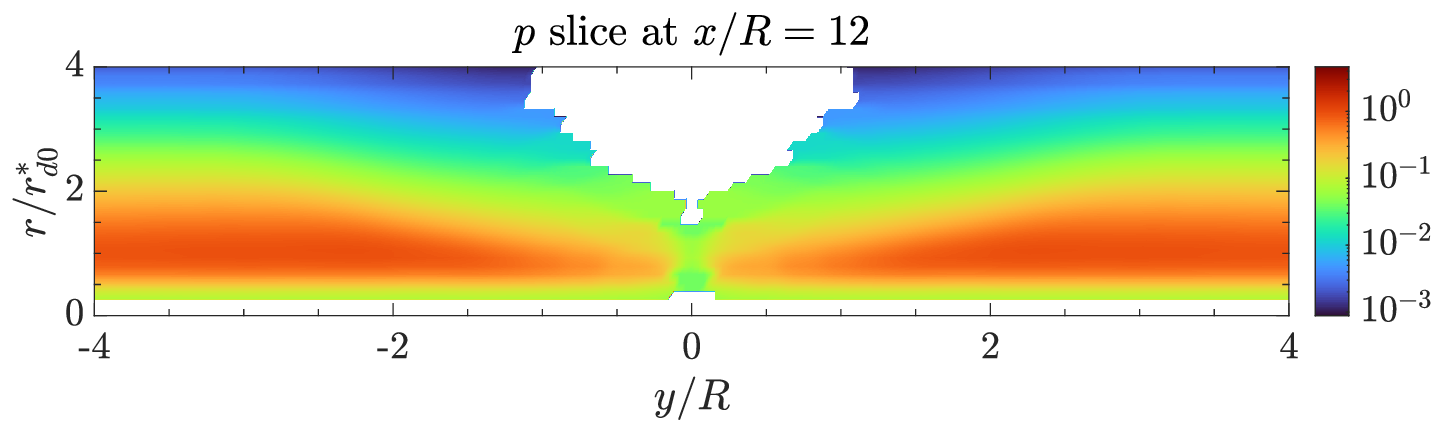}
        \caption{}
        \label{fig:9d}
    \end{subfigure}
    \begin{subfigure}[c]{0.495\textwidth}
        \includegraphics[width=\textwidth,trim={0 15 0 0}]{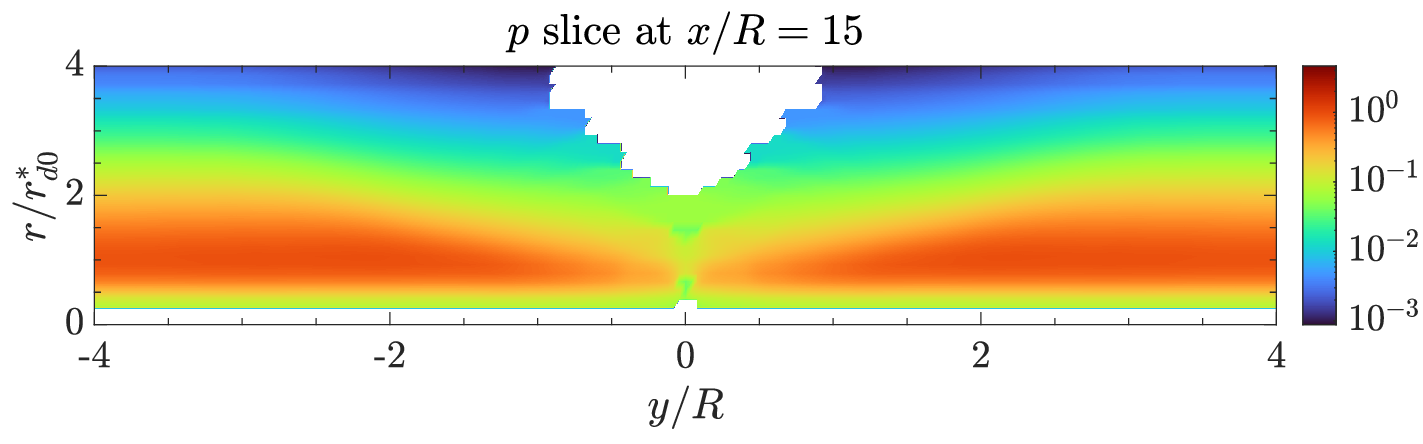}
        \caption{}
        \label{fig:9e}
    \end{subfigure}
    \begin{subfigure}[c]{0.495\textwidth}
        \includegraphics[width=\textwidth,trim={0 15 0 0}]{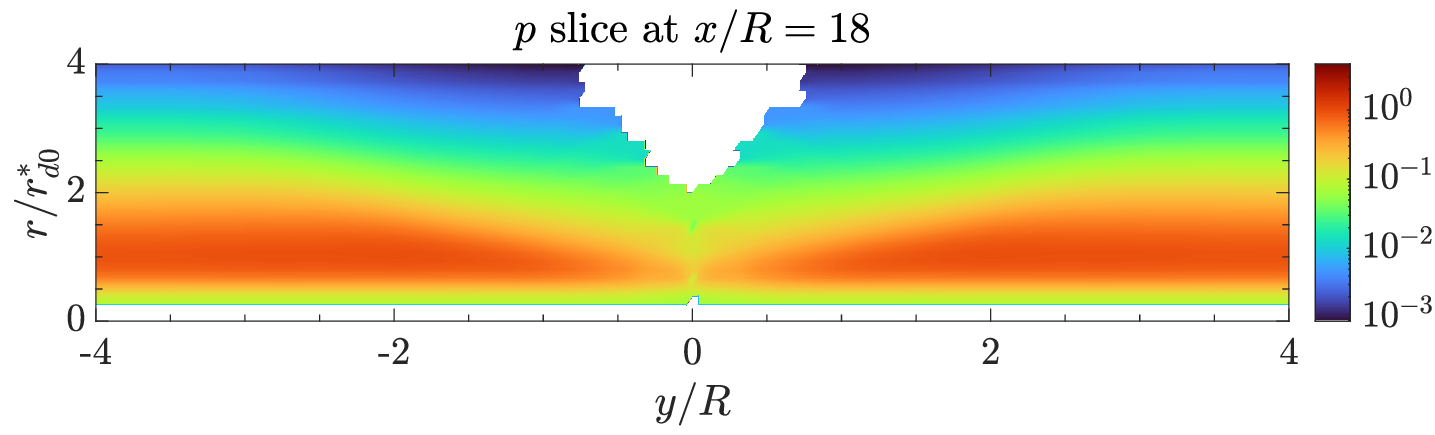}
        \caption{}
        \label{fig:9f}
    \end{subfigure}
    \caption{
        Slices of the probability density field ${{p}}$ obtained using kernel regression for polydisperse steady-state flow around a cylinder at $Re = 20$ for selected $x$ locations: \eqref{fig:9a} ${x} / R = 3$; \eqref{fig:9b} ${x} / R = 6$; \eqref{fig:9c} ${x} / R = 9$; \eqref{fig:9d} ${x} / R = 12$; \eqref{fig:9e} ${x} / R = 15$; \eqref{fig:9f} ${x} / R = 18$.
    }
    \label{fig:9}
\end{figure*}

The kernel regression procedure again enables the detail of the droplet distribution across the spatial dimensions and range of droplet sizes to be obtained, and in this case the variation in droplet size distribution in the ${y}$ direction is shown in Figure \ref{fig:9} as slices in the ${y}$ direction to demonstrate this. Since the flow is steady-state, the distribution behaviour that is observed across the slices varies in a straightforward manner, with the width of the wake behind the cylinder decreasing as the distance from the cylinder at ${x} / R = 0$ is increased. In terms of the droplet size distribution, it can clearly be seen that at a given distance from the cylinder, the width of the wake is less for smaller droplets, agreeing with the behaviour previously observed in Figure \ref{fig:6} for monodisperse droplets at selected values of $St$. Indeed, it is possible to extract the spatial number density field at a given droplet radius from the reconstructed field $p$, and choosing the values of ${r}$ corresponding to $St = 0.1$, $1$, and $10$ reproduces the distributions displayed in Figures \eqref{fig:3a}, \eqref{fig:3c}, and \eqref{fig:3e} respectively. Furthermore, at any point in the domain, it can be seen in Figure \ref{fig:9} that there is a clear distribution across the range of droplet sizes, which is reflective of the initial size distribution specified by Eq.~\eqref{eq:init-size-distribution}.

\begin{figure*}[!ht]
    \begin{subfigure}[c]{0.495\textwidth}
        \includegraphics[width=\textwidth,trim={0 0 0 0}]{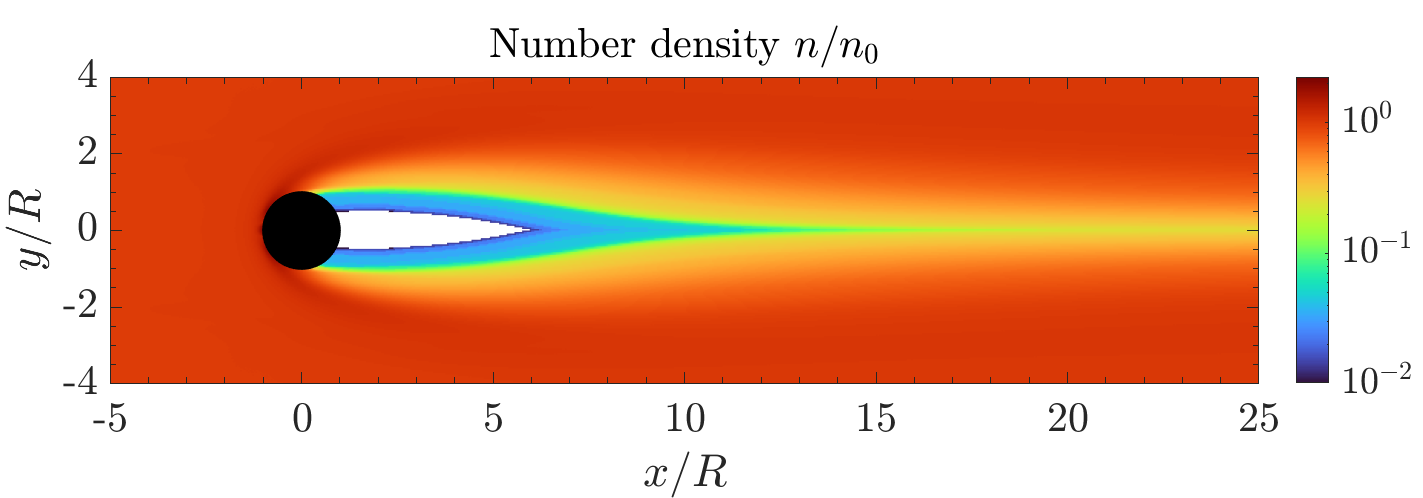}
        \caption{}
        \label{fig:10a}
    \end{subfigure}
    \begin{subfigure}[c]{0.495\textwidth}
        \includegraphics[width=\textwidth,trim={0 0 0 0}]{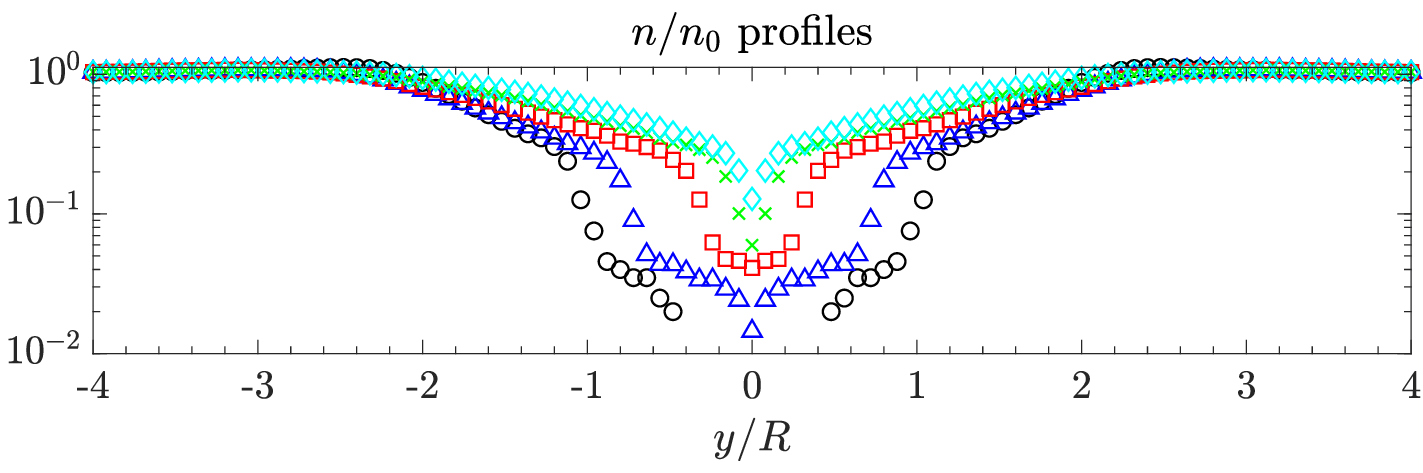}
        \caption{}
        \label{fig:10b}
    \end{subfigure}
    \begin{subfigure}[c]{0.495\textwidth}
        \includegraphics[width=\textwidth,trim={0 0 0 0}]{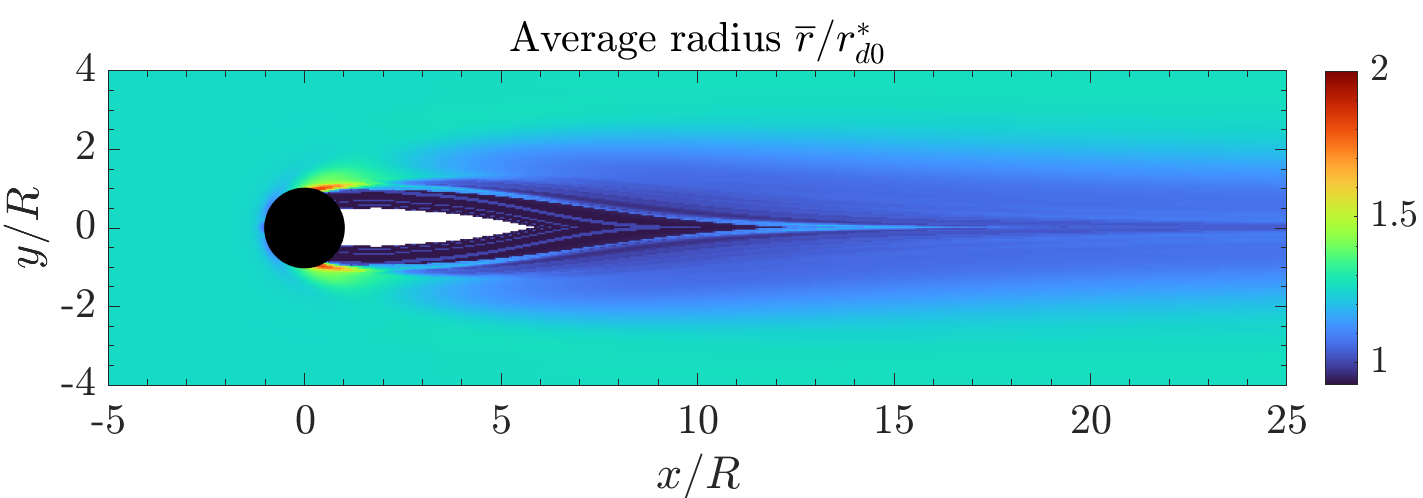}
        \caption{}
        \label{fig:10c}
    \end{subfigure}
    \begin{subfigure}[c]{0.495\textwidth}
        \includegraphics[width=\textwidth,trim={0 0 0 0}]{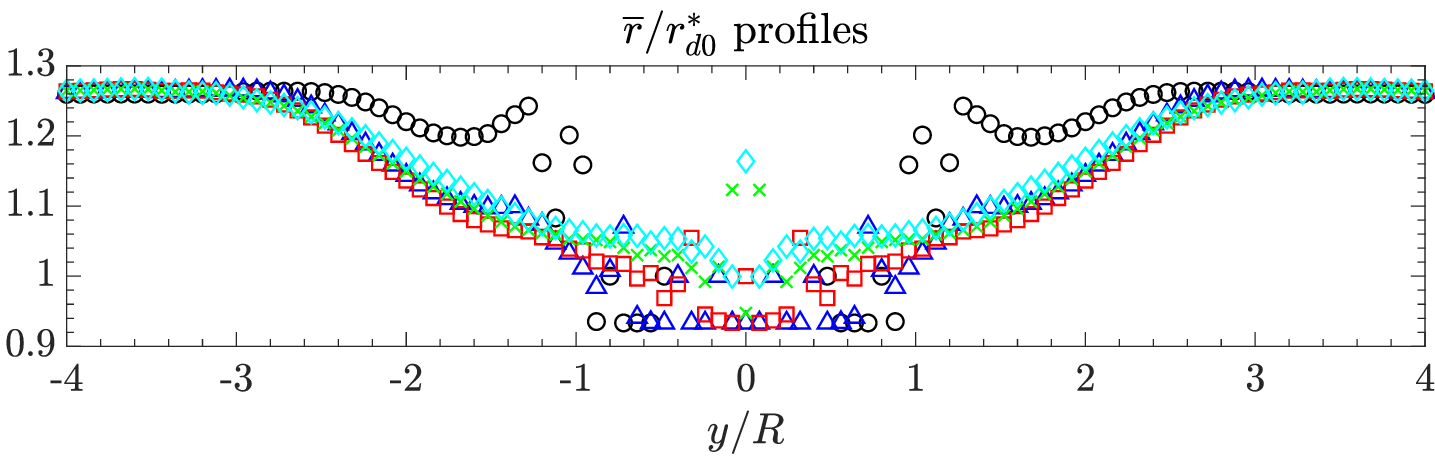}
        \caption{}
        \label{fig:10d}
    \end{subfigure}
    \begin{subfigure}[c]{0.495\textwidth}
        \includegraphics[width=\textwidth,trim={0 0 0 0}]{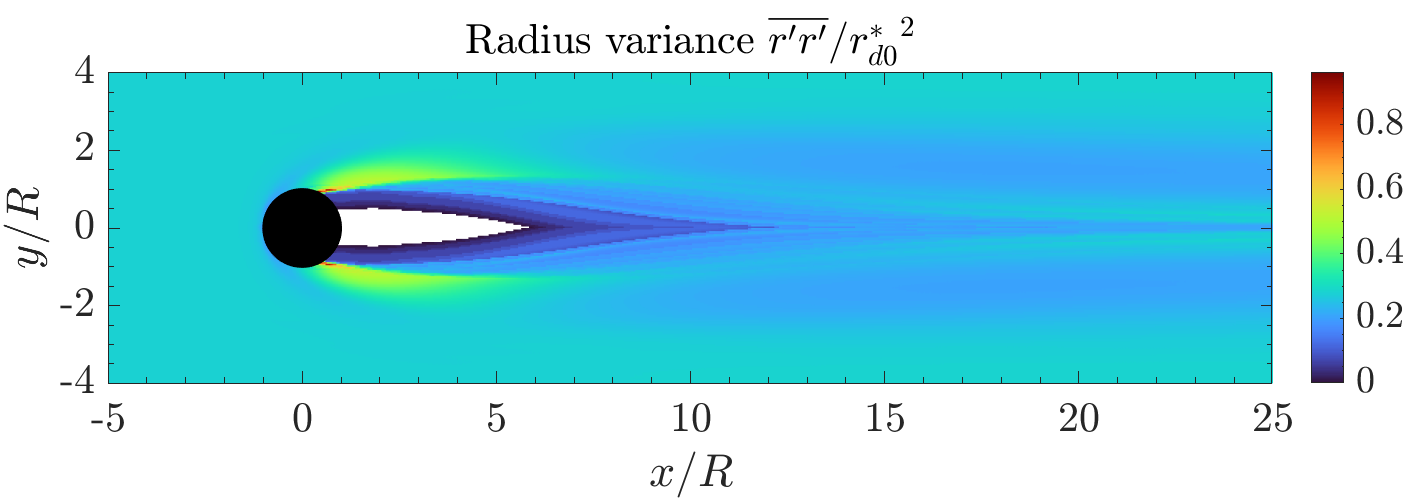}
        \caption{}
        \label{fig:10e}
    \end{subfigure}
    \begin{subfigure}[c]{0.495\textwidth}
        \includegraphics[width=\textwidth,trim={0 0 0 0}]{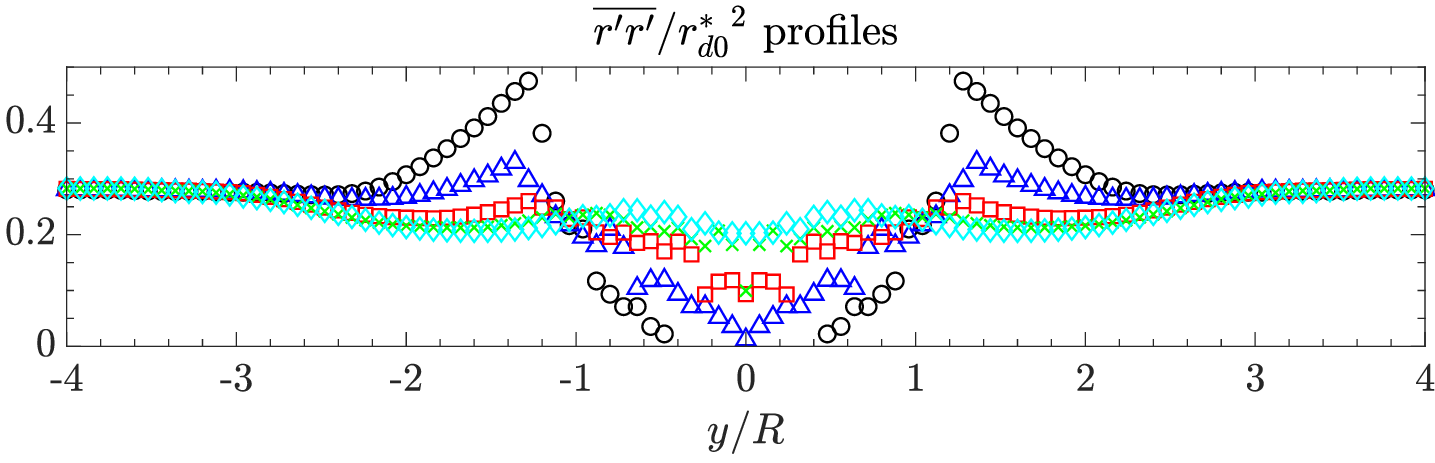}
        \caption{}
        \label{fig:10f}
    \end{subfigure}
    \caption{
        Reconstruction of the averaged field variables using kernel regression for steady-state polydisperse droplet flow around a cylinder at $Re = 20$: \eqref{fig:10a} Number density ${{n}} / n_0$; \eqref{fig:10c} Average radius $\overline{r} / r_{d0}^*$; \eqref{fig:10e} Radius variance $\overline{r^{\prime}r^{\prime}} / {r_{d0}^*}^2$; Profiles of the averaged field variables at selected ${x}$ locations:
        {\color{black}$\boldsymbol{\bigcirc}$} ${x} / R = 3$, {\color{blue}$\boldsymbol{\bigtriangleup}$} ${x} / R = 6$,
        {\color{red}$\boldsymbol{\Box}$} ${x} / R = 9$,
        {\color{green}$\boldsymbol{\times}$} ${x} / R = 12$,
        {\color{cyan}$\boldsymbol{\diamond}$} ${x} / R = 15$;
        \eqref{fig:10b} ${{n}} / n_0$; \eqref{fig:10d} $\overline{r} / r_{d0}^*$; \eqref{fig:10f} $\overline{r^{\prime}r^{\prime}} / {r_{d0}^*}^2$.
    }
    \label{fig:10}
\end{figure*}

The number density ${{n}}$ is displayed in Figure \eqref{fig:10a}, and shows the behaviour of the droplet spatial distribution across all droplet sizes. It is seen that the wake region behind the cylinder with no droplets present corresponds to that obtained in Figure \eqref{fig:3a} for the smallest droplet size of $St = 0.1$. Additionally, the number density varies smoothly and in general gradually when moving from the centreline ${y} / R = 0$ to the edges of the domain. This can be seen in the profiles of ${{n}}$ at selected ${y}$ locations displayed in Figure \eqref{fig:10b}, and shows the variation in number density at different distances behind the cylinder. The average radius distribution $\overline{r}$ shown in Figure \eqref{fig:10c} illustrates that larger droplets accumulate as the droplet field splits into two distinct regions when moving past the cylinder, whilst the region of the droplet field that is downstream of the cylinder wake only contains the smallest droplets. The profiles of $\overline{r}$ in Figure \eqref{fig:10d} show that at these selected $x$ locations the average droplet size does not vary much within the range of droplet sizes, and is uniformly lower towards the centreline $y / R = 0$. Both the distribution and profiles of variance in droplet radius $\overline{r^{\prime}r^{\prime}}$, as depicted in Figures \eqref{fig:10e} and \eqref{fig:10f} respectively, are qualitatively similar to the behaviour of $\overline{r}$, with the highest variance observed at the point where the droplet field parts to pass around the cylinder.

\subsubsection{Flow around a cylinder: transient case (Re = 100)} \label{sec:fpc-poly-transient}

For the transient regime of polydisperse droplet flow around a cylinder, $Re = 100$ is used as before in Section \ref{sec:fpc-mono-transient}, and otherwise the configuration is identical to that in Section \ref{sec:fpc-poly-steady}. In terms of the kernel regression procedure, the probability density field is accumulated onto a uniform Cartesian grid with a spatial spacing of $\Delta {x} / R = \Delta {y} / R = 0.04$ and radial spacing $\Delta {r} / r_{d0}^* = 0.133$ as before, with the initial smoothing lengths again specified as $h_{\bm{x}0} = \Delta x_{d0}$ and $h_{r0} = 0.667 \, \Delta r_{d0}$.
\begin{figure*}[!ht]
    \begin{subfigure}[c]{0.495\textwidth}
        \includegraphics[width=\textwidth,trim={0 15 0 0}]{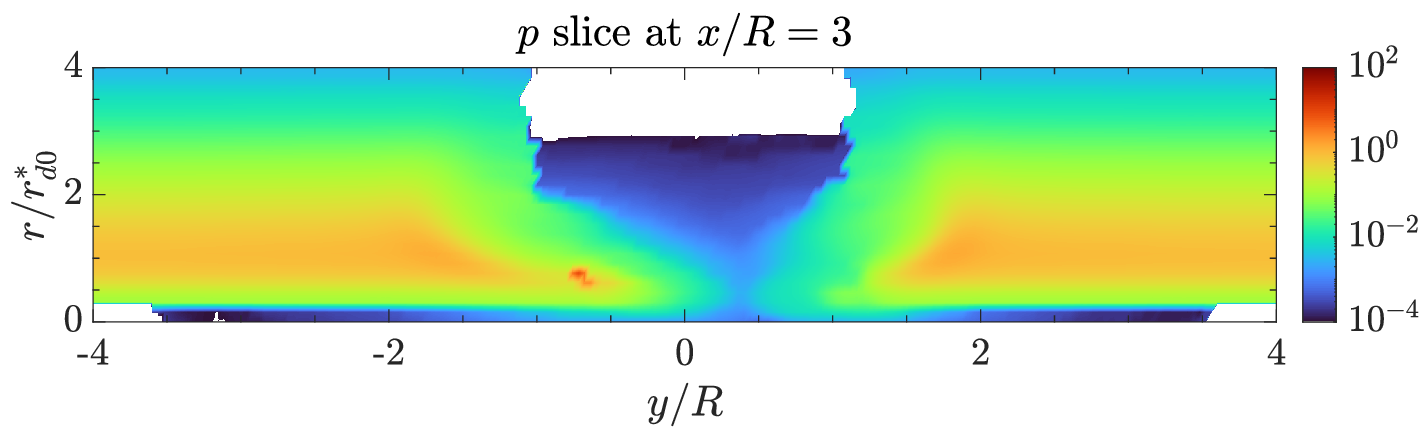}
        \caption{}
        \label{fig:11a}
    \end{subfigure}
    \begin{subfigure}[c]{0.495\textwidth}
        \includegraphics[width=\textwidth,trim={0 15 0 0}]{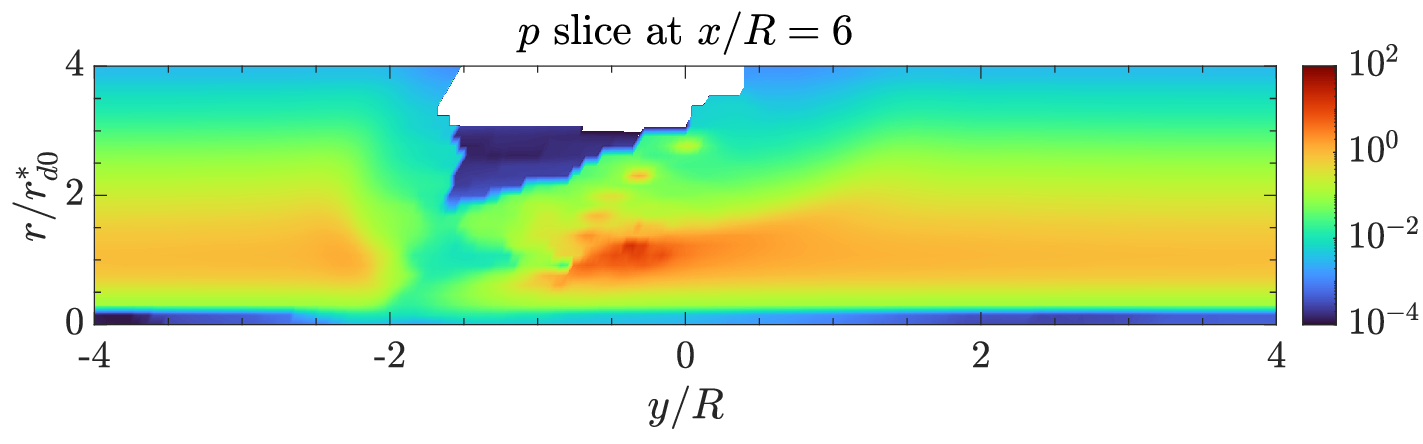}
        \caption{}
        \label{fig:11b}
    \end{subfigure}
    \begin{subfigure}[c]{0.495\textwidth}
        \includegraphics[width=\textwidth,trim={0 15 0 0}]{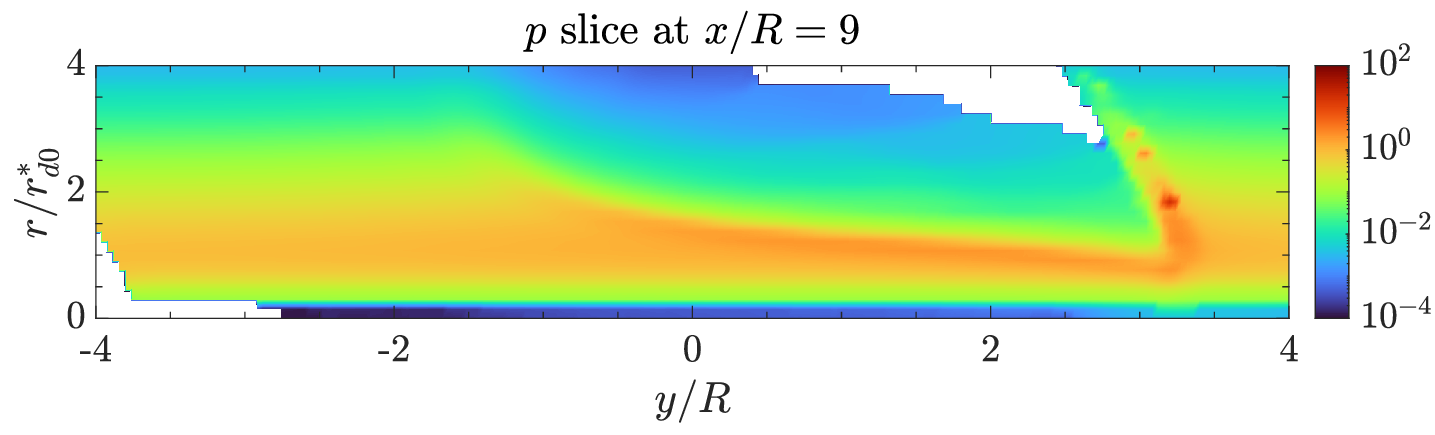}
        \caption{}
        \label{fig:11c}
    \end{subfigure}
    \begin{subfigure}[c]{0.495\textwidth}
        \includegraphics[width=\textwidth,trim={0 15 0 0}]{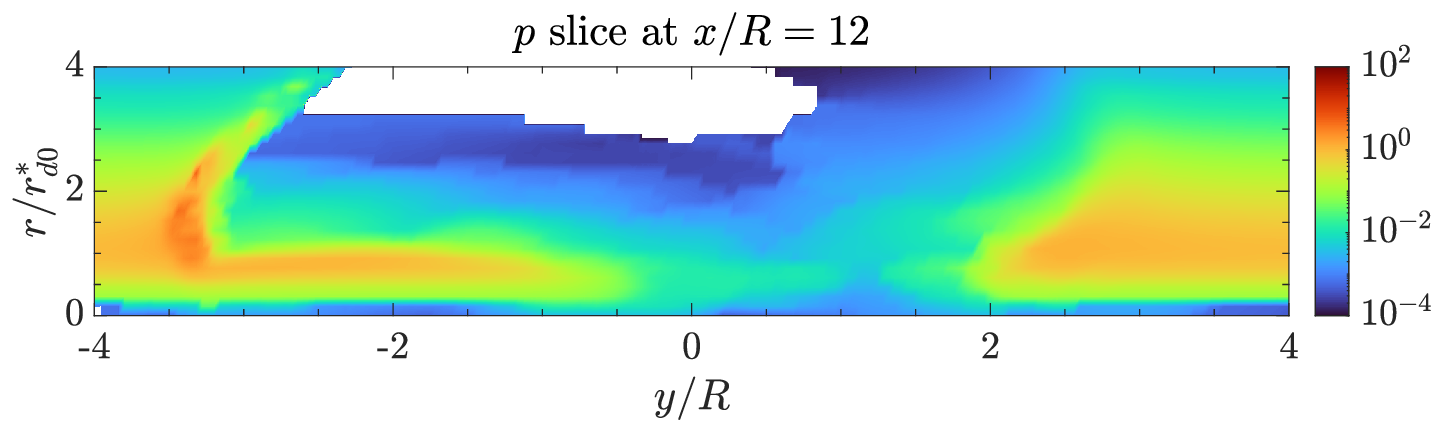}
        \caption{}
        \label{fig:11d}
    \end{subfigure}
    \begin{subfigure}[c]{0.495\textwidth}
        \includegraphics[width=\textwidth,trim={0 15 0 0}]{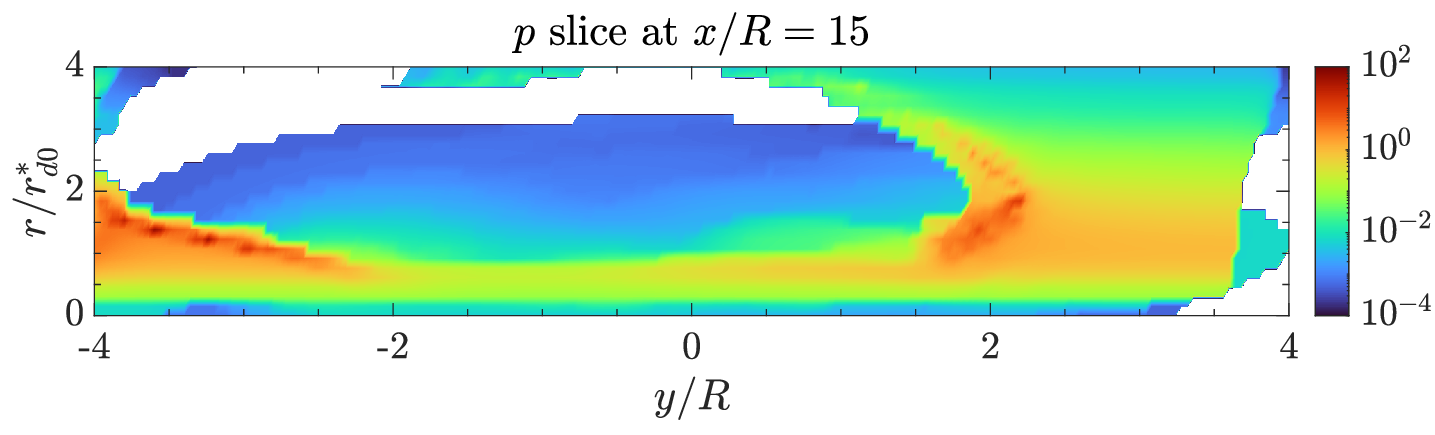}
        \caption{}
        \label{fig:11e}
    \end{subfigure}
    \begin{subfigure}[c]{0.495\textwidth}
        \includegraphics[width=\textwidth,trim={0 15 0 0}]{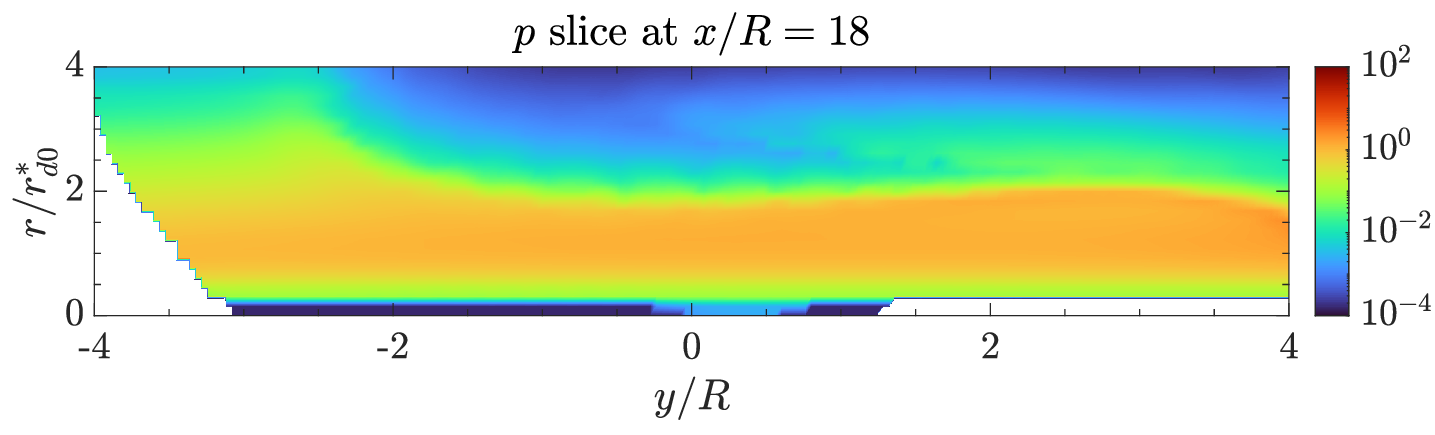}
        \caption{}
        \label{fig:11f}
    \end{subfigure}
    \caption{
        Slices of the probability density field ${{p}}$ obtained using kernel regression for polydisperse transient flow around a cylinder at $Re = 100$ and time $t = 30$ for selected $x$ locations: \eqref{fig:11a} ${x} / R = 3$; \eqref{fig:11b} ${x} / R = 6$; \eqref{fig:11c} ${x} / R = 9$; \eqref{fig:11d} ${x} / R = 12$; \eqref{fig:11e} ${x} / R = 15$; \eqref{fig:11f} ${x} / R = 18$.
    }
    \label{fig:11}
\end{figure*}

Again it is primarily the droplet spatial distribution which is of interest, and following the previous section the variation in droplet size distribution in the ${y}$ direction is shown in
Figure \ref{fig:11} as slices in the ${y}$ direction at $t = 30$. Whilst the probability density distribution in Figure \eqref{fig:11a} displays only minor deviations from being symmetrical at a distance of ${x} / R = 3$ behind the cylinder, at increasing distances this symmetry breaks down as the vortex street becomes established. At a sufficient distance, the probability density distribution at a given slice then exhibits regions with either no droplets or a high probability density depending where in the periodic cycle of the flow the slice is located, for example at ${x} / R = 15$ as displayed in Figure \eqref{fig:11e}. Due to the transient nature of the flow, the effect of specifying the initial probability density using Eq.~\eqref{eq:init-size-distribution} is less apparent than in the previous steady-state example, however the highest probability density values are still found among the smaller droplet sizes, with fewer larger droplets occurring. A wake is seen to persist for some distance behind the cylinder for only the largest droplets, but eventually ceases to exist once the periodic flow has become fully established. As with the steady-state case, the spatial number density field at a given droplet radius can be extracted from the reconstructed field ${{p}}$, whereby selecting the values of ${r}$ corresponding to $St = 0.1$, $1$, and $10$ retrieves the distributions in Figures \eqref{fig:7a}, \eqref{fig:7c}, and \eqref{fig:7e} respectively.

\begin{figure*}[!ht]
    \begin{subfigure}[c]{0.495\textwidth}
        \includegraphics[width=\textwidth,trim={0 0 0 0}]{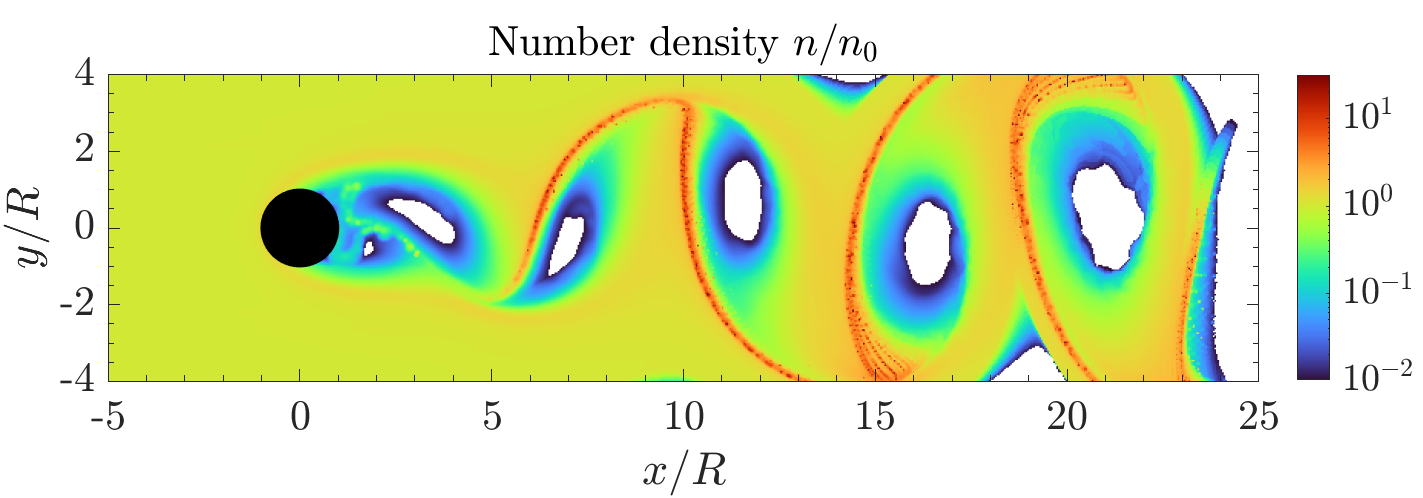}
        \caption{}
        \label{fig:12a}
    \end{subfigure}
    \begin{subfigure}[c]{0.495\textwidth}
        \includegraphics[width=\textwidth,trim={0 0 0 0}]{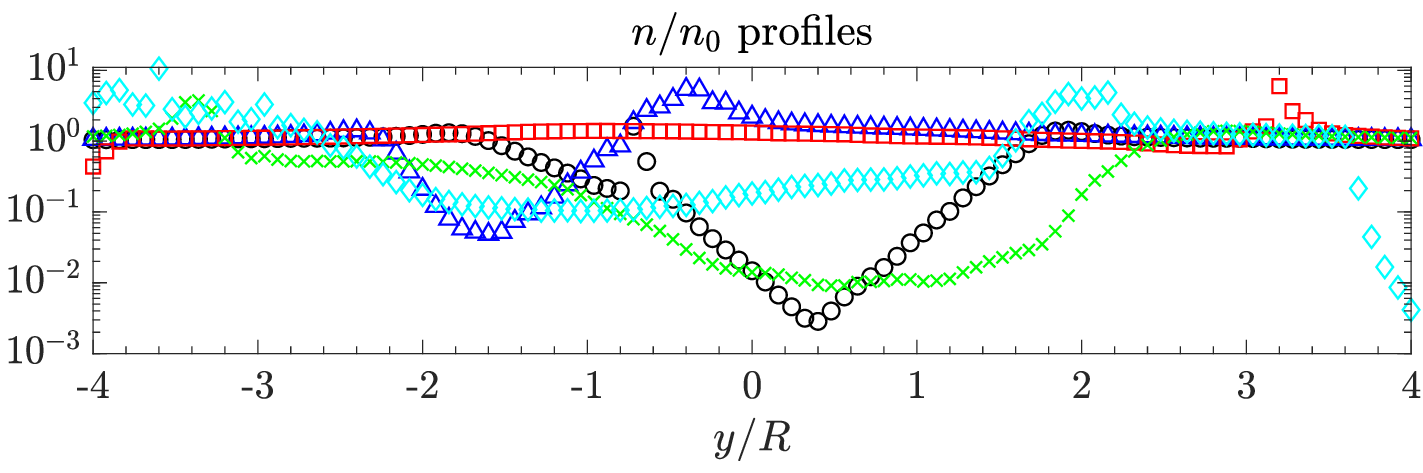}
        \caption{}
        \label{fig:12b}
    \end{subfigure}
    \begin{subfigure}[c]{0.495\textwidth}
        \includegraphics[width=\textwidth,trim={0 0 0 0}]{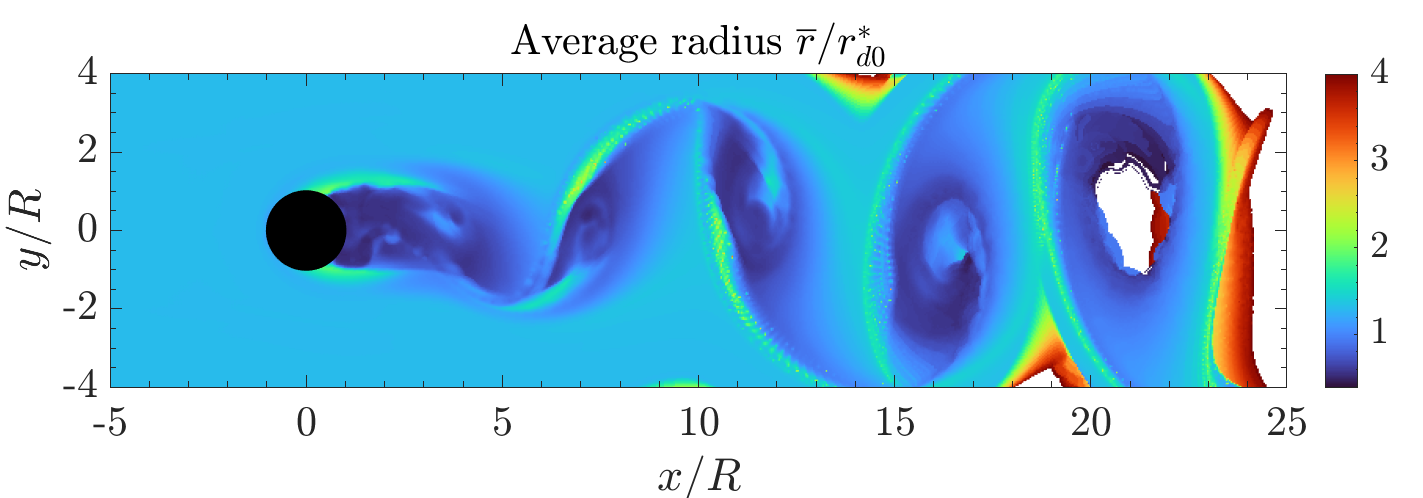}
        \caption{}
        \label{fig:12c}
    \end{subfigure}
    \begin{subfigure}[c]{0.495\textwidth}
        \includegraphics[width=\textwidth,trim={0 0 0 0}]{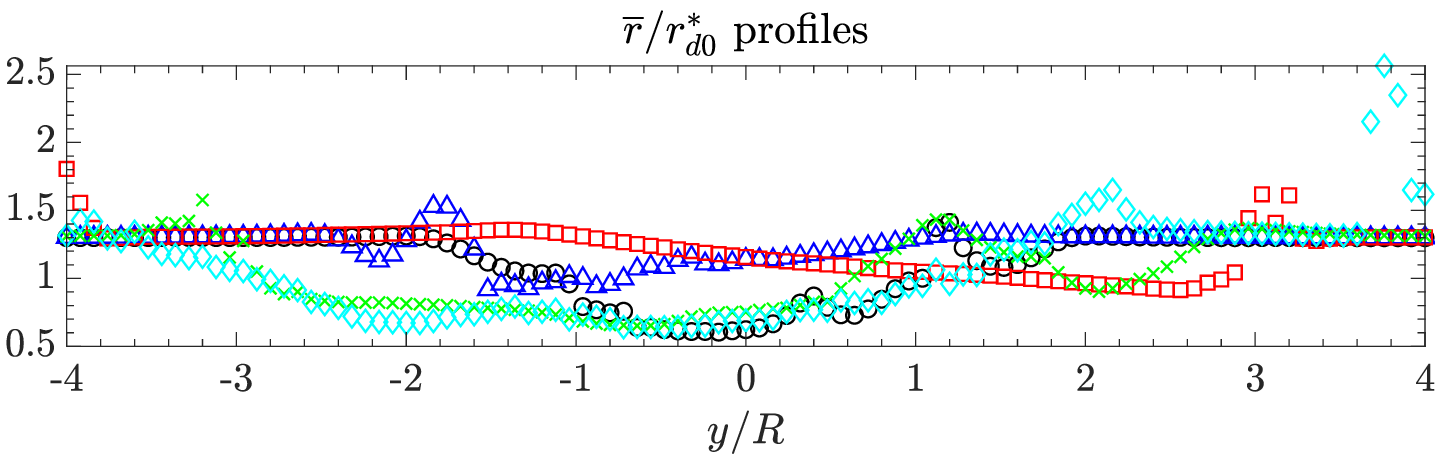}
        \caption{}
        \label{fig:12d}
    \end{subfigure}
    \begin{subfigure}[c]{0.495\textwidth}
        \includegraphics[width=\textwidth,trim={0 0 0 0}]{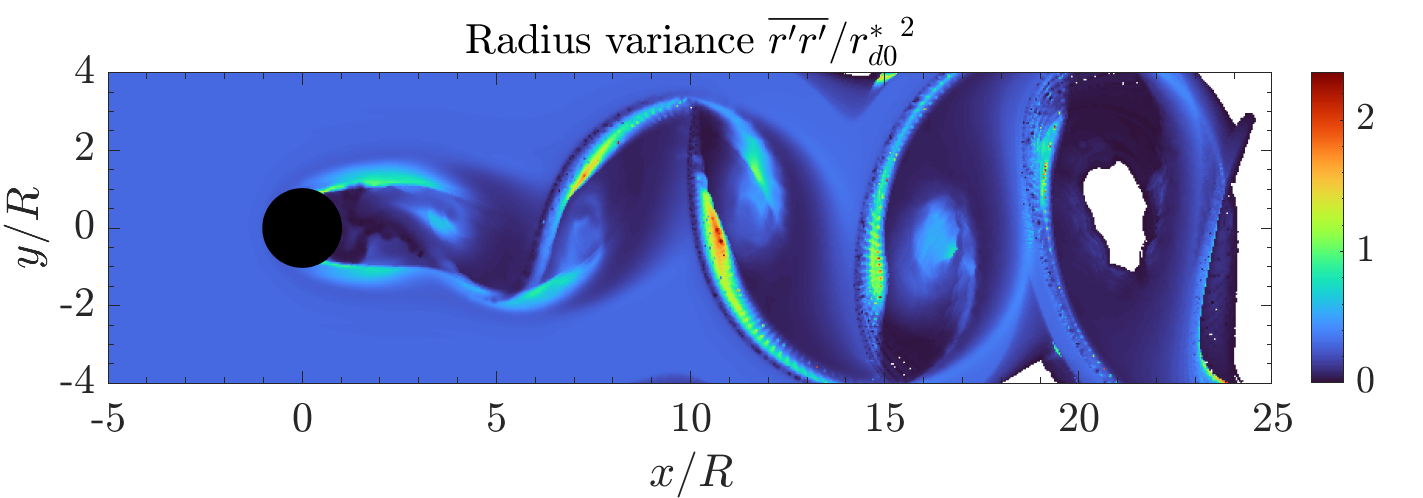}
        \caption{}
        \label{fig:12e}
    \end{subfigure}
    \begin{subfigure}[c]{0.495\textwidth}
        \includegraphics[width=\textwidth,trim={0 0 0 0}]{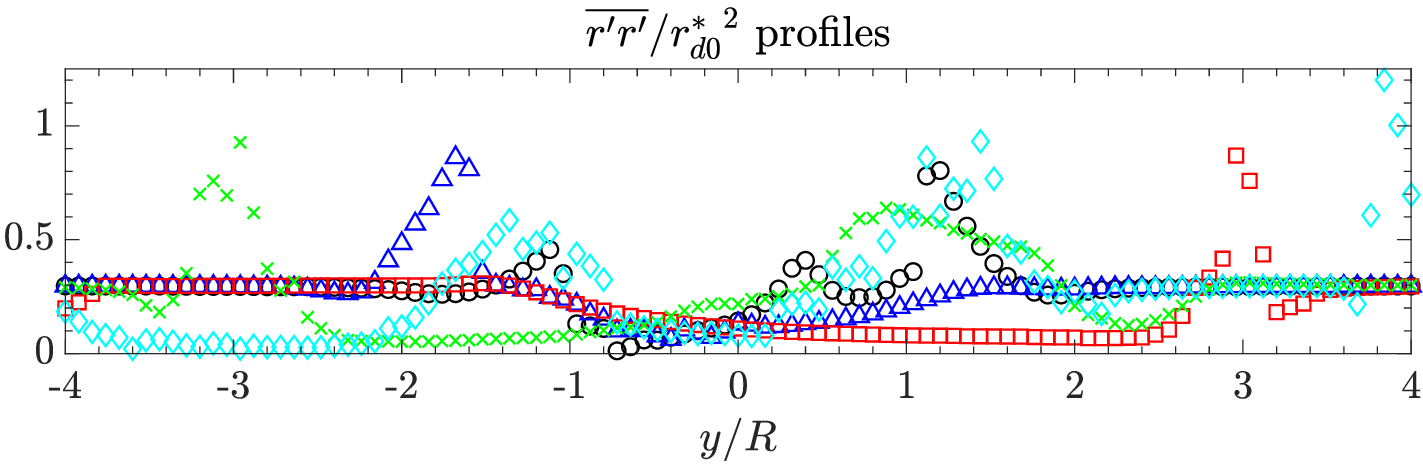}
        \caption{}
        \label{fig:12f}
    \end{subfigure}
    \caption{
        Reconstruction of the averaged field variables using kernel regression for transient polydisperse droplet flow around a cylinder at $Re = 100$ and time $t = 30$: \eqref{fig:12a} Number density ${{n}} / n_0$; \eqref{fig:12c} Average radius $\overline{r} / r_{d0}^*$; \eqref{fig:12e} Radius variance $\overline{r^{\prime}r^{\prime}} / {r_{d0}^*}^2$; Profiles of the averaged field variables at selected ${x}$ locations:
        {\color{black}$\boldsymbol{\bigcirc}$} ${x} / R = 3$, {\color{blue}$\boldsymbol{\bigtriangleup}$} ${x} / R = 6$,
        {\color{red}$\boldsymbol{\Box}$} ${x} / R = 9$,
        {\color{green}$\boldsymbol{\times}$} ${x} / R = 12$,
        {\color{cyan}$\boldsymbol{\diamond}$} ${x} / R = 15$;
        \eqref{fig:12b} ${{n}} / n_0$; \eqref{fig:12d} $\overline{r} / r_{d0}^*$; \eqref{fig:12f} $\overline{r^{\prime}r^{\prime}} / {r_{d0}^*}^2$.
    }
    \label{fig:12}
\end{figure*}

Figure \eqref{fig:12a} displays the number density ${{n}}$ at $t = 30$, exhibiting the cumulative effect of all droplet sizes on the droplet spatial distribution. It is observed that all droplet sizes are ejected from the vortices by virtue of their inertia, with clustering occurring along curves which are distinct for different droplet sizes. This is clearly an accumulation of the behaviour previously displayed in Figures \eqref{fig:7a}, \eqref{fig:7c}, and \eqref{fig:7e} for different droplet sizes in the monodisperse case, and exemplifies the ability of kernel regression to account for this range of behaviour in the process of reconstructing the number density field. The variation in number density at different ${x}$ profiles is displayed in Figure \eqref{fig:12b}, and it is seen that the behaviour of a given profile depends largely on its location with respect to the periodic cycle of the flow. For example, the number density at ${x} / R = 9$ is largely uniform for this cross-section of the droplet field, however the profile closer to the cylinder at ${x} / R = 3$ shows much higher variation as it passes through one of the vortices at this particular snapshot in time. The distribution and profiles of the average droplet radius $\overline{r}$ at $t = 30$ are shown in Figures \eqref{fig:12c} and \eqref{fig:12d} respectively, and evidence that larger droplets accumulate along the distinct curves between the vortices, with only smaller droplets existing in the vicinity of the vortices. As with the steady-state case in Figure \ref{fig:10}, the distribution and profiles of variance in droplet radius $\overline{r^{\prime}r^{\prime}}$ at $t = 30$ and illustrated in Figures \eqref{fig:12e} and \eqref{fig:12f} respectively are qualitatively similar to the behaviour of $\overline{r}$ in this case.

\section{Conclusions} \label{sec:conclusions}

This paper has presented a novel methodology for reconstructing the droplet number density field using the probability density along trajectories calculated by the generalised Fully Lagrangian Approach (gFLA; the Fully Lagrangian Approach taking into account the effect of droplet evaporation), and the technique of kernel regression to accumulate the contributions from individual droplets onto an Eulerian grid. This has been applied to a range of steady-state and transient flow configurations for both monodisperse and polydisperse droplets, and it has been demonstrated that the kernel regression procedure can reliably produce a smooth representation of the droplet probability density field and average field variables both in space and across the range of droplet sizes.

This approach has several advantages. The strongest benefit of kernel regression is making use of the meshfree nature of the method to define the kernel using the Jacobian tensor from the gFLA, so that individual droplet contributions are extrapolated to the surrounding Eulerian region in accordance with the local structure of the droplet number density field. This results in a consistent procedure that enables the retention of a high level of detail in the number density field from a relatively small sample of droplets, in contrast to a conventional Cloud-In-Cell approach which employs elementary averaging and therefore requires both more trajectories and a more frequent injection rate to reliably reproduce the number density field. It is demonstrated that kernel regression yields a stable result with $10^3$ times fewer droplet realisations than the CIC approach, and is therefore able to achieve the potential reduction in the computational cost required to obtain a smooth yet qualitatively accurate continuum representation for inertial droplet field variables. The informed use of data in such a way is becoming increasingly important, as the storage of datasets becomes impractical for highly resolved simulations due to the amount of information involved. Another consequence of being a meshfree method is that discontinuities in the droplet field are automatically dealt with by kernel regression, such as at the edge of the droplet region, and in this manner arbitrarily complex flows can be treated without requiring prior knowledge of the extent of the domain which is occupied by droplets.

Additionally, varying the size of the kernel in this way means that areas of sparse droplet number density can be effectively dealt with. Particle-based methods such as smoothed particle hydrodynamics require a minimum number of particles to satisfy the normalisation condition of the kernel, limiting their use to weakly compressible flows in which the number density of particles is roughly constant. In contrast, kernel regression satisfies the normalisation condition of the kernel regardless of how many particles are within the compact support, making the approach applicable to voids in which there are relatively few particles. Furthermore, specification of the kernel using the Jacobian provides a means of obtaining a smooth representation for regions containing few droplets, as the low number density associated with droplets in these regions results in the compact support of the kernel being proportionately large, and extending to cover the areas between sparsely distributed droplets.

Computationally, the procedure can be implemented by considering the contributions a given droplet makes to gridpoints within its compact support, rather than having to identify all the droplets which contribute at a given gridpoint. When the reconstruction is performed upon a Cartesian grid this avoids the need for a droplet search algorithm, and is therefore relatively efficient. It is also conceptually trivial to extend the kernel regression procedure to account for the droplet size distribution by extending the dimensionality of the kernel, and reconstructing the probability density field on a higher-dimensional grid.

The representation of the number density field produced by kernel regression is, however, a locally constant estimator, and is unable to predictively extrapolate beyond the range of number densities which are sampled along trajectories. This is a shortcoming that is also inherent in linear interpolation, and means that in areas of rapid variation in number density, kernel regression will not return the exact value of the maximum or minimum local number density unless there is a droplet at that point. Furthermore, by its nature kernel regression is a statistical procedure, and is able to achieve a smooth representation of the droplet field only at the expense of some loss of absolute accuracy. This generally also occurs in regions of rapid variation in number density, with the smoothing characteristic of kernel regression meaning that for a given smoothing length and number of droplets only a certain level of variation can be captured. In practice this is most restrictive at a low droplet inertia, as has been observed in this work for low $St$, whilst the accuracy remains good for higher values of $St$. In general, kernel regression requires far fewer droplets than conventional direct trajectory methods to achieve a smooth Eulerian representation of the droplet phase at a reasonable level of accuracy, however increasing the number of droplet trajectories that are sampled will improve the accuracy if needed.


The stability of kernel regression and its versatility across a variety of flow configurations clearly demonstrate the efficacy of this procedure, and its suitability for upscaling and application to more general spray systems. In particular, the ability to accurately determine the complete droplet size distribution and its associated statistics at any spatial location within the droplet field make this methodology relevant to the simulation of many industrial and environmental systems, in which detailed knowledge of the droplet behaviour is paramount to the control and optimisation of such processes.


\section*{Acknowledgments}

The authors are grateful to the UKRI Future Leaders Fellowship (Grant MR/T043326/1) for their financial support, and would also like to acknowledge Professor A. N. Osiptsov for his helpful suggestions.


\subsection*{Financial disclosure}

None reported.

\subsection*{Conflict of interest}

The authors declare no potential conflict of interests.

%
%
%

\appendix

\section{Verification of the Lagrangian droplet phase continuity equation for evaporating droplets} \label{sec:gfla-derivation} 

Consider droplets with position $\bm{x}_d(t)$, velocity $\bm{v}_d(t)$ and radius $r_d(t)$ for which the evolution of trajectories and evaporation are governed by Eqs.~\eqref{eq:part-rad-eom}. The general continuum representation of the droplet phase can then be described in terms of the one-particle distribution function $w (\bm{x},\bm{v},r,t)$ by the collisionless kinetic equation \cite{Osiptsov2000}
\begin{equation} \label{eq:kinetic-equation}
\frac{\partial w}{\partial t} = - \frac{\partial}{\partial \bm{x}} \cdot \left[ \bm{v} w \right] - \frac{\partial}{\partial \bm{v}} \cdot \left[ \bm{f} w \right] - \frac{\partial}{\partial r} \left[ \varphi w \right] \, .
\end{equation}
To obtain the droplet phase continuity equation in the subset of phase space given by $\boldsymbol{\xi} = (\bm{x},r)$, the droplet mean field variables defined as the velocity-averaged moments of $w (\bm{x},\bm{v},r,t)$ are considered. Specifically, the probability density $p (\bm{x},r,t)$ and average velocity $\overline{\bm{v}} (\bm{x},r,t)$ are defined as
\begin{subequations}
    \label{eq:velocity-averaged-moments}
    \begin{align}
        p (\bm{x},r,t) & = \int_{\bm{v}} w (\bm{x},\bm{v},r,t) \, d\bm{v} \, , \\
        \overline{\bm{v}} (\bm{x},r,t) & = \frac{1}{p (\bm{x},r,t)} \int_{\bm{v}} \bm{v} w (\bm{x},\bm{v},r,t) \, d\bm{v} \, .
    \end{align}
\end{subequations}
Then integrating the kinetic equation \eqref{eq:kinetic-equation} over $\bm{v}$ yields
\begin{equation} \label{eq:gfla-part-continuity}
\frac{\partial p}{\partial t} = - \frac{\partial}{\partial \bm{x}} \cdot \left[ \overline{\bm{v}} p \right] - \frac{\partial}{\partial r} \left[ \overline{\varphi} p \right] \, ,
\end{equation}
where the velocity derivative term in Eq.~\eqref{eq:kinetic-equation} vanishes as it is assumed that $w \rightarrow 0$ as $\bm{v} \rightarrow \pm\boldsymbol{\infty}$, and the velocity-averaged evaporation rate $\overline{\varphi} (\bm{x},r,t)$ is defined as
\begin{equation}
\overline{\varphi} (\bm{x},r,t) = \frac{1}{p (\bm{x},r,t)} \int_{\bm{v}} \varphi (\bm{x},\bm{v},r,t) w (\bm{x},\bm{v},r,t) \, d\bm{v} \, .
\end{equation}
Eq.~\eqref{eq:gfla-part-continuity} is the Eulerian form of the droplet phase continuity equation that governs the transport of $p (\bm{x},r,t)$. To proceed to the Lagrangian form of Eq.~\eqref{eq:gfla-part-continuity}, it is appropriate to consider the phase space trajectories $\bm{z}_d = ( \bm{x}_d, r_d )$ in $\boldsymbol{\xi}$ that evolve according to the average velocity $\overline{\bm{v}}$ and velocity-averaged evaporation rate $\overline{\varphi}$
\begin{subequations}
    \label{eq:part-eom-evap-pvf}
    \begin{align}
        \dot{\bm{x}}_d & = \overline{\bm{v}} ( \bm{x}_d, r_d, t ) \, , & \bm{x}_d (t_0) & = \bm{x}_0 \, , \\
        \dot{r}_d & = \overline{\varphi} ( \bm{x}_d, r_d, t ) \, , & r_d (t_0) & = r_0 \, .
    \end{align}
\end{subequations}
The corresponding system form of Eqs.~\eqref{eq:part-eom-evap-pvf} is then given by
\begin{align} \label{eq:part-eom-evap-averaged-phase-space}
    \dot{\bm{z}}_d & = \overline{\bm{S}} ( \bm{z}_d, t ) \, , & \bm{z}_d (t_0) & = \boldsymbol{\xi}_0 \, ,
\end{align}
where $\overline{\bm{S}} (\boldsymbol{\xi},t) = \left( \overline{\bm{v}} (\bm{x},r,t), \overline{\varphi} (\bm{x},r,t) \right)$. Thus Eq.~\eqref{eq:gfla-part-continuity} can be written in the phase space $\boldsymbol{\xi}$ as
\begin{equation} \label{eq:gfla-part-continuity-phase-space}
\frac{\partial }{\partial t} p (\boldsymbol{\xi},t) = - \frac{\partial}{\partial \boldsymbol{\xi}} \cdot \left[ \overline{\bm{S}} (\boldsymbol{\xi},t) p (\boldsymbol{\xi},t) \right] \, .
\end{equation}
The appropriate form of the Lagrangian derivative corresponding to motion along the phase space trajectories $\bm{z}_d$ is given by
\begin{equation} \label{eq:lagrangian-derivative}
\frac{D}{Dt} := \frac{\partial}{\partial t} + \overline{\bm{S}} (\boldsymbol{\xi},t) \cdot \frac{\partial}{\partial \boldsymbol{\xi}} \, ,
\end{equation}
whereupon Eq.~\eqref{eq:gfla-part-continuity-phase-space} can now be interpreted along the phase space trajectory $\bm{z}_d$ and written as
\begin{equation}
\frac{D}{Dt} p (\bm{z}_d,t) = - p (\bm{z}_d,t) \frac{\partial}{\partial \boldsymbol{\xi}} \cdot \overline{\bm{S}} (\bm{z}_d,t) \, .
\label{eq:part-continuity-phase-space-Lagrangian}
\end{equation}
Eq.~\eqref{eq:part-continuity-phase-space-Lagrangian} can be formally solved along the phase space trajectory $\bm{z}_d$, and along with the initial condition $p (\bm{z}_d (t_0),t_0) = p (\boldsymbol{\xi}_0,t_0)$ yields the solution
\begin{equation} \label{eq:Lagangian-number-density-solution-phase-space}
p (\bm{z}_d,t) = p (\boldsymbol{\xi}_0,t_0) \exp \left[ - \int_{t_0}^{t} \frac{\partial}{\partial \boldsymbol{\xi}} \cdot \overline{\bm{S}} (\bm{z}_d^{\prime}, t^{\prime}) \, d t^{\prime} \right] \, ,
\end{equation}
where $\bm{z}_d^{\prime} = \bm{z}_d (t^{\prime})$, and it is assumed that $p (\boldsymbol{\xi}_0,t_0) > 0$ in accordance with the probability density $p (\boldsymbol{\xi},t)$ being interpreted as a strictly positive quantity. Re-introducing the interpretations $\boldsymbol{\xi} = (\bm{x},r)$, $\overline{\bm{S}} = (\overline{\bm{v}}, \overline{\varphi})$ leads to
the solution in the familiar physical variables
\begin{equation} \label{eq:Lagrangian-number-density-solution}
p (\bm{x}_d,r_d,t) = p (\bm{x}_0,r_0,t_0) \exp \left[ - \int_{t_0}^{t} \frac{\partial}{\partial \bm{x}} \cdot \overline{\bm{v}} (\bm{x}_d^{\prime},r_d^{\prime},t^{\prime}) + \frac{\partial}{\partial r} \cdot \overline{\varphi} (\bm{x}_d^{\prime}, r_d^{\prime}, t^{\prime}) \, d t^{\prime} \right] \, ,
\end{equation}
where $\bm{x}_d^{\prime} = \bm{x}_d (t^{\prime})$ and $r_d^{\prime} = r_d (t^{\prime})$. Eq.~\eqref{eq:Lagrangian-number-density-solution} is the solution for the probability density $p (\bm{x}_d,r_d,t)$ as it evolves along trajectories. However, this assumes that droplet trajectories are governed by the velocity averaged Eqs.~\eqref{eq:part-eom-evap-pvf} through the velocity field $\overline{\bm{v}} (\bm{x},r,t)$ and evaporation rate $\overline{\varphi} (\bm{x},r,t)$. Now consider the Jacobian tensor defined by
\begin{equation} \label{eq:Jacobian-phase-space}
\bm{J} (\boldsymbol{\xi}_0,t) = \frac{\partial \bm{z}_d (t)}{\partial \boldsymbol{\xi}_0} \, .
\end{equation}
Then taking the partial derivative $\partial / \partial \boldsymbol{\xi}_0$ of the phase space trajectory $\bm{z}_d$ described by Eq.~\eqref{eq:part-eom-evap-averaged-phase-space}, the evolution of $\bm{J} (\boldsymbol{\xi}_0,t)$ is determined by
\begin{equation}
\label{eq:Jacobian-evolution-phase-space}
\dot{\bm{J}} (\boldsymbol{\xi}_0,t) = \frac{\partial \overline{\bm{S}}}{\partial \boldsymbol{\xi}} (\bm{z}_d,t) \cdot \bm{J} (\boldsymbol{\xi}_0,t) \, .
\end{equation}
Consider also the Jacobian determinant $\det \left( \bm{J} (\boldsymbol{\xi}_0,t) \right)$. Then Jacobi's formula from linear algebra states that
\begin{equation} \label{eq:Jacobis-formula}
\frac{d}{dt} \det \left( \bm{J} (\boldsymbol{\xi}_0,t) \right) = \det \left( \bm{J} (\boldsymbol{\xi}_0,t) \right) \, \text{tr} \left( \bm{J}^{-1} (\boldsymbol{\xi}_0,t) \cdot \dot{\bm{J}} (\boldsymbol{\xi}_0,t) \right) \, .
\end{equation}
Utilising the evolution equation \eqref{eq:Jacobian-evolution-phase-space} for $\bm{J} (\boldsymbol{\xi}_0,t)$ therefore leads to
\begin{align}
    & \quad \frac{d}{dt} \det \left( \bm{J} (\boldsymbol{\xi}_0,t) \right) \nonumber \\
    & = \text{tr} \left( \bm{J}^{-1} (\boldsymbol{\xi}_0,t) \cdot \frac{\partial \overline{\bm{S}}}{\partial \boldsymbol{\xi}} (\bm{z}_d,t) \cdot \bm{J} (\boldsymbol{\xi}_0,t) \right) \, \det \left( \bm{J} (\boldsymbol{\xi}_0,t) \right) \, , \nonumber \\
    & = \text{tr} \left( \bm{J} (\boldsymbol{\xi}_0,t) \cdot \bm{J}^{-1} (\boldsymbol{\xi}_0,t) \cdot \frac{\partial \overline{\bm{S}}}{\partial \boldsymbol{\xi}} (\bm{z}_d,t) \right) \, \det \left( \bm{J} (\boldsymbol{\xi}_0,t) \right) \, , \nonumber \\
    & = \text{tr} \left( \frac{\partial \overline{\bm{S}}}{\partial \boldsymbol{\xi}} (\bm{z}_d,t) \right) \, \det \left( \bm{J} (\boldsymbol{\xi}_0,t) \right) \, , \nonumber \\
    & = \frac{\partial}{\partial \boldsymbol{\xi}} \cdot \overline{\bm{S}} (\bm{z}_d,t) \, \det \left( \bm{J} (\boldsymbol{\xi}_0,t) \right) \, ,
    \label{eq:Jacobian-determinant-evolution-phase-space}
\end{align}
where the third line follows since the trace operator is invariant under cyclic permutations of products of arguments, and the final line is due to the trace of the gradient being equivalent to the divergence. Eq.~\eqref{eq:Jacobian-determinant-evolution-phase-space} can be formally solved along the phase space trajectory $\bm{z}_d$, which presents the general solution
\begin{equation} \label{eq:Jacobian-determinant-general-solution-phase-space}
\lvert \det \left( \bm{J} (\boldsymbol{\xi}_0,t) \right) \rvert = \lvert C \rvert \exp \left[ \int_{t_0}^{t} \frac{\partial}{\partial \boldsymbol{\xi}} \cdot \overline{\bm{S}} (\bm{z}_d^{\prime}, t^{\prime}) \, d t^{\prime} \right] \, ,
\end{equation}
in which no restriction is made upon the sign of the arbitrary constant $C$. Since $C$ is determined from the initial condition on $\det \left( \bm{J} (\boldsymbol{\xi}_0,t) \right)$, this allows for the case where $\det \left( \bm{J} (\boldsymbol{\xi}_0,t) \right)$ is negative to be accounted for in the description provided by Eq.~\eqref{eq:Jacobian-determinant-general-solution-phase-space}. Applying the general initial condition of $\det \left( \bm{J} (\boldsymbol{\xi}, t_0) \right) = 1$ (which follows from the definition of $\bm{J}$ in Eq.~\eqref{eq:Jacobian-phase-space} and the initial condition upon $\bm{z}_d$ in Eq.~\eqref{eq:part-eom-evap-averaged-phase-space} at time $t_0$) then yields
\begin{equation} \label{eq:Jacobian-determinant-solution-phase-space}
\lvert \det \left( \bm{J} (\boldsymbol{\xi}_0,t) \right) \rvert = \exp \left[ \int_{t_0}^{t} \frac{\partial}{\partial \boldsymbol{\xi}} \cdot \overline{\bm{S}} (\bm{z}_d^{\prime}, t^{\prime}) \, d t^{\prime} \right] \, .
\end{equation}
Re-introducing the interpretations $\xi = (\bm{x},r)$, $\overline{\bm{S}} = (\overline{\bm{v}}, \overline{\varphi})$ leads to the equivalent form in physical variables
\begin{equation}
\lvert \det \left( \bm{J} (\bm{x}_0,r_0,t) \right) \rvert = \exp \left[ \int_{t_0}^{t} \frac{\partial}{\partial \bm{x}} \cdot \overline{\bm{v}} (\bm{x}_d^{\prime}, r_d^{\prime}, t^{\prime}) + \frac{\partial}{\partial r} \cdot \overline{\varphi} (\bm{x}_d^{\prime}, r_d^{\prime}, t^{\prime}) \, d t^{\prime} \right] \, .
\label{eq:Jacobian-determinant-solution}
\end{equation}
Comparing the solutions for $p (\bm{x}_d,r_d,t)$ and $\det \left( \bm{J} (\bm{x}_0,r_0,t) \right)$ in Eqs.~\eqref{eq:Lagrangian-number-density-solution} and \eqref{eq:Jacobian-determinant-solution} respectively, it is seen that conservation of mass along trajectories in $(\bm{x},r)$ space can be expressed as
\begin{equation} \label{eq:Lagrangian-COM-equate}
\frac{p (\bm{x}_d,r_d,t)}{p (\bm{x}_0,r_0,t_0)} = \frac{1}{\lvert \det \left( \bm{J} (\bm{x}_0,r_0,t) \right) \rvert} \, ,
\end{equation}
from which the Lagrangian form of the continuity equation expressed in Eq.~\eqref{eq:Lagrangian-COM} immediately follows.

Note that since this formulation hinges upon use of droplet trajectories described by the velocity field $\overline{\bm{v}} (\bm{x},r,t)$, the resultant continuum description inherently assumes the invertibility of $\bm{J} (\bm{x}_0,r_0,t)$ at a given point in time, that is $\det \left( \bm{J} (\bm{x}_0,r_0,t) \right) \ne 0$. Physically this corresponds to the single-valuedness of droplet velocities at a given point $\bm{x}$, which in reality is only representative of fluid tracer droplets in the limit $St \rightarrow 0$. Consequently, this description is unable to qualitatively account for the crossing trajectories effect of sufficiently inertial droplets at finite $St$, for which occurrences of $\det \left( \bm{J} (\bm{x}_0,r_0,t) \right) = 0$ are observed along the envelope of folds in the droplet velocity field, and the probability density $p (\bm{x}_d,r_d,t)$ becomes singular. Similarly, $\det \left( \bm{J} (\bm{x}_0,r_0,t) \right)$ becomes negative once the associated trajectory crosses the envelope of a fold. However, these phenomena do not violate the physical correctness of Eq.~\eqref{eq:Lagrangian-COM} as a conservation law, and indeed do not preclude using values of $\det \left( \bm{J} (\bm{x}_0,r_0,t) \right)$ calculated from simulation data of inertial droplet trajectories. This can be understood by considering the geometric meaning of $\det \left( \bm{J} (\bm{x}_0,r_0,t) \right)$ as the volume scaling factor of the linear transformation described by the Jacobian $\bm{J} (\bm{x}_0,r_0,t)$, with the associated sign simply showing whether the transformation preserves or reverses orientation. In the context of calculating $p (\bm{x}_d,r_d,t)$ the orientation of the elemental volume is not relevant, and therefore using $\lvert \det \left( \bm{J} (\bm{x}_0,r_0,t) \right) \rvert$ to evaluate the probability density by means of Eq.~\eqref{eq:Lagrangian-COM} remains physically consistent in the case that $\det \left( \bm{J} (\bm{x}_0,r_0,t) \right) < 0$.

\section{Demonstration of the kernel domain size varying proportionately to the Jacobian determinant} \label{sec:kernel-domain-detJ}

In order to account for the high degree of variation in the number density field for inertial droplets, it is desired that reconstruction using a kernel-based method is able to vary the size of the kernel domain of support according to the local droplet number density. Within the FLA methodology, the physical interpretation of the Jacobian $\bm{J}^{\bm{x}\bm{x}}$ as the Eulerian-Lagrangian transformation provides the necessary information regarding the extent of the local Eulerian droplet field over which the influence of an individual droplet acts. It is therefore necessary to specify the kernel such that its domain of support is proportionate to $\lvert \det \left( \bm{J}^{\bm{x}\bm{x}} \right) \rvert$.

For the specific choice of the multivariate Gaussian kernel in Eq.~\eqref{eq:kernel-structured}, it can be shown that the domain size for a compact support is proportional to $\sqrt{\det (\bm{H})}$. Then for the positive semi-definite specification of $\bm{H} = h_0^2 \, \bm{J}^{\bm{x}\bm{x}} \cdot {\bm{J}^{\bm{x}\bm{x}}}^\top$ in Eq.~\eqref{eq:h-structured-kernel}, we have that
\begin{align}
    \sqrt{ \det (\bm{H})} = \sqrt{ \det \left( h_0^2 \, \bm{J}^{\bm{x}\bm{x}} \cdot {\bm{J}^{\bm{x}\bm{x}}}^\top \right) } 
    = \sqrt{ h_0^{2d} \det \left( \bm{J}^{\bm{x}\bm{x}} \right) \det \left( {\bm{J}^{\bm{x}\bm{x}}}^\top \right) } 
    = h_0^{d} \sqrt{ \det \left( \bm{J}^{\bm{x}\bm{x}} \right) \det \left( {\bm{J}^{\bm{x}\bm{x}}} \right) } 
    = h_0^{d} \lvert \det \left( \bm{J}^{\bm{x}\bm{x}} \right) \rvert \, .
\end{align}
Thus the domain of support for the kernel is proportionate to $\lvert \det \left( \bm{J}^{\bm{x}\bm{x}} \right) \vert$ as required, meaning that the kernel is able to adaptively scale in accordance with the local Eulerian-Lagrangian transformation so that the spatial structures of the droplet field, which accompany the variation in number density, are captured.

\bibliography{bibfile}%

%
%

\end{document}